\documentclass[12pt,epsfig]{report}
\usepackage[dvips]{graphics}
\usepackage{epsfig}
\usepackage{latexsym}
\begin{document}
\begin{center}
\pagenumbering{roman}
\thispagestyle{empty}
{\Huge \bf Some Studies in 
\vskip 0.3cm
Noncommutative Quantum Field}
\vskip 0.3cm
{\Huge \bf Theories}
\vskip 2.0 true cm
Thesis Submitted for the degree of\\
Doctor of Philosophy (Science)\\
of \\
WEST BENGAL  UNIVERSITY OF TECHNOLOGY
\vskip 3.0 true cm
 2008 \\
{\large \bf SUNANDAN GANGOPADHYAY}\\
S. N. Bose National Centre for Basic Sciences\\
JD Block, Sector III \\
Salt Lake City \\
Kolkata  700098\\
India
\end{center}
\newpage
\centerline{\large \bf CERTIFICATE FROM THE SUPERVISOR}
\noindent This is to certify that the thesis entitled
{\bf $``$Some Studies in Noncommutative Quantum Field Theories"}
submitted by  {\bf Sri Sunandan Gangopadhyay} who got his name
registered on {\bf 19.12.2005} for the award of Ph.D.(Science)
degree of West Bengal University of Technology, absolutely based upon his
own work under the supervision of {\bf Dr. Biswajit Chakraborty}
and that neither this thesis nor any part of it has been
submitted for any degree/diploma or any other academic
award anywhere before.\\
 
\noindent{\bf  Biswajit Chakraborty}\\
    Reader\\
    Department of Theoretical Sciences,\\
    S.N.Bose National Centre 
    for Basic Sciences, \\
    Salt Lake, Kolkata, India.

\newpage
\thispagestyle{empty}
.
\vskip 6.0cm
\begin{center}
{\Huge \bf  Dedicated}
\vskip 0.5 cm
{\Huge \bf to }
\vskip 0.5cm
{ \Huge \bf   My Parents}
\vskip 0.5 cm
{\Huge \bf and }
\vskip 0.5cm
{\Huge \bf   Grand Parents}
\end{center}
\newpage
\vskip 2.0cm
\begin{center}
{\large \bf ACKNOWLEDGEMENTS}
\end{center} 

\noindent With great pleasure, I express my deep sense of gratitude to my 
thesis advisor, Dr. Biswajit Chakraborty for his guidance and 
encouragement. I also acknowledge with sincere gratitude, 
the guidance I received from Prof. F.G. Scholtz 
and am grateful to him for always being
there to help me in my work. I record my sincere thanks 
to Dr. Biswajit Chakraborty for helping me throughout 
the course of this work and 
the timely completion of this thesis is a  
result of his support. It is also my duty and joy to thank the 
family of Dr. Chakraborty for their hospitality during my visits 
for professional discussions.
       
\noindent I am grateful to Prof. S. Dattagupta, 
ex. Director of Satyendra Nath Bose National Centre for
Basic Sciences (SNBNCBS), for giving me the opportunity to do research here.
I am thankful to Prof. R. Banerjee, Academic Programme Coordinator, 
SNBNCBS, for the academic help rendered to me.

\noindent I thank all the academic and administrative staff of SNBNCBS 
for helping
me in many ways. In particular, I am thankful to the Library staff
for the excellent assistance provided to me.
          
\noindent It is my pleasure to thank my friends Mr. Arindam Ghosh Hazra
and Mr. Anirban Saha 
who with their help and support in both academic
and personal matters made my research an experience I cherish much.

\noindent Finally and most importantly, I express my whole hearted gratitude 
to my family members. It is the love and unflinching
support of my parents and grand parents that enabled me to pursue 
this line of study which finally culminated in this thesis. 
I dedicate this thesis to them. 
\newpage
\tableofcontents
\chapter{Introduction and Overview}
\pagenumbering{arabic}
Noncommutative geometry \cite{connes} is presently one of the most
important areas of investigation. From a purely
mathematical point of view,  noncommutative geometry amounts 
to a program of unification
in mathematics under the aegis of the quantum apparatus, i.e. the theory of
operators and of $C^{*}$-algebras. There has been an explosion of 
intense research these days by some of the world's leading mathematicians,
and a variety of applications starting from the reinterpretation of the
phenomenological Standard Model of particle physics as a new spacetime 
geometry, to the quantum Hall effect, strings, renormalization and more in 
quantum field theory.
\vskip 0.003cm
\noindent The development of Noncommutative 
Quantum Field theories historically
starts with Heisenberg's observation (in a letter he wrote
to Pierls in the late 1930 \cite{pauli}) on the possibility of introducing
uncertatinty relations for coordinates, as a way to avoid singularities 
of the electron self energy. Pierls made use of these
ideas eventually in his work related to the Landau level problem.
Heisenberg also commented on this possibility to Pauli who then
involved Oppenheimer in the discussion \cite{pauli1}.
Finally it was Hartland Snyder, a student of Oppenheimer who first
formalised this idea in an article 
on {\it{Quantised Space time}} \cite{snyder} entirely devoted to this subject.
Almost immediately, C.N. Yang reacted to this paper and published a letter
to the Editor of the Physical Review \cite{yang} 
extending Snyder's treatment to the case of curved space (in
particular  de Sitter space). Then in 1948, Moyal addressed the problem
using Wigner phase-space distribution functions and he introduced
what is known as the Moyal star product, a noncommutative associative
product, in order to discuss the mathematical structure
of quantum mechanics \cite{moyal}.
For a simple classical system like a particle moving on a real line,
the construction of the star product can be motivated by considering
the set of Weyl ordered phase-space operators and its isomorphism
to the set of classical phase-space functions. (See \cite{zachos} for a
review.) This result has also been shown later through a geometric approach
by Berezin \cite{berezin}, Batalin and Tyutin \cite{batalin}. 
\vskip 0.003cm
\noindent The contemporary success of the renormalisation 
program shadowed these ideas
for a while. However, the ideas of noncommutative 
geometry were once again revived
in the 1980's by the mathematicians Connes, Woronowicz and Drinfel'd,
who generalised the notion of a differential structure to the 
noncommutative setting \cite{connes}. Just as it is possible
to give many differential structures to a given topological space, it
is possible to define many differential calculi over a given algebra.
Along with the introduction of a generalized 
integral \cite{connes1}, this permits
one in principle to define the action of a Yang-Mills 
field on a large class of noncommutative geometries.
\vskip 0.003cm
\noindent More concrete evidence for spacetime noncommutativity
came from string theory, which at present 
is arguably the most promising candidate
for a quantum theory of gravity. Strings having a finite intrinsic length
scale $l_{s}$, can be used as probes of short distance structure.
Hence, distances smaller than $l_{s}$ are not possible to observe.
In fact, based on the analysis of very high-energy
string scattering amplitudes \cite{venez, mende, amati}, 
string-modified Heisenberg uncertainty relations 
have been postulated in the form:
\begin{eqnarray}
\Delta x= \frac{\hbar}{2}\left(\frac{1}{\Delta p}+l_{s}^2\Delta p\right).
\label{O.1}
\end{eqnarray}
It is easy to see that one recovers the usual
quantum mechanical result in the limit $l_{s}\rightarrow 0$.
\vskip 0.003cm
\noindent The seminal paper of Seiberg and Witten 
\cite{switten} identified limits
in which the entire string dynamics can be described in terms of a
minimally coupled (supersymmetric) 
gauge theory on a noncommutative space. Their
analysis leads to an equivalence between ordinary gauge fields and 
noncommutative gauge fields, realized by a change of variables 
that can be described
explicitly. This change of variables 
(commonly known as the Seiberg-Witten (SW) map in the literature) is checked
by comparing the ordinary Dirac-Born-Infeld theory 
with its noncommutative counterpart.
\vskip 0.003cm
\noindent The central theme of this thesis is to study some
aspects of noncommutative quantum mechanics 
and noncommutative quantum field theory. We explore
how noncommutative structures can emerge 
and study the consequences of such structures
in various physical models. The outline of this thesis is as follows.
\vskip 0.003cm
\noindent $\bullet$ We present a review of noncommutative
quantum mechanics in chapter 2 where we discuss 
the procedure of Weyl quantization
which is an useful technique 
for translating an ordinary field
theory into a noncommutative one.
The Weyl operators are introduced and the Weyl-Wigner correspondence 
is derived. We then move on to present an alternative route to the
star product formalism following \cite{zachos}. 
\vskip 0.003cm
\noindent $\bullet$ In chapter 3, as a ``warm-up exercise", we demonstrate how 
noncommuting structures can be obtained in the first place by exploiting the
reparametrization symmetry of particle models. Studies has been going on for
some time in this 
direction and it has been observed that an 
important role in this context is played
by change of variables which provide a map 
among the commutative and noncommutative
structures. However, a precise underying principle
on which such maps are based was found to be missing. 
We have made a thorough study giving a 
systematic formulation of such maps, where they are essentially 
gauge/reparametrization transformations.
\vskip 0.003cm
\noindent $\bullet$ As we have mentioned earlier, the SW map has played
a central role in the analysis of noncommutative 
quantum field theories as it 
provides a map from the noncommutative 
to the commutative space, while preserving
the gauge invariance. On the other hand, 
issues related to the violation
of Lorentz symmetry in noncommutative 
relativistic systems have become important
and studies have been done using noncommutative 
variables or with their equivalent
commutative counterpart obtained by SW map.
\vskip 0.003cm
\noindent In chapter 4, we have carried out investigations in this line
by constructing an effective $U(1)$ gauge invariant theory for a 
noncommutative nonrelativistic model, 
where the Schr\"{o}dinger field is coupled to a 
$U(1)_{\star}$ gauge field in $2+1$-dimensions, using the first order
SW map. We study how this effective theory can be cast in the form of
usual Schr\"{o}dinger action with 
interaction terms of noncommutative  origin. We then explore the 
Galilean symmetry of the model in details and 
observe a violation of the above symmetry in our model. 
This violation is shown to be a noncommutative effect.
As an application of our effective model,
we have also computed the Hall conductivity and find that there is no
correction due to noncommutativity.
\vskip 0.003cm
\noindent $\bullet$ In chapter 5, we go through a detailed study
of noncommutative quantum mechanics. Here we carry out the
construction of a one parameter family of interacting noncommuting 
physically equivalent Hamiltonians (i.e. Hamiltonians having the
same spectrum). We have been
able to perform this construction exactly (to all orders in the 
noncommutative parameter $\theta$) and analytically 
in two dimensions for a free particle and
a harmonic oscillator in the presence of a constant magnetic field. We 
then investigate the implications
of the SW map in this context in details. Finally, we work out
an approximate duality between interacting commutative and weakly interacting 
noncommutative Hamiltonians for harmonic oscillator potentials.
\vskip 0.003cm
\noindent $\bullet$ In chapter 6, we take up the quantum Hall
system which has been an important area of application
of two dimensional noncommutative quantum systems.
Here, we discuss the role that interactions play 
in the noncommutative structure
that arises when the relative coordinates of two interacting particles are
projected onto the lowest Landau level. It is shown that the interactions
in general renormalize the noncommutative parameter away from the
non-interacting value $\frac{1}{B}$. The effective noncommutative parameter
is in general also angular momentum dependent. An heuristic argument,
based on the noncommutative coordinates, is given to find the filling
fractions at incompressibilty, which are in 
general renormalized by the interactions, 
and the results are consistent with known results 
in the case of singular magnetic fields.
\vskip 0.003cm
\noindent $\bullet$ The twist approach to noncommutative 
quantum field theory has recently gained a lot of popularity. 
As mentioned earlier, breaking of Lorentz invariance 
following from the choice of a 
particular noncommutative matrix $\theta$ 
have become important in noncommutative relativistic systems.
The twist approach was proposed as a way to 
circumvent this problem. 
It was triggered by the realization that it is 
possible to twist the coproduct of the universal 
envelope $U(\mathcal{P})$ of the Poincar\'e algebra, 
which is a Hopf algebra, 
such that it is compatible with the $\star$-product.
Two interesting consequences follow from the twisted
implementation of the Poincar\'e group. Firstly, the
IR/UV mixing is no longer there which implies that the
high and low energy sectors decouple, in contrast to
the untwisted formulation. The second important consequence
is an apparent violation of Pauli's exclusion principle.
\vskip 0.003cm
\noindent
In chapter 7, we show the twisted Galilean invariance of the noncommutative
parameter, even in presence of spacetime noncommutativity.
The deformed algebra of the Schr\"odinger field is then obtained in 
configuration and momentum space by studying the action of the 
twisted Galilean group on the nonrelativistic limit of the 
Klein-Gordon field and can be extended in a straightforward manner
for the Dirac field also. This deformed algebra is used to compute 
the two particle correlation function to study the possible 
extent to which the previously proposed violation of the Pauli 
principle may impact at low energies. It is concluded that any 
possible effect is probably well beyond detection at current energies.    
 
\vskip 0.003cm
\noindent $\bullet$
Finally, we end up with conclusions in chapter 8.


\newpage
\noindent This thesis is based on the following publications.
\begin{enumerate}
\item Seiberg-Witten map and Galilean symmetry violation in a 
noncommutative planar system 
 \cite{bcsgas}\\
B. Chakraborty, {\bf{S. Gangopadhyay}}, A. Saha,\\
{\it  Phys. Rev. D} {\bf{70}} (2004) 107707.
\item  Noncommutativity and reparametrization symmetry
\cite{rbbcsg}\\
R. Banerjee, B. Chakraborty, {\bf{S. Gangopadhyay}}, \\
{\it  J. Phys. A. }{\bf 38} (2005) 957. 
\item  Dual families of noncommutative quantum systems
\cite{fgsbcsgagh}\\
F.G. Scholtz, B. Chakraborty, {\bf{S. Gangopadhyay}}, A.G. Hazra, \\
{\it Phys. Rev. D} {\bf 71} (2005) 085005. 
\item  Interactions and noncommutativity in quantum Hall systems
\cite{qhepaper}\\
F.G. Scholtz, B. Chakraborty, {\bf{S. Gangopadhyay}}, J. Govaerts,\\
{\it J. Phys. A} {\bf 38} (2005) 9849.
\item Twisted Galilean symmetry and the Pauli principle
at low energies \cite{bcsgaghfgs}\\
B. Chakraborty, {\bf{S. Gangopadhyay}}, A.G. Hazra, F.G. Scholtz, \\
{\it J. Phys. A} {\bf 39} (2006) 9557.
\item  Lie algebraic  noncommuting structures from 
reparametrization symmetry
\cite{sgangopadhyay}\\
{\bf{S. Gangopadhyay}},\\
{\it J. Math. Phys} {\bf 48} (2007) 052302.
\end{enumerate}



\chapter{Review of Noncommutative Quantum Mechanics and Introduction
to Star product}


\section{Weyl quantization and Groenewold-Moyal product}
The idea behind spacetime noncommutativity
is very much inspired by the foundations of quantum mechanics.
Within the framework of canonical quantization, Weyl introduced an elegant
prescription for associating a quantum operator to a classical function of the
phase-space variables~\cite{weylbook}. 
This programme leads to a deep conceptual revolution because
the emphasis on group-theoretical methods provides a scheme where
Weyl systems can be considered in the first place \cite{sudarshan}
and classical mechanics is eventually recovered.
Further, this technique provides a systematic way
to describe noncommutative spaces in general and to study field theories
defined thereon. In this section we shall introduce this formalism which will
play a central role in most of our subsequent analysis. It is also
worthwhile to mention that
Weyl quantization works for very general type of
commutation relations\footnote{In the following section,
we have drawn freely from \cite{szaborj}. Some of the intermediate steps
in the derivation of the key results has been worked out in details.}.

\subsection{Weyl Operators}
Let us consider the commutative algebra of 
(possibly complex-valued) functions
on $D$-dimensional Euclidean space $\mathcal R^{D}$, with the usual
pointwise multiplication of functions defined as the product. 
We will assume that all fields defined
on $\mathcal R^{D}$ live in an appropriate Schwartz space of functions
of sufficiently
rapid decrease at infinity~\cite{rieffelschwartz}, 
i.e. those functions whose
derivatives vanish at 
infinity in both position and momentum
space to arbitrary order. This condition can be characterized, 
for example, by the requirements
\begin{equation}
\sup_{x}\,\Bigl(1+|x|^2\Bigr)^{k+n_{1}+\dots+n_D}\,
\Bigl|\partial_1^{n_1}\cdots
\partial_D^{n_D}f(x)\Bigr|^2~<~\infty
\label{Schwartzcondn}
\end{equation}
for every set of integers $k,n_i\in\mathcal{Z}_+$,
where $\partial_i=\partial/\partial
x^i$. In that case, the algebra of functions 
may be given the structure of a
Banach space by defining the $L^\infty$-norm
\begin{equation}
\|f\|^{~}_\infty=\sup_x\,\Bigl|f(x)\Bigr| \ .
\label{Linftynorm}
\end{equation}
The Schwartz condition also implies that any function $f(x)$ may be described
by its Fourier transform
\begin{equation}
\tilde f(k)=\int_{-\infty}^{+\infty} d^{D}x~f(x)e^{-ik_ix^i}
\label{Fourierf}
\end{equation}
with $\tilde f(-k)=\tilde f(k)^*$ whenever $f(x)$ is real-valued. 
We now define a
noncommutative space by replacing the local
coordinates $x^i$ of $\mathcal{R}^D$ by Hermitian operators 
$\hat x^i$ obeying the
commutation relations:
\begin{equation}
[\hat{x}^{i}, \hat{x}^{j}]=i\theta^{ij}.
\label{commrelation}
\end{equation}
The noncommutative algebra of operators is then generated by $\hat{x}^i$. 
A one-to-one correspondence between the algebra of fields on $\mathcal{R}^D$
and this ring of operators is provided by Weyl quantization, 
and it may be thought of as an analog of the operator-state
correspondence of local quantum field theory. 
Given the function $f(x)$ and its
corresponding Fourier coefficients (\ref{Fourierf}), one can 
introduce its {\it Weyl symbol} by
\begin{equation}
\hat{\mathcal W}[f]=\int_{-\infty}^{+\infty}\frac{d^{D}k}{(2\pi)^D}
\tilde{f}(k)e^{ik_i\hat x^i} 
\label{Weylopdef}
\end{equation}
where we have chosen the symmetric Weyl operator ordering prescription.
\vskip 0.003cm
\noindent For example, choosing $f(x)=e^{ik_{i}x^{i}}$, 
eq.(s) (\ref{Fourierf}) 
and (\ref{Weylopdef}) leads to:
\begin{eqnarray}
\hat{\mathcal W}[e^{ik_{i}x^{i}}]&=&\int_{-\infty}^{+\infty}
\frac{d^{D}k^{\prime}d^{D}y}{(2\pi)^D}e^{ik^{\prime}_{i}(\hat x^i-y^i)} 
e^{ik_i y^i}\nonumber\\
&=&\int_{-\infty}^{+\infty}d^{D}k^{\prime}e^{ik^{\prime}_{i}\hat x^i}
\delta^{(D)}(k_{i}-k^{\prime}_{i})\nonumber\\
&=&e^{ik_{i}\hat{x}^{i}}.
\label{exam}
\end{eqnarray}
Note that the Weyl operator $\hat{\mathcal{W}}[f]$
is Hermitian if $f(x)$ is real-valued.
Using eq.(\ref{Fourierf}), one can write eq.(\ref{Weylopdef}) 
in terms of an explicit map $\hat\Delta(x)$
between operators and fields to get
\begin{eqnarray}
\hat{\mathcal W}[f]=\int_{-\infty}^{+\infty} d^Dx~f(x)\,\hat\Delta(x)
\label{WeylDelta}
\end{eqnarray}
where,
\begin{eqnarray}
\hat\Delta(x)&=&\int_{-\infty}^{+\infty}
\frac{d^{D}k}{(2\pi)^D}e^{ik_i(\hat x^i-x^i)}.
\label{Deltadef}
\end{eqnarray}
The operator (\ref{Deltadef}) is Hermitian,
$\hat\Delta(x)^\dagger=\hat\Delta(x)$, and it describes a mixed basis for
operators and fields on spacetime. In this way we may interpret the field
$f(x)$ as the coordinate space representation of the Weyl operator 
$\hat{\mathcal W}[f]$.
Note that in the commutative case $\theta^{ij}=0$, the map (\ref{Deltadef})
reduces trivially to a delta-function $\delta^D(\hat x-x)$ and
$\hat{\mathcal W}[f]|_{\theta=0}=f(\hat x)$. But generally, by the
Baker-Campbell-Hausdorff (BCH) formula, for $\theta^{ij}\neq0$ it is a highly
non-trivial field operator.\\
To proceed further, we now introduce ``derivatives'' of operators 
through an anti-Hermitian linear
derivation $\hat\partial_{i}$ defined by the commutation relations
\begin{eqnarray}
\left[\hat\partial_i\,,\,\hat x^j\right]=\delta_i^{~j}~~~~~~,~~~~~~\left[\hat
\partial_i\,,\,\hat\partial_j\right]=0.
\label{derivdef}
\end{eqnarray}
Then after a little algebra, it is straightforward to show that
\begin{eqnarray}
\left[\hat\partial_i\,,\,\hat\Delta(x)\right]=-\partial_i\,\hat\Delta(x) 
\label{partialDelta}
\end{eqnarray}
which upon integration by parts in eq.(\ref{WeylDelta}) leads to
\begin{eqnarray}
\left[\hat\partial_{i}\,,\,\hat{\mathcal W}[f]\right]&=&
\int_{-\infty}^{+\infty} d^Dx~\partial_{i}f(x)\,\hat\Delta(x)=
\hat{\mathcal W}[\partial_{i}f].
\label{integration}
\end{eqnarray}
Now using eq.(s) (\ref{Deltadef}), (\ref{derivdef}), 
(\ref{partialDelta}) and the 
BCH-formula
\begin{eqnarray}
e^{\hat{A}}e^{\hat{B}}=e^{\hat{A}+\hat{B}+\frac{1}{2}[\hat{A}, \hat{B}]},~
[\hat{A}, \hat{B}]=c
\label{bch}
\end{eqnarray}
(where, $c$ is a number) we find that the computation of 
$e^{v^i\hat\partial_{i}}\,\hat\Delta(x)\,e^{-v^i\hat\partial_i}$ leads to: 
\begin{eqnarray}
e^{v^i\hat\partial_{i}}\,\hat\Delta(x)\,e^{-v^i\hat\partial_i}&=&
\int_{-\infty}^{+\infty}\frac{d^{D}k}{(2\pi)^D}
e^{v^{i}\hat{\partial}_{i}}e^{ik_{i}\hat{x}^i}
e^{-v^{j}\hat{\partial}_{j}}e^{-ik_{i}x^i}
\nonumber\\
&=&\int_{-\infty}^{+\infty}\frac{d^{D}k}{(2\pi)^D}
e^{ik_{i}[\hat{x}^i-(x^i-v^i)]}
\nonumber\\
&=&\hat\Delta(x+v).
\label{Deltatransl}
\end{eqnarray}
Eq.(\ref{Deltatransl}) implies that translation generators can be
represented by unitary operators $e^{v^{i}\hat\partial_{i}}$ 
($v\in{\mathcal R}^D$).
The property (\ref{Deltatransl}) also 
implies that any cyclic trace tr defined on
the algebra of Weyl operators has the feature that $\,\hat\Delta(x)$ is
independent of $x\in{\mathcal R}^D$. 
From eq.(\ref{WeylDelta}) it follows that the trace
tr is uniquely given by an integration over spacetime
\begin{eqnarray}
tr~\hat{\mathcal W}[f]=\int_{-\infty}^{+\infty} d^Dx~f(x) 
\label{Tracedef}
\end{eqnarray}
where we have chosen the normalization $tr\hat\Delta(x)=1$. In this sense,
the operator trace tr is equivalent to integration over the noncommuting
coordinates $\hat x^i$. \vskip 0.003cm
\noindent With the above results at our hands, we compute 
the products of operators $\hat\Delta(x)$ 
at distinct points as follows. To begin with, let us 
observe that the BCH-formula (\ref{bch}) yields:
\begin{eqnarray}
e^{ik_i\hat x^i}~e^{ik_j'\hat x^j}=e^{-\frac i2\,\theta^{ij}k_ik_j'}
e^{i(k+k')_i\hat x^i}. 
\label{BCH}
\end{eqnarray}
This along with eq.(\ref{Deltadef}), leads to:
\begin{eqnarray}
\hat\Delta(x)\hat\Delta(y)&=&\int_{-\infty}^{+\infty}\frac{d^{D}k}{(2\pi)^D}
\frac{d^{D}k^{\prime}}{(2\pi)^D}e^{ik_{i}(\hat{x}^{i}-x^{i})}
e^{ik_{j}^{\prime}(\hat{x}^{j}-y^{j})}\nonumber\\
&=&\int_{-\infty}^{+\infty}\frac{d^{D}k}{(2\pi)^D}
\frac{d^{D}k^{\prime}}{(2\pi)^D}e^{i(k_{i}+k_{i}^{\prime})\hat{x}^{i}}
e^{-\frac{i}{2}\theta^{ij}k_{i}k_{j}^{\prime}}
e^{-ik_{i}x^{i}-ik_{i}^{\prime}y^{i}}\nonumber\\
&=&\int_{-\infty}^{+\infty}\frac{d^{D}k d^{D}k^{\prime}}{(2\pi)^{2D}}
\left[\int_{-\infty}^{+\infty}d^{D}z e^{i(k_{i}+k_{i}^{\prime})z^i}
\hat{\Delta}(z)\right]e^{-\frac{i}{2}\theta^{ij}k_{i}k_{j}^{\prime}}
e^{-ik_{i}x^{i}-k_{i}^{\prime}y^{i}}\nonumber\\
&=&\int_{-\infty}^{+\infty}\frac{d^{D}z}{(2\pi)^{2D}}\hat{\Delta}(z)
\int_{-\infty}^{+\infty}d^{D}k e^{ik_{i}(z^{i}-x^{i})}
\int_{-\infty}^{+\infty}d^{D}k^{\prime}
e^{ik_{j}^{\prime}(z^{j}-y^{j})}
e^{-\frac{i}{2}\theta^{ij}k_{i}k_{j}^{\prime}}\nonumber\\
&=&\int_{-\infty}^{+\infty}\frac{d^{D}z}{(2\pi)^{D}}\hat{\Delta}(z)
\int_{-\infty}^{+\infty} d^{D}k e^{ik_{i}(z^{i}-x^{i})}\delta^{(D)}
(\frac{1}{2}k_{i}\theta^{ij}-a^{j})
\label{new3}
\end{eqnarray}
where in the third line, we have used
\begin{eqnarray}
e^{ik_{i}\hat{x}^{i}}&=&\int_{-\infty}^{+\infty}
d^{D}z \hat{\Delta}(z)e^{ik_{i}z^{i}}.
\label{new3a}
\end{eqnarray}
If $\theta$ is an invertible matrix (this necessarily requires that the
spacetime dimension $D$ be even), then the
delta function integration over the momentum $k$ in eq.(\ref{new3}) 
can be explicitly carried out to get
\begin{eqnarray}
\hat\Delta(x)\,\hat\Delta(y)=\frac1{\pi^D|\det\theta|}\,
\int_{-\infty}^{+\infty}
d^Dz~\hat\Delta(z)~\exp^{-2i(\theta^{-1})_{ij}(x-z)^i(y-z)^j} \ .
\label{Delta2prod}
\end{eqnarray}
It follows from eq.(\ref{Delta2prod}), by the use of 
the trace normalization and the antisymmetry of
$\theta^{-1}$,
that the operators
$\hat\Delta(x)$ (for $x\in{\mathcal R}^D$) form an orthonormal set
\begin{eqnarray}
tr~\Bigl(\hat\Delta(x)\,\hat\Delta(y)\Bigr)=\delta^D(x-y) \ .
\label{Deltaortho}
\end{eqnarray}
This, along with eq.(\ref{WeylDelta}), implies that the transformation
$f(x)\stackrel{\hat\Delta(x)}{\longmapsto}\hat{\mathcal W}[f]$ 
is invertible with inverse given by:
\begin{eqnarray}
f(x)=tr~\Bigl(\hat{\mathcal W}[f]\,\hat\Delta(x)\Bigr) \ .
\label{inverseDelta}
\end{eqnarray}
The function $f(x)$ obtained in this way from a quantum operator is usually
called a {\it Wigner distribution function}~\cite{wigner}. 
Therefore, the map $\hat\Delta(x)$ provides 
a one-to-one correspondence between Wigner fields and
Weyl operators\footnote{An explicit formula for eq.(\ref{Deltadef}) in
terms of parity operators can be found in \cite{roye, JMGBContemp}.}. 
This is usually referred in the literature 
as the {\it Weyl-Wigner correspondence}.

\subsection{The Star-Product}

We are now in a position to derive the form of the star product.
We begin by considering the product of two Weyl 
operators $\hat{\mathcal W}[f]$ 
and $\hat{\mathcal W}[g]$
corresponding to functions $f(x)$ and $g(x)$. 
From eq.(s) (\ref{WeylDelta}),
(\ref{Delta2prod}) and (\ref{Deltaortho}) 
it follows that the coordinate space
representation of their product 
can be written (for invertible $\theta$) as
\begin{eqnarray}
tr~\Bigl(\hat{\mathcal W}[f]\,\hat{\mathcal W}[g]\,\hat\Delta(x)\Bigr)
&=&tr\left[\int_{-\infty}^{+\infty}d^Dy~d^Dz~\hat\Delta(y)\hat\Delta(z)
f(y)\,g(z)\hat\Delta(x)\right]
\nonumber\\
&=&\int_{-\infty}^{+\infty}\frac{d^Dy~d^Dz~d^Dw}{\pi^D|det\theta|}
f(y)\,g(z)\exp^{-2i(\theta^{-1})_{ij}(y-w)^i(z-w)^j}
tr(\hat\Delta(w)\hat\Delta(x))\nonumber\\
&=&\frac1{\pi^D|\det\theta|}\,
\int_{-\infty}^{+\infty} d^Dy~d^Dz~f(y)\,g(z)~
\exp^{-2i(\theta^{-1})_{ij}(x-y)^i(x-z)^j}\nonumber\\
&=&\int_{-\infty}^{+\infty}\frac{d^Dk~d^Dk^{\prime}}{(2\pi)^D}\tilde{f}(k)
\tilde{g}(k^{\prime})e^{ik_{i}x^{i}}e^{-\frac{i}{2}
\theta^{ij}k_{i}k_{j}^{\prime}}e^{ik^{\prime}_{j}x^{j}}\nonumber\\
&=&f(x)e^{\frac{i}{2}\stackrel{\leftarrow}{\partial}_{i}\theta^{ij}
\stackrel{\rightarrow}{\partial}_{j}}g(x)\nonumber\\
&\equiv& (f\star g)(x)
\label{Weylprodcoord}
\end{eqnarray}
where we have used 
eq.(s) (\ref{Fourierf}), (\ref{Weylopdef}) and (\ref{BCH}) 
and introduced the {\it Groenewold-Moyal
star-product}~\cite{moyal}.
On the other hand 
\begin{eqnarray}
tr~\Bigl(\hat{\mathcal W}[f\star g]\hat\Delta(x)\Bigr)
&=&tr\left[\int_{-\infty}^{+\infty}d^{D}z \hat\Delta(z)
(f\star g)(z)\hat\Delta(x)\right]\nonumber\\
&=&\int_{-\infty}^{+\infty}d^{D}z (f\star g)(z)~tr~\Bigl(\hat\Delta(z)
\hat\Delta(x)\Bigr)\nonumber\\
&=&\int_{-\infty}^{+\infty}d^{D}z (f\star g)(z)\delta^{(D)}(z-x)\nonumber\\
&=& (f\star g)(x).
\label{Weyls1}
\end{eqnarray}
From eq.(s) (\ref{Weylprodcoord}) and (\ref{Weyls1}), we finally obtain
the celebrated {\it{Weyl-Wigner correspondence}}
\begin{eqnarray}
\hat{\mathcal W}[f]\,\hat{\mathcal W}[g]=\hat{\mathcal W}[f\star g]. 
\label{Weylstar}
\end{eqnarray}

\section{Another approach to star product formalism}
\label{starprod}
\label{mcs1}
\begin{sloppypar}
In this section we present an alternative approach to 
the basic ideas of star product formalism
essentially following \cite{zachos}.
We consider the case of a particle moving on a real line ${\cal{R}}^1$ as
an illustrative example.
Clearly the classical phase-space ($x, p$) is the two dimensional space 
${\cal{R}}^2$.  
An arbitrary phase-space function $f(x, p)$ can  be written as
\begin{eqnarray}
f(x, p)&=& \int_{-\infty}^{+\infty} dx^{\prime} d p^{\prime} \delta (x - 
x^{\prime}) \delta (p - p^{\prime})  f(x^{\prime} , p^{\prime})\nonumber\\ 
&=&  \frac{1}{(2\pi)^2}\int_{-\infty}^{+\infty} dx^{\prime} d p^{\prime} 
d\tau d\sigma e^{i[\tau (
x -x^{\prime}) + \sigma (p - p^{\prime})]} f(x^{\prime} , p^{\prime})
\label{bat1} 
\end{eqnarray}
where the integral representation 
\begin{equation}
\delta (x - x^{\prime}) =  \frac{1}{2\pi} \int_{-\infty}^{+\infty} d\tau
e^{i\tau (x -x^{\prime})}
\label{bat2}
\end{equation} 
of the Dirac delta function $\delta (x -x^{\prime})$ and a similar 
representation for $ \delta (p - p^{\prime})$ are used. 
At the quantum level, the operator analogues $\hat{x}, \hat{p}$ of $x, p$ 
obey the Heisenberg-Weyl Lie algebra 
\begin{equation}
[\hat{x}, \hat{p}] = i\hbar \quad, \quad [\hat{x}, \hat{x}] = 0\quad,
\quad [\hat{p}, \hat{p}] = 0 
\label{bat3}
\end{equation}
and exp$[i(\tau \hat{x} + \sigma \hat{p} )]$ 
is a particular element of the
corresponding Lie group.
\vskip 0.003cm
\noindent As we have seen in the earlier section, 
Weyl's  prescription \cite{weylbook} 
for arriving at the operator $\hat{f}(\hat{x}, \hat{p})$ 
corresponding to the phase-space function $f(x, p)$ 
(taken to have a polynomial form)  
consists of rewriting eq.(\ref{bat1}) with the replacements 
$x \rightarrow \hat{x}, p \rightarrow \hat{p}$ to get:
\begin{equation} 
\hat{f}(\hat{x},\hat{p}) = \frac{1}{(2\pi)^2}\int_{-\infty}^{+\infty} 
dx^{\prime} d p^{\prime}
d\tau d\sigma e^{i[\tau (
\hat{x} -x^{\prime}) + \sigma (\hat{p} - p^{\prime})]} 
f(x^{\prime} , p^{\prime}).
\label{bat4}
\end{equation} 
An equivalent prescription due to Batalin and Tyutin 
\cite{batalin} is to define\footnote{The choice of the 
origin $x=p=0$ for evaluating 
$\hat{f}(\hat{x},\hat{p})$ is not mandatory. 
The operator $\hat{f}(\hat{x},\hat{p})$, 
evaluated at different points, 
are in fact related by canonical transformations \cite{batalin}.}
\begin{equation}
\hat{f}(\hat{x},\hat{p}) = e^{[\hat{x} \partial_x + 
\hat{p}\partial_p ]} f(x, p)|_{x=p=0}~.
\label{4+1}
\end{equation}
We however continue with the prescription (\ref{bat4}) 
for the time being.
\vskip 0.003cm
\noindent Now using the mapping (\ref{bat4}), one can obtain the
phase-space function (also called the classical kernel) 
of the  operator product $\hat{f}\hat{g}$ of two
phase-space operators $\hat{f}$ and $\hat{g}$ from the corresponding kernels
$f$ and $g$ respectively. For that one has to express $\hat{g}(\hat{x}, 
\hat{p})$ just in the manner of $\hat{f}$ in eq.(\ref{bat4}). 
One can then write
\begin{eqnarray}
\hat{f}\hat{g}&=&
\frac{1}{(2\pi)^4} \int_{-\infty}^{+\infty} {d\xi d\eta d\xi' d\eta'
dx' dx'' dp' dp''}   f(x',p') g(x'',p'')\nonumber\\
&&\hskip 2.8cm \times\exp i(\xi ( {\hat{ p}}-p')+\eta ( {\hat{ x}}-x'))
\exp i(\xi' ( {\hat{ p}}-p'')+\eta' ( {\hat{ x}}-x''))\quad
\quad\quad\quad\quad\quad\quad\quad\quad \nonumber \\
&=&\frac{1}{(2\pi)^4}\int_{-\infty}^{+\infty}d\xi~d\eta~d\xi^{\prime}
~d\eta^{\prime}~
dx^{\prime}~dx^{\prime\prime}~dp^{\prime}~dp^{\prime\prime} 
~f(x',p') g(x'',p'')
\exp i\left( (\xi +\xi')  {\hat{ p}}+(\eta+\eta')  {\hat {x}}\right)
\nonumber \\
&&\hskip 3.2cm\times ~\exp\left(-\xi p'-\eta x'-\xi'
p''-\eta'x'' +{\hbar\over 2} (\xi\eta'-\eta\xi') \right).
\label{bat5}
\end{eqnarray}
Changing integration variables to
\begin{equation}
\xi'\equiv {2\over \hbar} (x-x'), \quad
\xi\equiv \tau-  {2\over \hbar} (x-x'), \quad
\eta'\equiv {2\over \hbar} (p'-p), \quad
\eta\equiv \sigma- {2\over \hbar} (p'-p)
\label{bat6}
\end{equation} 
reduces the above integral to
$$ 
{\hat{ f}}{\hat{g}}= \frac{1}{(2\pi)^2}\int_{-\infty}^{+\infty}
d\tau d\sigma
dx~dp~\exp i \left(\tau ( {\hat{ p}}-p)+\sigma ( {\hat{x}}-x)\right)$$
\begin{equation} 
\times
\left\{\int_{-\infty}^{+\infty} dp'
dp''  dx' dx''  ~f(x',p')~g(x'',p'')
~\left[{1\over  (\pi \hbar)^2 }\exp \left(\frac{-2i}{\hbar}
\left( p(x'-x'') + p'(x''-x)+p''(x-x') \right )\right)\right]\right\}.
\label{bat7}
\end{equation}
We consider now the exponential 
inside the square bracket in the above equation:  
$$
 {1\over  (\pi \hbar)^2 }\exp \left(\frac{-2i}{\hbar}
\left( p(x'-x'') + p'(x''-x)+p''(x-x') \right )\right) $$
$$ ={1\over  (\pi \hbar)^2 } \exp \left(\frac{i}{\hbar}
\left(- 2(p' - p)(x''- x) + 2(x' - x)(p''- p) \right) \right) $$
$$ = \frac{1}{(2\pi)^2}\int_{-\infty}^{+\infty}
d\lambda~d\mu~\delta (x' - x - {\mu \hbar \over 2}) \delta (p' - p + 
{\lambda \hbar \over 2}) \exp \left({i} \left(\lambda (x''-x) + \mu (p''- p)
\right )\right) $$ 
$$ =~~~~\frac{1}{(2\pi)^4} \int_{-\infty}^{+\infty}
d\lambda~d\mu~d\alpha~d\beta~\exp i[\alpha(x' - x) + \beta (p' - p)]
\exp {i \hbar \over 2}{(\stackrel{\leftarrow }
{\partial}_x \stackrel{\rightarrow }
{\partial }_{p}-\stackrel{\leftarrow }{\partial }_{p}
\stackrel{\rightarrow }{\partial }_{x})}$$ $$ \times\exp \left({i}
\left(\lambda (x''-x) + \mu (p''- p)
\right )\right) $$
where the representation (\ref{bat2}) is used. 
With the aid of the above relation,
the integral in the curly bracket  in eq.(\ref{bat7}) can be written as
$$
{1 \over ({2\pi})^4
} \int_{-\infty}^{+\infty} d\lambda d\mu d\alpha d\beta d x' d p' dx'' dp''  
 \exp i[\alpha(x' - x) + \beta (p' - p)] 
\exp i\hbar{(\stackrel{\leftarrow }{\partial}_x \stackrel{\rightarrow }
{\partial }_{p}-\stackrel{\leftarrow }{\partial }_{p}
\stackrel{\rightarrow }{\partial }_{x})/2} $$
$$\times \exp \left({i}
 \left(\lambda (x''-x) + \mu (p''- p)
\right )\right)
~f(x',p')~g(x'',p'')$$
\begin{equation}
=  f(x,p) ~e^{{i \hbar \over 2}(\stackrel{\leftarrow }{\partial }_{x}
\stackrel{\rightarrow }{\partial }_{p}-\stackrel{\leftarrow }{\partial }_{p}
\stackrel{\rightarrow }{\partial }_{x})/2}~ g(x,p).
\label{bat8}
\end{equation} 
Hence the composition rule is given by: 
\begin{equation} 
{\hat{ f}}{\hat{g}}= \frac{1}{(2\pi)^2}\int_{-\infty}^{+\infty}d\tau d\sigma
dx dp \exp [i \left(\tau ( {\hat{ p}}-p)+\sigma ( {\hat{x}}-x)
\right)]
(f\star g)(x, p)
\label{bat9}
\end{equation}
where the  $\star$ product  is defined as   
\begin{equation}
f(x,p) \star g(x,p) \equiv f(x,p) ~e^{{i\hbar \over 2}
(\stackrel{\leftarrow }{\partial }_{x}
\stackrel{\rightarrow }{\partial }_{p}-\stackrel{\leftarrow }{\partial }_{p}
\stackrel{\rightarrow }{\partial }_{x})}~ g(x,p).
\label{10}
\end{equation}
Thus the fact that $f(x,p) \star g(x,p)$ is the  the classical kernel
of $\hat{f}\hat{g}$ has been established \cite{groenewold}. 
\end{sloppypar}

\chapter{Noncommutative structures from Reparametrization symmetry}
In the previous chapter, we have discussed the 
basic formalism on which the foundations of noncommutative quantum field 
theory is based. Now we shall first demonstrate how noncommutative
space-space or (spacetime) structures can arise 
from reparametrization symmetry
of particle models.
Investigations in this line has been carried out recently 
in simple particle models and it has been observed that noncommutative
structures emerge by 
suitable change of variables providing 
a map among the commutative and noncommutative structures 
\cite{np}, \cite{dh}, \cite{pahorvathy}, \cite{pin}. 
However, these studies lack a precise underlying principle 
on which such maps are based. 
One of the motives of our work is 
to provide a systematic formulation of such maps 
\cite{rbbcsg}. 
In the models that we discuss here, these maps are essentially 
gauge/reparametrization transformations.
\vskip 0.003cm
\noindent  To start with, we first consider the case of the
nonrelativistic (NR) free particle in details. 
Interestingly, even though the model 
does not have any natural reparametrization symmetry, 
we can introduce it by hand and then exploit it in order
to reveal the various noncommuting structures. 
As other examples, we consider the 
free relativistic particle as well as its interaction with a 
background electromagnetic field.
\vskip 0.003cm
\noindent The methodology that we adopt is to 
utilize the reparametrization invariance of the model
to find a non-standard gauge 
in which the spacetime and/or space-space coordinates become noncommuting. 
We also show that the variable redefinition relating the nonstandard and 
standard gauges is a gauge transformation.\\
\noindent The structure of the angular momentum operator is then
studied in some details.
A gauge independent expression is obtained, which therefore does not 
require
any central extension in the non-standard gauge. 
\vskip 0.003cm
\noindent Another important point to note is that the structures 
that we obtain are Lie-algebraic in the case of the 
NR free particle, but not so in 
its relativistic counterpart. However, in \cite{sgangopadhyay} we have
shown that there exists some special choice of the
reparametrization parameter for which one can obtain noncommuting
space-space structures falling in
the Lie-algebraic category even in the relativistic case.
We emphasize that these Lie-algebraic structures
may be useful in giving explicit forms of the
star products and SW maps (discussed in \cite{madore})
by reading off the structure constants of the algebra.\\
\noindent Moreover, there exists solutions of $\epsilon$ for which
the noncommutativity between spatial coordinates vanish, but the spacetime
algebra still remains noncommutative.
\vskip 0.003cm
\noindent Finally, there are two appendices in this chapter. In appendix A, 
we demonstrate the connection between
Dirac brackets (DB) in the axial and radiation gauges using suitable gauge
transformations. In appendix B, we show using the symplectic formalism,
the connection between integral curves and the 
equations of motion in the time
reparametrized version. 
This also indicates how constraints come into picture
naturally in the time-reparametrized formulation.
\section{Particle models}
Let us start from the action for a point particle in classical mechanics
\begin{equation}
S[x(t)] = \int_{t_{1}}^{t_{2}} dt L\left(x, \dot{x}\right)
\quad; \quad \dot{x} = \frac{dx}{dt}~ .
\label{z1}
\end{equation}
It is easy to rewrite the above form of the action in a time-reparametrized 
invariant form by elevating the status of time $t$ 
to that of an additional variable,
along with $x$, in the configuration space as
\begin{equation}
S[x(\tau), t(\tau) ] = \int_{\tau_{1}}^{\tau_{2}} 
d\tau \dot{t} L\left(x, \frac{\dot{x}}{\dot{t}}\right) 
= \int_{\tau_{1}}^{\tau_{2}} d\tau L_{\tau}
\left(x, \dot{x}, t, \dot{t}\right)
\label{ncr2}
\end{equation}
where,
\begin{equation}
L_{\tau}(x, \dot{x}, t, \dot{t}) = \dot{t}L\left(x, \frac{\dot{x}}{\dot{t}}
\right)  \quad;\quad \dot{t} = \frac{dt}{d\tau}  
\label{Z3}
\end{equation}
and $\tau$ is the new evolution parameter which can be taken to be an 
arbitrary monotonically increasing function of time $t$.
The canonically conjugate momenta corresponding 
to the coordinates $t$ and $x$ 
are given by:
\begin{eqnarray}
p_{t} &=& \frac{\partial L_{\tau}}{\partial \dot{t}} =  
L\left(x, \frac{\dot x}{\dot t}\right) + \dot{t}
\frac{\partial L\left(x, \frac{\dot x}{\dot t}\right)}{\partial \dot t}
\nonumber\\
&=& L\left(x, \frac{dx}{dt}\right) - \frac{dx}{dt}
\frac{\partial L(x, dx/dt)}{\partial (dx/dt)} = -H
\label{Z4}
\end{eqnarray}
\begin{equation}
p_{x} = \frac{\partial L_{\tau}}{\partial \dot{x}}~. 
\label{Z5}
\end{equation}
Now for a time-reparametrized theory, 
the canonical Hamiltonian (using eq.(s) (\ref{Z4}, \ref{Z5})) vanishes:
\begin{equation}
H_{\tau} = p_{t}\dot{t} + p_{x}\dot{x} - L_{\tau} = \dot{t}(H + p_{t}) = 0.
\label{N6}
\end{equation}
As a particular case of eq.(\ref{z1}), we start from the action of a 
free NR particle in one dimension
\begin{equation}
S = \int dt \frac{1}{2}m\left(\frac{dx}{dt}\right)^{2}~.
\label{N1}
\end{equation}
We rewrite the above form of the action in a time-reparametrized 
invariant form as in eq.(\ref{ncr2}):
\begin{equation}
S = \int d\tau L_{\tau}(x, \dot{x}, t, \dot{t})
\label{N2}
\end{equation}
where,
\begin{equation}
L_{\tau}(x, \dot{x}, t, \dot{t}) = 
\frac{m}{2}\frac{\dot{x}^2}{\dot{t}}\qquad; 
\qquad \dot{x} = \frac{dx}{d\tau} \quad, \quad\dot{t} = \frac{dt}{d\tau}~.  
\label{N3}
\end{equation}
Now the canonical momenta corresponding to the coordinates $t$ and $x$ are 
given by
\begin{equation}
p_{t} = \frac{\partial L_{\tau}}{\partial \dot{t}} 
= -\frac{m\dot{x}^2}{2\dot{t}^2}
\label{N4}
\end{equation}
\begin{equation}
p_{x} = \frac{\partial L_{\tau}}{\partial \dot{x}} = \frac{m\dot{x}}{\dot{t}}
\label{N5}
\end{equation}
which satisfy the standard canonical Poisson bracket (PB) relations
\begin{equation}
\{x,p_{x}\} = \{t, p_{t}\} = 1\quad;\quad
\{x, x\} = \{p_{x}, p_{x}\} = \{t, t\} = \{p_{t}, p_{t}\} 
= 0~.
\label{N500}
\end{equation}
The fact that the canonical Hamiltonian vanishes 
for a time-reparametrized theory can be easily checked 
using eq.(s) (\ref{N4}) and (\ref{N5}).
Also, the primary constraint in the theory, obtained from 
eq.(s) (\ref{N4}, \ref{N5}) is given by
\begin{equation}
\phi_{1} = p_{x}^2 + 2mp_{t} \approx 0
\label{N7}
\end{equation}
where $\approx 0$ implies equality in the 
``weak" sense \cite{diraclecture}. Clearly the spacetime
coordinate $x^{\mu}(\tau)$, ($\mu = 0, 1 ; x^0 = t, x^1 = x$), 
transforms as
a scalar under reparametrization:
\begin{eqnarray}
\tau \rightarrow \tau^{'} &=& \tau^{'}(\tau)\nonumber\\
x^{\mu}(\tau)\rightarrow x^{'\mu}(\tau^{'}) &=& x^{\mu}(\tau)~.
\label{N71}
\end{eqnarray}
Hence, the infinitesimal change 
in the spacetime coordinate ($\delta x^{\mu}(\tau)$) under an infinitesimal 
reparametrization transformation 
($\tau^{'} = \tau - \epsilon$), is given by
\begin{eqnarray}
\delta x^{\mu}(\tau) = x^{'\mu}(\tau) - x^{\mu}(\tau) = 
\epsilon \frac{dx^\mu}{d\tau}~. 
\label{N8}
\end{eqnarray}
Now we proceed to find the generator of this reparametrization transformation. 
To do this, we first write the variation in the Lagrangian $L_\tau$ (\ref{N3}) 
under the transformation (\ref{N8}) as a total derivative:
\begin{eqnarray}
\delta L_\tau = \frac{dB}{d\tau} \qquad; \qquad B = \frac{m\epsilon}{2} 
\frac{\dot x^2}{\dot t}~.
\label{N11}
\end{eqnarray}
The usual Noether's prescription 
can then be used to obtain the generator $G$ as
\begin{eqnarray}
G = p_{t}\delta t + p_{x}\delta x - B = \frac{\epsilon \dot t}{2m}\phi_{1}.
\label{N12}
\end{eqnarray}
It is easy to see (using eq.(\ref{N500})) 
that this generator reproduces the appropriate 
transformation (\ref{N8})
\begin{eqnarray}
\delta x^{\mu}(\tau) = \{x^{\mu}, G\} = \epsilon \frac{dx^\mu}{d\tau} 
\label{N12a}
\end{eqnarray}
which is in agreement with Dirac's treatment 
\cite{diraclecture}\footnote{In this treatment, the generator is a 
linear combination of the first class constraints. 
Since we have only one first class constraint $\phi_{1}$ in the theory, 
the gauge generator is proportional to $\phi_{1}$.}. 
Note that $x^{\mu}$'s are not gauge invariant variables in this case. 
This example shows that reparametrization symmetry can be identified with 
gauge symmetry.\\
We now fix the gauge symmetry by imposing a gauge condition. 
The standard choice is to identify the time coordinate 
$t$ with the parameter $\tau$ 
\begin{equation}
\phi_{2} = t - \tau \approx 0~.
\label{N13}
\end{equation}
A straightforward computation of the algebra 
between the constraints (\ref{N7}, \ref{N13})  
(using eq.(\ref{N500}) once again) leads to the following 
second class set with:
\begin{equation}
\phi_{ab} = \{\phi_{a}, \phi_{b}\} = -2m\epsilon_{ab}
\quad;\quad (a, b = 1, 2)
\label{N14}
\end{equation}
where, $\epsilon_{ab}$ is an anti-symmetric tensor 
with $\epsilon_{12} = 1$.\\
The next step is to compute the DB(s) defined as
\begin{equation}
\{A, B\}_{DB} = \{A, B\} - \{A, \phi_{a}\}(\phi^{-1})_{ab}\{\phi_{b}, B\} 
\label{N15}
\end{equation}
where $A$, $B$ are any pair of phase-space variables 
and $(\phi^{-1})_{ab} = (2m)^{-1}\epsilon_{ab}$ 
is the inverse of $\phi_{ab}$. It then follows
\begin{equation}
\{x, x\}_{DB} = \{p_{x}, p_{x}\}_{DB} = 0 \qquad;
\qquad \{x, p_{x}\}_{DB} = 1~.
\label{N16}
\end{equation}
The expected canonical bracket structure in the 
usual $2-d$ reduced phase-space comprising of 
variables $x$ and $p_{x}$ only is thus reproduced. The DB(s) imply 
a strong imposition of the second class constraints ($\phi_{a}$). 
Consistent with this, $\{t, x\}_{DB} = 0$ showing that 
there is no spacetime noncommutativity if a gauge-fixing 
condition like eq.(\ref{N13}) is chosen. 
A question which now arises naturally is whether spacetime (or space-space) 
noncommutativity can be obtained by imposing a suitable variant 
of the gauge fixing condition (\ref{N13}). 
Before answering this question, we emphasize at this point 
that the DB(s) between various 
gauges should be related by suitable gauge transformations\footnote{We 
show (see appendix A) how this is done for a 
free Maxwell theory where the DB between phase-space variables 
in radiation and axial gauges are related 
by appropriate gauge transformations.}. 
This idea will be useful in the sequel.
\vskip 0.003cm
\noindent In the present case, the same procedure, 
as done (in the appendix) for a free Maxwell theory, is adopted 
to get hold of a set of variables 
$x^{'}$, $t^{'}$ satisfying a noncommutative algebra
\begin{eqnarray}
\{t^{'}, x^{'}\}_{DB} = \theta
\label{N17}
\end{eqnarray}
with $\theta$ being constant. 
The transformations (\ref{N8}) are written in 
terms of phase-space variables after strongly 
implementing the constraint (\ref{N13}). In component notation, we then have:
\begin{eqnarray}
t^{'} = t + \epsilon
\label{N18}
\end{eqnarray}
\begin{eqnarray}
x^{'} = x + \epsilon\frac{dx}{d\tau} = x + \epsilon\frac{p_{x}}{m}~.
\label{N19}
\end{eqnarray}
Substitution of the above transformations 
in the L.H.S. of eq.(\ref{N17}) and using the 
Dirac algebra (\ref{N16}) for the unprimed variables, fixes $\epsilon$ 
to be:
\begin{eqnarray}
\epsilon = -\theta p_{x}~.
\label{N20}
\end{eqnarray}
This shows the desired gauge fixing condition to be
\begin{eqnarray}
t^{'} + \theta p_{x} - \tau \approx 0~.
\label{N21}
\end{eqnarray}
Now one can just drop the prime to rewrite eq.(\ref{N21}) as 
\begin{eqnarray}
t + \theta p_{x} - \tau \approx 0~.
\label{N2111}
\end{eqnarray}
As one might expect, a direct calculation of the DB in this gauge 
immediately reproduces the noncommutative structure $\{t, x\}_{DB} = \theta$.
\vskip 0.003cm
\noindent The analysis carried out above can be generalised trivially 
to higher $d + 1$-dimensional 
Galilean spacetime. In the case of $d \geq 2$, one can see that 
the above spacetime noncommutativity is of the form 
$\{x^{0}, x^i\}_{DB} = \theta^{0i}$; ($x^{0} = t$). 
This can be derived by writing the counterpart of the transformations 
(\ref{N18}, \ref{N19}) for $d \geq 2$ as:
\begin{eqnarray}
x^{'0} = x^{0} + \epsilon
\label{N18a}
\end{eqnarray}
\begin{eqnarray}
x^{'i} = x^{i} + \epsilon\frac{dx^{i}}{d\tau} 
= x^{i} + \epsilon\frac{p^{i}}{m}~.
\label{N19b}
\end{eqnarray}
Substituting back in the L.H.S. of $\{x^{'0}, x^{'i}\} = \theta^{0i}$, 
fixes $\epsilon$ to be:
\begin{eqnarray}
\epsilon = -\theta^{0i}p_{i}~.
\label{N19c}
\end{eqnarray}
The desired gauge fixing condition (dropping the prime) now becomes
\begin{eqnarray}
x^{0} + \theta^{0i}p_{i} - \tau \approx 0
\label{N19d}
\end{eqnarray}
which is the analogue of (\ref{N2111}).  
The space-space algebra for $d \geq 2$ is also noncommutative 
\begin{eqnarray}
\{x^{i}, x^{j}\}_{DB} = -\frac{1}{m}\left(\theta^{0i}p^{j}- 
\theta^{0j}p^{i} \right).
\label{N211}
\end{eqnarray}
The remaining non-vanishing DB(s) are
\begin{eqnarray}
\{x^{i}, p_{0}\}_{DB} = -\frac{p^{i}}{m}\qquad \{x^{i}, p_{j}\}_{DB} 
= {\delta^{i}}_{j}~.
\label{N211a}
\end{eqnarray}
The above forms of the DB(s) show a Lie-algebraic structure
for the brackets involving phase-space variables 
(with the inclusion of identity). 
Following \cite{madore}, an appropriate ``diamond product" can
be associated for this, 
in order to compose any pair of phase-space functions.
\vskip 0.003cm
\noindent We have thus systematically 
derived the non-standard gauge condition 
leading to a noncommutative algebra. Also, the change of variables 
mapping this noncommutative algebra with the usual (commutative) 
algebra is found to be a gauge transformation.
\vskip 0.003cm
\noindent There is yet another interesting way of deriving 
the Dirac algebra if one looks
at the symplectic two-form $\omega = dp_{\mu}\wedge dx^{\mu}$ and then 
simply impose the conditions on $p_{0}$ and $x^{0}$, 
for all cases discussed. We
consider the simplest case here. In $1+1$-dimension, 
the two-form $\omega$ can
be written as
\begin{eqnarray}
\omega = dp_{t}\wedge dt +  dp_{x}\wedge dx~.
\label{N211b}
\end{eqnarray}
Now imposing the condition on $p_{t}$ (\ref{N7}) and $t$ (\ref{N13}), we get:
\begin{eqnarray}
\omega = -\frac{p_{x}}{m}dp_{x}\wedge d\tau + dp_{x}\wedge dx~.  
\label{N211c}
\end{eqnarray}
Note that the first term on the right hand side of the above equation vanishes
as $\tau$ is not a variable in the configuration space. Now the inverse of the
components of the two-form yields the brackets (\ref{N16}). 
\vskip 0.003cm
\noindent In the non-standard gauge (\ref{N2111}),
the two-form $\omega$ reads
\begin{eqnarray}
\omega = dp_{t}\wedge dt - \frac{1}{\theta} dt\wedge dx  
\label{N211d}
\end{eqnarray}
once the condition on $p_{x}$ from eq.(\ref{N2111}) is imposed.
The inverse of the
components of the two-form can be computed in a 
straightforward way to obtain the  noncommutative structure
$\{t, x\} = \theta$.
The same procedure can be followed for the other cases
discussed in the chapter.
\vskip 0.003cm
\noindent The role of integral curves within 
this symplectic formalism \cite{sud} 
is discussed in appendix B.

\section{Free relativistic particle}
In this section we take up the case of a free relativistic particle 
and study how spacetime noncommutativity can arise in this case 
also through a suitably modified gauge fixing condition. 
We start with the standard reparametrization 
invariant action of a relativistic free particle which propagates 
in $d + 1$-dimensional ``target
 spacetime" 
\begin{equation}
S_{0} = -m\int d\tau \sqrt{-\dot{x}^{2}}
\label{1r}
\end{equation}
with spacetime coordinates  $x^{\mu}$, ${\mu} = 0, 1, ...d$, 
the dot denoting
differentiation with respect to the evolution parameter $\tau$, 
and the Minkowski metric is $\eta = diag(-1, 1, ..., 1)$. 
In contrast to the NR case, the action 
here is already in the reparametrized 
form with all $x^{\mu}$'s (including $x^{0} = t$) 
contained in the configuration space. 
The canonically conjugate momenta are given by
\begin{equation}
p_{\mu} = \frac{m\dot{x}_{\mu}}{\sqrt{-\dot{x}^{2}}}
\label{2r}
\end{equation}
and satisfy the standard PB relations
\begin{equation}
\{x^{\mu}, p_{\nu}\} = \delta^{\mu}_{\nu}  
\qquad;\qquad \{x^{\mu}, x^{\nu}\} = \{p^{\mu}, p^{\nu}\} = 0.
\label{3r}
\end{equation}
Taking the square of eq.(\ref{2r}), it is easy to see that
these are subject to the Einstein constraint
\begin{equation}
\phi_{1} = p^{2} + m^{2} \approx 0~.
\label{4r}
\end{equation}
The reparametrization symmetry of the problem 
(under which the action (\ref{1r}) is invariant) can now be used
together with the fact that $x^{\mu}(\tau)$ 
transforms as a  scalar under world-line reparametrization 
(\ref{N71}), to find the infinitesimal 
transformation of the spacetime coordinate (\ref{N8}).
As before, to derive the generator of the reparametrization invariance 
we write the variation in the Lagrangian as a total derivative:  
\begin{eqnarray}
\delta L &=& \frac{dB}{d\tau}\qquad;\qquad B = -m\epsilon\sqrt{-\dot{x}^2}~.
\label{4dr}
\end{eqnarray}
The generator of the infinitesimal transformation 
of the spacetime coordinate (\ref{N12a}) can then be obtained from the usual 
Noether's prescription\footnote {The factor of $1/2$ 
comes from symmetrization. To make this point clear, 
we must note that while computing $\{x^{\mu}, G\}$, 
an additional factor of 2 crops up from the bracket between $x^{\mu}$ 
and $\delta x_{\mu}$ as $\delta x_{\mu}$ is related to $p_{\mu}$ 
by the relations (\ref{N8}) and (\ref{2r}). 
The factor of $1/2$ is placed in order to cancel this additional 
factor of $2$.}
\begin{eqnarray}
G &=& \frac{1}{2}\left(p^{\mu}\delta x_{\mu} - B\right)\nonumber\\
&=& \frac{1}{2}
\left(p^{\mu}\epsilon\frac{dx_{\mu}}{d\tau} + m\epsilon\sqrt{-\dot{x}^2}
\right)\nonumber\\
&=& \frac{\epsilon\sqrt{-\dot{x}^2}}{2m}\phi_{1}
\label{5a}
\end{eqnarray}
where we have used eq.(s) (\ref{N8}, \ref{4dr}). 

\noindent  A gauge condition can now be imposed
to curtail the gauge freedom just 
as in the NR case. The standard choice is to 
identify the time coordinate $x^{0}$ with the parameter $\tau$ 
\begin{equation}
\phi_{2} = x^{0} - \tau \approx 0
\label{5dr}
\end{equation}
which is the analogue of eq.(\ref{N13}).
The constraints (\ref{4r}, \ref{5dr}) form a second class set with
\begin{equation}
\{\phi_{a}, \phi_{b}\} = 2p_{0}\epsilon_{ab}~.
\label{6r}
\end{equation}
The resulting non-vanishing DB(s) are
\begin{eqnarray}
\{x^{i}, p_{0}\}_{DB} = \frac{p^{i}}{p_{0}} \qquad;\qquad
\{x^{i}, p_{j}\}_{DB} 
= {\delta^{i}}_{j}
\label{8000}
\end{eqnarray}
which imposes the constraints $\phi_{1}$ and $\phi_{2}$ strongly. 
In particular,  we observe $\{x^0, x^i\}_{DB} = 0$, 
showing that there is no spacetime noncommutativity. 
This is again consistent with the fact that the constraint 
(\ref{5dr}) is now strongly imposed. 
Taking a cue from our previous NR example, 
we see that we must have a variant of eq.(\ref{5dr}) 
as a gauge fixing condition to get spacetime noncommutativity in the 
following form
\begin{eqnarray}
\{x^{'0}, x^{'i}\}_{DB} = \theta^{0i}
\label{9ar}
\end{eqnarray}
($\theta^{0i}$ being constants) where $x^{'\mu}$ 
denotes the appropriate 
gauge transforms of $x^{\mu}$ variables. 
The transformed variables $x^{'\mu}$ 
in terms of the variables  
$x^{\mu}$ can be determined by considering an
infinitesimal transformation (\ref{N8}) written in terms of 
phase-space variables as 
\begin{eqnarray}
x^{'0} = x^{0} + \epsilon \qquad;\qquad x^{'i} &=& x^{i} - 
\epsilon\frac{p^{i}}{p_{0}}
\label{9br}
\end{eqnarray}
where we have used the relation $\frac{dx^{i}}{d\tau} 
= -\frac{p^{i}}{p_{0}}$
obtained from eq.(\ref{2r}).
A simple inspection after substituting the above relations (\ref{9br}) 
back in eq.(\ref{9ar}) 
and using eq.(\ref{8000}), shows that $\epsilon$ 
is given by
\begin{eqnarray}
\epsilon = -\theta^{0i}p_{i}
\label{9dr}
\end{eqnarray}
which is identical to eq.(\ref{N19c}).
Hence the gauge transformed variables $x^{'\mu}$ (\ref{9br}) 
for the above choice of $\epsilon$ are given by:
\begin{eqnarray}
x^{'0} = x^{0} - \theta^{0i}p_{i}
\label{10er}
\end{eqnarray}
\begin{eqnarray}
x^{'i} &=& x^{i} + \theta^{0j}p_{j}\frac{p^{i}}{p_{0}}~.
\label{10fr}
\end{eqnarray}
The above set of transformations and the relation (\ref{8000}), 
leads to the following Dirac algebra between the primed variables
\begin{eqnarray}
\{x^{'0}, x^{'i}\}_{DB} = \theta^{0i} 
\label{10ar}
\end{eqnarray}
\begin{eqnarray}
\{x^{'i}, x^{'j}\}_{DB} = \frac{1}{p_{0}}
\left(\theta^{0i}p^{j} - \theta^{0j}p^{i}\right) 
\label{10br}
\end{eqnarray}
\begin{eqnarray}
\{x^{'i}, p^{'}_{0}\}_{DB} = \frac{p^{i}}{p_{0}}\qquad; 
\qquad\{x^{'i}, p^{'}_{j}\}_{DB} = {\delta^{i}}_{j}~. 
\label{10cr}
\end{eqnarray}
Note that unlike $x$'s, $p$'s are gauge invariant objects 
as $\{p^{\mu}, \phi\} = 0$; hence $p^{'}_{\mu} = p_{\mu}$.
\vskip 0.003cm
\noindent It is interesting to observe 
that the solution of the gauge parameter 
$\epsilon$ remains the same in both the relativistic case as 
well as the NR case. Also, $m$ in the NR case gets replaced by 
$-p_{0}$ in the relativistic case. With this identification, 
one can easily see that the complete Dirac algebra in the NR case 
goes over  to the corresponding algebra in the relativistic case. 
However, since $p_{0}$ does not have a vanishing bracket 
with all other phase-space
variables, its occurence in the denominators in eq.(s) 
(\ref{10br}, \ref{10cr})
shows that the bracket structure of the phase-space variables in the
relativistic case is no longer Lie-algebraic, 
unlike the NR case discussed
in the previous section.
\vskip 0.003cm
\noindent Furthermore, the modified gauge fixing condition is given by:
\begin{eqnarray}
\phi_{2} = x^{0} + \theta^{0i}p_{i} - \tau \approx 0\qquad, 
\qquad i = 1, 2, ...d~.
\label{9e9}
\end{eqnarray}
It is trivial to check that the constraints (\ref{4r}, \ref{9e9}) 
also form a second class pair as
\begin{eqnarray}
\{\phi_{a}, \phi_{b}\} = 2p_{0}\epsilon_{ab}~. 
\label{100r}
\end{eqnarray}
The set of non-vanishing DB(s) consistent with the strong imposition of the 
constraints (\ref{4r}, \ref{9e9}) reproduces the results 
(\ref{10ar}, \ref{10br}, \ref{10cr}).
Eq.(\ref{10cr}) is the same as in the standard gauge (\ref{5dr}), while 
eq.(\ref{10br}) implies non-trivial bracket relations 
among spatial coordinates upon imposition of the 
gauge fixing condition (\ref{9e9}). 
\vskip 0.003cm
\noindent It should be noted that the above gauge fixing condition 
(\ref{9e9}) was also given in \cite{pin}. 
Indeed a change of variables, 
which is different from eq.(s) (\ref{10er}, \ref{10fr}), 
is found there by inspection, using which the spacetime 
noncommutativity gets removed. However, the change of variables 
that we find here is related to a gauge transformation providing 
in turn a systematic derivation of the modified gauge condition 
and also spacetime noncommutativity. Moreover, the  
definition of the Lorentz generators (rotations and boosts) in (\cite{pin}) 
requires some additional terms (in the modified gauge) in order 
to have a closed algebra between the generators. 
In our approach, the definition of the Lorentz generators remains unchanged, 
simply because these are gauge invariant.
\vskip 0.003cm
\noindent The Lorentz generators (rotations and boosts) are defined as:
\begin{eqnarray}
M_{ij} = x_{i}p_{j} - x_{j}p_{i} 
\label{27r}
\end{eqnarray}
\begin{eqnarray}
M_{0i} = x_{0}p_{i} - x_{i}p_{0}. 
\label{28r}
\end{eqnarray}
As expected, they satisfy the usual algebra in both the unprimed and 
the primed coordinates as $M_{\mu\nu}$ and $p_{\mu}$ 
are both gauge invariant.
\begin{eqnarray}
\{M_{ij}, p_{k}\}_{DB} = \delta_{ik}p_{j} - \delta_{jk}p_{i} 
\label{29r}
\end{eqnarray}
\begin{eqnarray}
\{M_{ij}, M_{kl}\}_{DB} = \delta_{ik}M_{jl} - \delta_{jk}M_{il} 
+ \delta_{jl}M_{ik} - \delta_{il}M_{jk}
\label{29ar}
\end{eqnarray}
\begin{eqnarray}
\{M_{ij}, M_{0k}\}_{DB} = \delta_{ik}M_{0j} - \delta_{jk}M_{0i} 
\label{29br}
\end{eqnarray}
\begin{eqnarray}
\{M_{0i}, M_{0j}\}_{DB} = M_{ji}. 
\label{29cr}
\end{eqnarray}
However, the algebra between the space coordinates and the rotations, boosts
are different in the two gauges (\ref{5dr}, \ref{9e9}). 
This is not surprising as $x^{k}$ is not 
gauge invariant under gauge transformation. 
We find
\begin{eqnarray}
\{M_{ij}, x^{k}\}_{DB} = {\delta_{i}}^{k}x_{j} - {\delta_{j}}^{k}x_{i} 
\label{30r}
\end{eqnarray}
\begin{eqnarray}
\{M_{0i}, x^{j}\}_{DB} = x_{i}\frac{p^{j}}{p_{0}} - x_{0}{\delta_{i}}^{j} 
\label{31r}
\end{eqnarray}
\begin{eqnarray}
\{M_{ij}, x^{'k}\}_{DB} &=& \{M_{ij}, x^{k} + 
\theta^{0l}p_{l}\frac{p^{k}}{p_{0}}\}_{DB}\nonumber\\
&=& {\delta_{i}}^{k}x^{'}_{j} - 
{\delta_{j}}^{k}x^{'}_{i} + \frac{1}{p_{0}}
\left({\theta^{0}}_{i}p^{k}p_{j} 
- {\theta^{0}}_{j}p^{k}p_{i}\right) 
\label{32r}
\end{eqnarray}
\begin{eqnarray}
\{M_{0i}, x^{'j}\}_{DB} &=& \{M_{0i}, x^{j} + 
\theta^{0l}p_{l}\frac{p^{j}}{p_{0}}\}_{DB}\nonumber\\
&=& x^{'}_{i}\frac{p^{j}}{p_{0}} - x^{'}_{0}{\delta_{i}}^{j} 
- {\theta^{0}}_{i}p^{j} 
\label{33r}
\end{eqnarray}
where we have used eq.(\ref{10fr}) and the algebra $(\ref{8000})$.
The same results can also be obtained using 
the relations (\ref{10ar}, \ref{10br}, \ref{10cr}).
\vskip 0.003cm
\noindent Now we note that the gauge choice (\ref{9e9}) 
is not Lorentz invariant.
Yet the Dirac bracket procedure forces this constraint equation to be 
strongly valid in all Lorentz frames \cite{hanson}. 
This can be made consistent if and only if an infinitesimal 
Lorentz boost to a new frame\footnote{A similar treatment 
as in \cite{hanson} has been given in \cite{bcsingle} 
for a free relativistic particle coupled to Chern-Simons term.} 
\begin{eqnarray}
p^{\mu} \rightarrow p^{'\mu} = p^{\mu} + \omega^{\mu\nu}p_{\nu}
\label{34r}
\end{eqnarray}
is accompanied by a compensating infinitesimal gauge transformation
\begin{eqnarray}
\tau \rightarrow \tau^{'} = \tau + \Delta\tau.
\label{35r}
\end{eqnarray}
The change in $x^{\mu}$, upto first order, is therefore
\begin{eqnarray}
x^{'\mu}(\tau) & = &x^{\mu}(\tau^{'}) + \omega^{\mu\nu}x_{\nu}(\tau)
\nonumber\\
 & = & x^{\mu}(\tau) + \Delta\tau\frac{dx^{\mu}}{d\tau} 
+ \omega^{\mu\nu}x_{\nu}.
\label{36r}
\end{eqnarray}
In particular, the zeroth component is given by:
\begin{eqnarray}
x^{'0}(\tau) = x^{0}(\tau) + \Delta\tau\frac{dx^{0}}{d\tau} 
+ \omega^{0i}x_{i}.
\label{37r}
\end{eqnarray}
Since the gauge condition (\ref{9e9}) is 
$x^{0}(\tau) \approx \tau - \theta^{0i}p_{i}$, $x^{'0}(\tau)$ 
also must satisfy $x^{'0}(\tau) = (\tau - \theta^{0i}p^{'}_{i})$ 
in the boosted frame, which can now be written using eq.(\ref{34r}), as
\begin{eqnarray}
x^{'0}(\tau) &=& \tau - \theta^{0i}p^{'}_{i}\nonumber\\
 &=& \tau - \theta^{0i}p_{i} + \theta^{0i}\omega^{0i}p_{0}.
\label{38r}
\end{eqnarray}
Comparing with eq.(\ref{37r}) and using the gauge condition (\ref{9e9}), 
we can now solve for $\Delta \tau$ to get:
\begin{eqnarray}
\Delta\tau = \frac{\theta^{0i}\omega^{0i}p_{0} 
- \omega^{0i}x_{i}}{1 - \theta^{0i}\dot{p_{i}}}\qquad; 
\quad \dot p_{i} = \frac{dp_{i}}{d\tau}~.
\label{39r}
\end{eqnarray}
The spatial components of eq.(\ref{36r}) 
(for a pure boost) therefore satisfy
\begin{eqnarray}
\delta x^{j}(\tau) &=& x^{'j}(\tau) - x^{j}(\tau) 
= \Delta\tau\frac{dx^{j}}{d\tau} + \omega^{j0}x_{0}\nonumber\\
&=& \omega^{0i}\left(x_{i}\frac{p^{j}}{p_{0}} 
- x_{0}{\delta_{i}}^{j} - \theta^{0i}p^{j}\right).
\label{40r}
\end{eqnarray}
Hence we find that eq.(\ref{40r}) and eq.(\ref{33r}) 
are consistent with each other. 
However, note that in the above derivation 
we have taken $\theta^{0i}$ 
to be a constant. If we take $\theta^{0i}$ to transform as a tensor, 
then for a Lorentz boost to a new frame, it changes as
\begin{eqnarray}
\theta^{0i} \rightarrow \theta^{'0i} = \theta^{0i} + \omega^{0j}\theta^{ji} 
\label{41}
\end{eqnarray}
and the entire consistency program would fail. 
The ($1 + 1$)--dimensional case is special, since even 
if we take $\theta^{01}$ to transform as a tensor, 
this will not affect the consistency program as it 
remains invariant ($\theta^{'01} = \theta^{01}$) under Lorentz boost.
\vskip 0.003cm
\noindent Now in \cite{sgangopadhyay} we have shown 
that there exists some special 
values of the reparametrization parameter $\epsilon$ which 
leads to noncommuting structures falling in the Lie-algebraic
category \cite{madore}.
\vskip 0.003cm
\noindent Setting
\begin{eqnarray}
\epsilon=-\theta^{0k}p_{k}\frac{p_{0}}{m}
\label{lie1}
\end{eqnarray}
and using eq.(\ref{8000}) and eq.(\ref{9br}), we obtain the following 
algebra between the primed coordinates:
\begin{eqnarray}
\{x^{'i}, x^{'j}\}_{DB} = \frac{1}{m}\left(\theta^{0i}p^{j}
 - \theta^{0j}p^{i}\right) 
\label{lie2}
\end{eqnarray}
\begin{eqnarray}
\{x^{'0}, x^{'i}\}_{DB} = \frac{1}{m}(\theta^{0i}p_{0}+\theta^{0k}p_{k}
\frac{p^{i}}{p_{0}})
\label{lie3}
\end{eqnarray}
\begin{eqnarray}
\{x^{'i}, p^{'}_{0}\}_{DB} = \frac{p^{i}}{p_{0}}      \qquad;
\qquad\{x^{'i}, p^{'}_{j}\}_{DB} = {\delta^{i}}_{j}~. 
\label{lie4}
\end{eqnarray}
It is now important to observe that the noncommutativity in
the space-space coordinates (\ref{lie2}) has a Lie-algebraic structure
in phase-space (with the inclusion of identity) 
and not in spacetime. This is in contrast 
to the results derived for the relativistic free particle
where space-space noncommutativity 
(eq.(\ref{10br})) was not Lie-algebraic. 
\vskip 0.003cm
\noindent The above solution of $\epsilon$ (\ref{lie1})
shows that the desired
gauge fixing condition is given by:
\begin{eqnarray}
\phi_{3} = x^{0} + \theta^{0k}p_{k}\frac{p_{0}}{m} - \tau \approx 0,
\qquad k = 1, 2, ...d.
\label{lie5}
\end{eqnarray}
It is easy to check that the constraints (\ref{4r}, \ref{lie5}) 
once again form a second class pair (\ref{100r}).
The set of non-vanishing DB(s) consistent with the strong imposition
of the constraints (\ref{4r}, \ref{lie5}) reproduces the results
(\ref{lie2}, \ref{lie3}, \ref{lie4}).
\vskip 0.003cm
\noindent Another interesting choice of $\epsilon$ is the following: 
\begin{eqnarray}
\epsilon=-d_{k}\theta^{kl}p_{l}\frac{p_{0}}{m}
\label{lie6}
\end{eqnarray}
where, $d_{k}$ are arbitrary dimensionless constants.
\vskip 0.003cm
\noindent This yields (using eq.(\ref{8000}) and eq.(\ref{9br}))
the following algebra between the primed coordinates:
\begin{eqnarray}
\{x^{'i}, x^{'j}\}_{DB} = \frac{d_{k}}{m}\left(\theta^{ki}p^{j}
- \theta^{kj}p^{i}\right)
\label{lie7}
\end{eqnarray}
\begin{eqnarray}
\{x^{'0}, x^{'i}\}_{DB} = \frac{d_{k}}{m}\left(\theta^{ki}p_{0}+
\theta^{kl}p_{l}\frac{p_{i}}{p_{0}}\right)
\label{lie8}
\end{eqnarray}
\begin{eqnarray}
\{x^{'i}, p^{'}_{0}\}_{DB} = \frac{p_{i}}{p_{0}}      \qquad;
 \qquad\{x^{'i}, p^{'}_{j}\}_{DB} = {\delta^{i}}_{j}~. 
\label{lie9}
\end{eqnarray}
Once again we obtain a Lie-algebraic noncommutative structure in the
space-space sector. However, note that eq.(\ref{lie7}) is different from
eq.(\ref{lie2}) because the noncommutative parameter $\theta$ in
eq.(\ref{lie7}) has space indices in contrast to the spacetime indices
appearing in eq.(\ref{lie2}). The spacetime algebra is once again
not Lie-algebraic in form.
\vskip 0.003cm
\noindent The desired gauge fixing condition which lead to the above
DB(s) read:
\begin{eqnarray}
\phi_{4} = x^{0} + d_{k}\theta^{kl}p_{l}\frac{p_{0}}{m} - \tau \approx 0,
 \qquad k = 1, 2, ...d.
\label{lie10}
\end{eqnarray}
The algebra of the Lorentz generators
for the above choices of the
reparametrization parameter $\epsilon$ can be investigated in a 
similar way as for the relativistic free particle and once again
the internal consistency of our analysis can be established.
\vskip 0.003cm
\noindent Finally, there exists choices of $\epsilon$
for which the space-space algebra
can be made to vanish. The choices are:
\begin{eqnarray}
\epsilon=e_{k}\theta^{0k}\frac{p_{0}^2}{m}
\label{vanish1}
\end{eqnarray}
and
\begin{eqnarray}
\epsilon=-f_{kl}\theta^{kl}p_{0}
\label{vanish2}
\end{eqnarray}
where, $e_{k}$ and $f_{kl}$ are arbitrary dimensionless constants.
\vskip 0.003cm
\noindent The spacetime algebras however
do not vanish for the above values of
$\epsilon$ and are as follows:
\begin{eqnarray}
\{x^{\prime 0}, x^{\prime i}\}=\frac{2e_{k}}{m}\theta^{0k}p^{i}
\label{vanish3}
\end{eqnarray}
and
\begin{eqnarray}
\{x^{\prime 0}, x^{\prime i}\}=f_{kl}\theta^{kl}\frac{p^{i}}{p_{0}}~.
\label{vanish4}
\end{eqnarray}
\vskip 0.003cm
\noindent Let us now make certain observations.
Although, the relations (\ref{N211}), (\ref{10br}), 
(\ref{lie2}) and (\ref{lie7}) are reminescent of 
Snyder's algebra \cite{snyder}, there is a subtle
difference. This can be seen by noting that 
the right hand side of these relations
do not have the structure of an angular momentum
operator in their differential representation
(obtained by repacing $p_{j}$ by ($-i\partial_{j}$)) in contrast to  
Snyder's algebra. 
Further, the relations (\ref{lie2}) and (\ref{lie7})
has a similar structure to the commutation relations
describing the Lie-algebraic deformation of the
Minkowski space \cite{woro}, the only difference being that
momentum operators appear at the right hand side of the relations 
instead of the position operators.
\vskip 0.003cm
\noindent Now in the cases where the noncommutativity 
takes the canonical structure
($[\hat x^\mu,\hat x^\nu]=i\theta^{\mu\nu}$), one can infer 
the presence of non-locality
from the fact that two  localised functions
$f$ and $g$ having supports within a size
$\delta << \sqrt{||\theta||}$, yields a function $f\star g$
 which is non-vanishing over a much larger 
region of size $||\theta||/\delta$ \cite{szaborj}. 
One therefore expects a similar qualitative feature 
of non-locality arising from the ``diamond product" appropriate for the
Lie-bracket structure of noncommutativity in the NR case also.
This is further reinforced by the fact
that coordinate transformations (\ref{N18a}, \ref{N19b}) involve mixing of
coordinates and momenta. Since this mixing 
is present in the relativistic case
as well (eq.(s) (\ref{10er}, \ref{10fr})), 
it is expected to maintain the non-locality 
of the noncommutative theory, although an appropriate 
``diamond product" cannot be readily 
constructed because of the absence of a Lie-bracket structure. Also, 
the mixing
of coordinates and momenta is a natural consequence of our gauge 
conditions which essentially involve phase-space 
variables interpolating between the
commutative and noncommutative descriptions.
\vskip 0.003cm
\noindent Besides, spacetime noncommutativity arising from a relation
like eq.(\ref{10ar}), implies that the ``co-ordinate" time $\hat x^0$ cannot
be localised as any state will have a spread in the spectrum of $\hat x^0$.
This eventually leads to the failure of causality and violation of locality
in quantum field theory \cite{balachan, dran}.
\section{Interaction with background Electromagnetic Field}
In this section, we consider interactions with a background electromagnetic
field which still keeps the time reparametrization symmetry of the 
relativistic free particle intact. Before going over to the general case, 
we consider a constant background field. The interaction term to be 
added to $S_{0}$ is then
\begin{eqnarray}
S_{F} = -\frac{1}{2}\int d\tau F_{\mu\nu}x^{\mu}\dot{x^{\nu}}
\label{42r}
\end{eqnarray}
where, $F_{\mu\nu}$ is a constant field strength tensor.
The canonical momenta are given by
\begin{eqnarray}
\Pi_{\mu} = p_{\mu} + \frac{1}{2}F_{\mu\nu}x^{\nu}
\label{43r}
\end{eqnarray}
where, $p_{\mu}$ is given by eq.(\ref{2r}).
The Einstein constraint 
(\ref{4r}) which is the first class constraint of the theory once again
follows from the reparametrization symmetry of the model. 
The PB(s) are\footnote{These relations follow from 
the basic canonical algebra $\{x_{\mu}, \Pi^{\nu}\} = 
\delta_{\mu}^{\nu}~;~ 
\{x_{\mu}, x_{\nu}\} = \{\Pi_{\mu}, \Pi_{\nu}\} = 0$.}
\begin{eqnarray}
\{x^\mu, p_{\nu}\} = \delta^{\mu}_{\nu} \quad;\quad \{x^\mu, x^{\nu}\} 
= 0 \quad;\quad \{p_\mu, p_{\nu}\} = -F_{\mu\nu}.
\label{44r}
\end{eqnarray}
Note that $p_{\mu}$ does not have zero PB with the 
constraint (\ref{4r}) anymore and thus is not gauge invariant. 
Now to obtain the generator of reparametrization symmetry, 
we again exploit the infinitesimal transformation of the 
spacetime coordinate given by eq.(\ref{N8}). 
Proceeding exactly as in the earlier sections, we write the variation 
of the Lagrangian in a total derivative form as:
\begin{eqnarray}
\delta L &=& \frac{dB}{d\tau}\qquad ; \qquad B 
= -m\epsilon\sqrt{-\dot{x}^2} 
- \frac{\epsilon}{2}F_{\mu\nu}x^{\mu}\frac{dx{^\nu}}{d\tau}~.
\label{45r}
\end{eqnarray}
Then the generator is obtained from usual Noether's prescription 
(as it was done for the case of the free relativistic particle), 
by making use of eq.(\ref{43r}) to get
\begin{eqnarray}
G &=& \frac{1}{2}\left(\Pi^{\mu}\delta x_{\mu} - B\right)\nonumber\\
&=&\frac{\epsilon\sqrt{-\dot{x}^2}}{2m}\left[\Pi^{\mu}p_{\mu}
+m^{2}+\frac{1}{2}F_{\mu\nu}x^{\mu}p^{\nu}\right]\nonumber\\
&=&\frac{\epsilon\sqrt{-\dot{x}^2}}{2m}\phi_{1}
\label{46r}
\end{eqnarray}
where, $\phi_{1} = p^2 + m^2 \approx 0$ 
is the first class constraint (\ref{4r}).
This clearly generates the infinitesimal transformation 
of the spacetime coordinate (\ref{N12a}).
Hence we have again shown that the generator is indeed proportional 
to the first class constraint which is in conformity with Dirac's treatment. 
Further, the relation between reparametrization symmetry and 
gauge symmetry becomes evident once more.  
Now the gauge/reparametrization symmetry can be fixed by imposing a 
gauge condition. The standard choice is given by eq.(\ref{5dr}). 
The constraints (\ref{4r}, \ref{5dr}) form a second class set 
with the PB(s) between them given by eq.(\ref{6r}). 
So the non-vanishing DB(s) are given by eq.(\ref{8000}) and
\begin{eqnarray}
\{p_{i}, p_{j}\}_{DB} = -F_{ij}\qquad;\qquad \{p_{0}, p_{i}\}_{DB} 
= F_{ij}\frac{p_{j}}{p_{0}}~.
\label{46ar}
\end{eqnarray}
\vskip 0.003cm
\noindent To obtain noncommutativity between the primed set of spacetime 
coordinates  (\ref{9ar}), we first observe that the zeroth component 
and spatial components of eq.(\ref{N8}) (in the standard gauge (\ref{5dr})) 
leads to eq.(\ref{9br})
where we have used the relation $\frac{dx^{i}}{d\tau} 
= -\frac{p_{i}}{p_{0}}$ obtained from eq.(\ref{2r}).
Using the relations (\ref{9ar}, \ref{9br}) fixes the value of
$\epsilon$, which, in view of the non-vanishing bracket (\ref{46ar}), 
turns out to be
\begin{eqnarray}
\epsilon = -\theta^{0j}P_{j}
\label{47r}
\end{eqnarray}
where,
\begin{eqnarray}
P_{\mu} = p_{\mu} + F_{\mu\nu}x^{\nu}
\label{48r}
\end{eqnarray}
is gauge invariant since $\{P_{\mu}, p_{\nu}\} = 0$. 
As a simple consistency check, we observe that 
the solution (\ref{47r}) reduces to the 
free particle solution (\ref{9dr}) for vanishing electromagnetic field. 
It should also be noted that the non-vanishing DB(s) involving 
$P_{\mu}$ in the standard gauge (\ref{5dr}) are given by:
\begin{eqnarray}
\{x^{i}, P_{j}\}_{DB} = {\delta^{i}}_{j} 
\quad;\quad \{P_{\mu}, P_{\nu}\}_{DB} 
= F_{\mu\nu}\quad;\quad \{x^{i}, P_{0}\}_{DB} = \frac{p^{i}}{p_{0}}~.
\label{50r}
\end{eqnarray}
\vskip 0.003cm
\noindent The set of transformations 
relating the unprimed and primed coordinates can now be written down 
using eq.(s) (\ref{9br}) and (\ref{47r}):
\begin{eqnarray}
x^{'0} = x^{0} - \theta^{0i}P_{i}
\label{58r}
\end{eqnarray}
\begin{eqnarray}
x^{'i} = x^{i} - \theta^{0j}P_{j}\frac{dx^{i}}{d\tau} = x^{i} 
+ \theta^{0j}P_{j}\frac{p^{i}}{p_{0}}
\label{59r}
\end{eqnarray}
where we have used the relation $\frac{p^{'}_{j}}{p^{'}_{0}} 
= -\frac{dx^{'}_{j}}{d\tau}$ since $\frac{dx^{0}}{d\tau} = 1$ 
in the old gauge (\ref{5dr}).
From the above set of transformations and 
the relations (\ref{8000}, \ref{46ar}, \ref{50r}), 
we compute the DB(s) between the primed variables:
\begin{eqnarray}
\{x^{'0}, x^{'i}\}_{DB} &=& \theta^{0i}
\label{60r}
\end{eqnarray}
\begin{eqnarray}
\{x^{'i}, x^{'j}\}_{DB} &=& \frac{1}{p_{0}}\left(\theta^{0i}p^{j} 
- \theta^{0j}p^{i}\right)\nonumber\\
&=& \frac{1}{p^{'}_{0}}\left(\theta^{0i}p^{'j} - \theta^{0j}p^{'i}\right) 
+ O(\theta^2).
\label{61r}
\end{eqnarray}
In order to express the variables on the R.H.S. in terms 
of primed ones\footnote{Note that, since $P_{\mu}$ (eq.(\ref{48r})) 
is gauge invariant, $P^{'}_{\mu} = P_{\mu}$.}, 
use has been made of eq.(\ref{59r}) to get:
\begin{eqnarray}
\frac{p^{'}_{j}}{p^{'}_{0}} = \frac{p_{j}}{p_{0}} 
- \theta^{0k}P_{k}\frac{d}{d\tau}\left(\frac{p_{j}}{p_{0}}\right) 
+ O(\theta^2).
\label{64ar}
\end{eqnarray}
\vskip 0.003cm
\noindent Observe that the change of variables (\ref{58r}, \ref{59r}) 
leading to the algebra among the primed variables, 
are basically infinitesimal gauge transformations 
that are valid to first order in the 
reparametrization parameter $\epsilon$. 
Moreover, from eq.(\ref{47r}) it follows that $\epsilon$ 
is proportional to $\theta$. Hence, the Dirac algebra 
(\ref{60r}, \ref{61r}) between the primed variables are also valid 
upto order $\theta$. However, it turns out that these 
results are actually exact, 
as we shall now show below.
\vskip 0.003cm
\noindent As before, it is possible to write 
down the modified gauge condition 
from the solution (\ref{47r}) for $\epsilon$ as
\begin{eqnarray}
\phi_{2} = x^{0} + \theta^{0i}P_{i} - \tau \approx 0, \qquad i = 1, 2, ...d.
\label{64aaar}
\end{eqnarray}
The constraints $(\ref{4r}, \ref{64aaar})$ again form a second class set 
with the PB(s) between them being given by (\ref{6r}). 
So we recover the previous DB(s) (\ref{60r}, \ref{61r}) 
between spacetime coordinates  $x^\mu$.
\vskip 0.003cm
\noindent Finally we consider the coupling of the relativistic free particle 
to an arbitrary electromagnetic field. 
As before the action is reparametrization invariant. 
Here we replace eq.(\ref{42r}) by
\begin{eqnarray}
S_{F} = -\int d\tau A_{\mu}(x)\dot{x^{\mu}}.
\label{65r}
\end{eqnarray}
The choice $A_{\mu} = -\frac{1}{2}F_{\mu\nu}x^{\nu}$ for constant 
$F_{\mu\nu}$ reproduces the action (\ref{42r}).
The Einstein constraint (\ref{4r}) and PB(s) (\ref{44r}) again follow.
The canonical momenta are given by:
\begin{eqnarray}
\Pi_{\mu} = p_{\mu} - A_{\mu}
\label{66r}
\end{eqnarray}
where, $p_{\mu}$ is defined by eq.(\ref{2r}). 
The gauge symmetry can be fixed by imposing a gauge condition. 
The standard choice is given by eq.(\ref{5dr}). 
The constraints (\ref{4r}, \ref{5dr}) form a second class set with the 
PB(s) between them again given by eq.(\ref{6r}). 
So the non-vanishing DB(s) 
are given by eq.(s) (\ref{8000}) and (\ref{46ar}).
As before, exploiting the reparametrization symmetry of the problem, 
the infinitesimal transformation of the spacetime coordinate 
is given by eq.(\ref{N8}) which leads to eq.(\ref{9br}) 
in the standard gauge (\ref{5dr})
(where we have again used the relation 
$\frac{dx^{i}}{d\tau} = -\frac{p^{i}}{p_{0}}$ obtained from eq.(\ref{2r})).
Demanding noncommutativity between the primed set of spacetime 
coordinates by imposing the condition (\ref{9ar}) 
and using the relation (\ref{9br})
leads to: 
\begin{eqnarray}
\{x^{0} + \epsilon, x^{i} - \epsilon\frac{p^{i}}{p_{0}}\}_{DB} = \theta^{0i}
\label{6800r}
\end{eqnarray}
which fixes the value of
$\epsilon$ to be
\begin{eqnarray}
\epsilon = -\theta^{0j}p_{j} + O(\theta^{2}).
\label{69r}
\end{eqnarray}
Here we are content with expression linear in $\theta$ 
as a gauge invariant $P_{\mu}$ (counterpart of eq.(\ref{48r})) 
cannot be defined here.
\vskip 0.003cm
\noindent A gauge condition 
(which is the same as eq.(\ref{9e9})) can be identified once again
leading to noncommutativity between spacetime coordinates. 
The computation of the DB
between the spacetime coordinates in this gauge gives:
\begin{eqnarray}
\{x^{0}, x^{i}\}_{DB} = 
\frac{\theta^{0i}}{1 + \theta^{0j}F_{j\mu}\frac{p^{\mu}}{p_{0}}}
\label{690r}
\end{eqnarray}
which has already been given in \cite{pin}.
One can easily see that to the linear order 
in $\theta$, the above result goes to eq.(\ref{9ar}).

\section{Summary}
We have discussed an approach whereby both space-space as well as spacetime
noncommutative stuctures are obtained in a particular (non-standard gauge)
in models having reparametrization invariance. These structures are obtained
by calculating either DB(s) or symplectic brackets and the results
agree. We have also shown that 
the noncommutative results in the non-standard gauge
and the commutative results in the standard gauge are 
gauge transforms of each other. In other words, equivalent physics
is described by working either with the usual brackets
or the noncommuting brackets.
We feel our approach is conceptually cleaner
and more elegant than those \cite{pin} where such change of variables 
are found by inspection leading to ambiguities in the definition 
of physical (gauge invariant) variables and apparently lacking any connection
with the symmetries of the problem. 
For instance, the angular momentum operator gets modified
in distinct gauges, by appropriate inclusion of extra terms, 
so that the closure property is satisfied. 
In our approach, on the contrary, the angular momentum
remains invariant since the change of variables
is just a gauge transformation. Consequently we do not find  
these extra terms appearing.
We also feel that the present approach 
could be useful in illuminating the role
of variable changes used for relating the commuting and noncommuting
descriptions in field theory.
\section{Appendix A}
Here we would like to show how the 
DB(s) for any pair of variables, computed for Coulomb and axial gauges, 
are connected through gauge transformations. 
For that we consider the action of free Maxwell theory 
\begin{eqnarray}
S = -\frac{1}{4}\int d^{4}x F_{\mu\nu}F^{\mu\nu}.
\label{Ap1}
\end{eqnarray}
The first class constraints of the theory responsible for 
generating gauge transformations are 
\begin{eqnarray}
\pi_{0}(x) \approx 0  \quad;\quad \partial_{i}\pi_{i}(x) \approx 0~.
\label{Ap2}
\end{eqnarray}
The above set
of constraints can be rendered second class by gauge fixing. 
We first consider the Coulomb gauge given by:
\begin{eqnarray}
A_{0} \approx 0 \qquad; \qquad\partial_{i}A_{i}(x) \approx 0. 
\label{Ap3}
\end{eqnarray}
The DB computed between $A_{i}$, $\Pi_{j}$ 
in this gauge yields the familiar 
transverse delta function \cite{diraclecture}, \cite{hanson}:
\begin{eqnarray}
\{A_{i}(x), \Pi_{j}(y)\}^{(c)}_{DB} &=& -\left(\delta_{ij} 
-\frac{\partial_{i}\partial_{j}}{\partial^2}\right)\delta (x - y)
\nonumber\\
&=& - \delta^{T}_{ij}\delta (x - y) 
\label{Ap4}
\end{eqnarray}
where the superscript $c$ denotes the Coulomb gauge. 
\vskip 0.003cm
\noindent The corresponding DB in axial gauge 
$A_{3} \approx 0$ and $(\Pi_{3} - \partial_{3}A_{0}) 
\approx 0$ \footnote{This follows by demanding 
time conservation of the gauge; i.e., 
$\partial_{0}A_{3} = \partial_{0}A_{3} - 
\partial_{3}A_{0} + \partial_{3}A_{0} = -\Pi_{3} + \partial_{3}A_{0} 
\approx 0$.}is given by \cite{diraclecture}, \cite{hanson}:
\begin{eqnarray}
\{A_{i}(x), \Pi_{j}(y)\}^{(a)}_{DB} &=& -\delta_{ij}\delta (x - y) 
+ \delta_{3j}\frac{\partial_{i}}{\partial_{3}}\delta (x - y).
\label{Ap8}
\end{eqnarray}
Now the gauge field configurations $A^{(a)}_{i}$ and 
$A^{(c)}_{i}$ are connected by the gauge transformation 
\begin{eqnarray}
A^{(a)}_{i} = A^{(c)}_{i} + \partial_{i}\Lambda
\label{Ap5}
\end{eqnarray}
where $\Lambda$ is the gauge transformation parameter.
Imposing $A_{3}^{(a)} = 0$ (axial gauge), fixes the value 
of $\Lambda$ to be
\begin{eqnarray}
\Lambda = -\frac{1}{\partial_{3}}A^{(c)}_{3}
\label{Ap6}
\end{eqnarray}
so that
\begin{eqnarray}
A^{(a)}_{i} = A^{(c)}_{i} - \frac{\partial_{i}}{\partial_{3}}A^{(c)}_{3}.
\label{Ap7}
\end{eqnarray}
On the other hand, $\Pi_{i}$ is gauge invariant, 
$\Pi_{i}^{(a)} = \Pi_{i}^{(c)}$. Hence, we have:
\begin{eqnarray}
\{A_{i}(x), \Pi_{j}(y)\}_{DB}^{(a)} = \{A_{i}(x)- 
\frac{\partial_{i}}{\partial_{3}}A_{3}(x), \Pi_{j}(y)\}_{DB}^{(c)}
\label{Ap70}
\end{eqnarray}
Using the Coulomb gauge result (\ref{Ap4}), 
the axial gauge algebra (\ref{Ap8}) is correctly reproduced.
\section{Appendix B}
We develop the symplectic formalism in this appendix and show the 
connection between integral curves and the Hamilton's equations of motion in
the time-reparametrized version. 
\vskip 0.003cm
\noindent Let $Q = R \times Q_{0}$, ($Q_{0} 
= {q^{i}(t), i = 1,2,...,n}$), 
be a $n + 1$--dimensional configuration space 
which includes time $t$. The
corresponding phase-space $\Gamma$ is $2n + 2$--dimensional 
with coordinates $(t, q^{i}, p_{t}, p_{i})$. 
A function $F(t, q^{i}, p_{t}, p_{i})$ 
on this phase-space is defined as follows:
\begin{eqnarray}
F(t, q^{i}, p_{t}, p_{i}) = p_{t} + H_{0}(q^{i}, p_{i}).
\label{ap8}
\end{eqnarray}
Also let $\tilde \theta  = p_{t}dt + p_{i}dq^{i}$ be a $1$-form on $\Gamma$.
Now let $\Sigma$ be a sub-manifold of $\Gamma$ 
defined by $F(t, q^{i}, p_{t}, p_{i}) = 0$.
Restricting $\tilde \theta$ to $\Sigma$, we get:
\begin{eqnarray}
\tilde \theta|_{\Sigma} = -H_{0}(q^{i}, p_{i})dt + p_{i}dq^{i}. 
\label{ap9}
\end{eqnarray}
An arbitrary tangent vector $\vec{X}$ to a curve in $\Sigma$ is given by:
\begin{eqnarray}
\vec{X} = u\frac{\partial}{\partial t} + v^{j}(q^{i}, p_{i})
\frac{\partial}{\partial q^{j}} + f_{j}(q^{i}, p_{i})
\frac{\partial}{\partial p_{j}}
\label{ap10}
\end{eqnarray}
with $u$, $v^{j}$ and $f_{j}$'s being arbitrary coefficients.
\vskip 0.003cm
\noindent Demanding that the $2$-form 
$\tilde \omega = d\tilde\theta|_{\Sigma}$ 
is degenerate, i.e., $\exists$ $\vec{X} \not= 0$, 
such that upon contraction, the one-form 
$\tilde \omega (\vec{X}) = 0$, 
we immediately obtain the following equations:
\begin{eqnarray}
f_{i} + u\frac{\partial H_{0}}{\partial q^{i}} = 0
\label{ap11}
\end{eqnarray}
\begin{eqnarray}
-v_{i} + u\frac{\partial H_{0}}{\partial p_{i}} = 0~.
\label{ap12}
\end{eqnarray}
Hence eq.(\ref{ap10}) can be written as
\begin{eqnarray}
\vec{X} = u\left(\frac{\partial}{\partial t} + 
\frac{\partial H_{0}}{\partial p_{i}}
\frac{\partial}{\partial q^{i}} 
- \frac{\partial H_{0}}{\partial q^{i}}
\frac{\partial}{\partial p_{i}}\right).
\label{ap12a}
\end{eqnarray}
Now recall that an integral curve of a vector field is a curve
such that the tangent at any point to this curve gives 
the value of the vector field at that point.
\vskip 0.003cm
\noindent In general, any tangent vector field $\vec{X}$ 
to a family of curves,
parametrised by $\tau$, in the space $\Sigma$ can be 
written as
\begin{eqnarray}
\vec{X} &=& \dot{x}^{\mu}\partial_{\mu} 
\quad; \quad \dot{x}^{\mu} = \frac{dx^{\mu}}{d\tau}\nonumber\\
&=& \dot{t}\frac{\partial}{\partial t} + \dot{q}^{i}
\frac{\partial}{\partial q^{i}} + \dot{p}_{i}
\frac{\partial}{\partial p_{i}}~. 
\label{ap12b}
\end{eqnarray}
The equations of the integral curves 
(obtained by comparing eq.(s) (\ref{ap12a}, \ref{ap12b})) are given by:
\begin{eqnarray}
\dot{q}^{i} = u\frac{\partial H_{0}}{\partial p_{i}} \quad,
\quad \dot t = u\quad,\quad \dot{p}_{i} 
= -u\frac{\partial H_{0}}{\partial q^{i}}~.
\label{ap12c}
\end{eqnarray}
Note that we recover 
the usual Hamiltonian equations of motion in the $t = \tau$ gauge. 
It is the parameter $u$ which is responsible 
for inducing the time reparametrization invariance.
\vskip 0.003cm
\noindent Now we consider the example of a NR particle 
in $1 + 1$-dimension, the Hamiltonian of which reads: 
\begin{eqnarray}
H_{0} = \frac{p_{x}^2}{2m}~.
\label{ap130}
\end{eqnarray}
In $1 + 1$-dimension, the equations of the integral curves (\ref{ap12c}) 
can be rewritten as 
\begin{eqnarray}
\dot{x} = u\frac{\partial H_{0}}{\partial p_{x}}\quad,\quad\dot{t} 
= u\quad,\quad \dot{p_{x}} = -u\frac{\partial H_{0}}{\partial x}~.
\label{ap18}
\end{eqnarray}
Substituting the form of the Hamiltonian (\ref{ap130}) in eq.(\ref{ap18}), 
we obtain:
\begin{eqnarray}
p_{x} = \frac{m\dot x}{\dot t} = m\frac{dx}{dt} = constant
\label{ap19}
\end{eqnarray}
which is the equation of the integral curve. 
Note that the above form of the canonical 
momentum is independent of the parameter $u$. 
This establishes a connection between the integral curve on 
$\Sigma$ and the canonical momenta. 
Also from eq.(s) (\ref{ap8}, \ref{ap130}), we have:
\begin{eqnarray}
p_{t} + \frac{p_{x}^2}{2m}= 0
\label{ap20}
\end{eqnarray}
which is nothing but the first class constraint (\ref{N7}) in the 
time reparametrized version of the NR particle.
Hence, the constraint of the time
reparametrized theory is also obtained from the integral curve. 
The connection between the integral curves and the constraints
for the other models discussed in the chapter can be
shown in a similar way following the above approach.

\chapter{Seiberg-Witten map and violation of Galilean 
symmetry in a noncommutative planar system}
As we have mentioned earlier, motivated by string theory,
noncommutative spacetimes have drawn considerable attention
in field theories \cite{switten}, \cite{szaborj}, \cite{schom}, 
\cite{doug}, quantum mechanics 
\cite{gamboa, mezincescu, tur, tur1, lor, ho, pol, muthu, bar, bcsgas}
as well as for their phenomenological implications
\cite{carroll}, \cite{rizzo}, \cite{lebed}, \cite{dine}, \cite{carone},
\cite{tur1}, \cite{lor}. 
One of the most interesting things in noncommutative field theories is that
even the $U(1)_{\star}$ gauge group has non-Abelian-like characterestics 
such as self-interactions.

\noindent On the other hand,
investigations towards violation of Lorentz symmetry 
in noncommutative systems steming from a fundamental 
length scale provided 
by noncommutative parameter $\theta$ have gained considerable momentum 
in recent literature. It is generally assumed 
that there is  no spacetime noncommutativity ($\theta^{0i} = 0$), 
in order to avoid any non-unitarity in 
quantum field theory based on it. Another reason 
for assuming $\theta^{0i} = 0 $ is to avoid higher 
order time derivatives in the action\footnote{In a series 
of fundamental papers Doplicher et.al \cite{dop1, dop11} 
have however shown in complete 
generality that one can construct unitary 
quantum field theory even when $\theta^{0i}\neq 0$.}. 
Clearly, the condition  $\theta^{0i} = 0 $ spoils the manifest 
Lorentz symmetry right in the beginning 
and this is true, irrespective of whether 
one works with the original theory involving noncommutative 
variables or with an equivalent effective theory in 
terms of ordinary commutative variables obtained by 
SW map \cite{switten}, \cite{sw}, 
whenever applicable\footnote{By the term ``applicable", 
we mean that SW map can be only applied 
to fields which transform appropriately under gauge 
transformation in presence of a gauge symmetry. 
For example, SW map cannot be applied to  
real scalar fields, which do not transform under 
(local) gauge transformation.}. These two methods 
of analysis need not always be equivalent. 
For example, the IR problem found in noncommutative field 
theory \cite{mats, mats10} is not present in the commutative 
variable approach \cite{bicsu}, revealing an 
equivalence at best on a perturbative level. 
For the latter method, one can, for example, 
consider the action of $U(1)_{\star}$ Maxwell gauge theory, 
which when rewritten in terms of commutative variables 
using SW map, develops certain $\theta$-dependent terms, 
in addition to the standard ones, which are manifestly 
Lorentz invariant (non-invariant) if $\theta^{\mu \nu}$ 
transforms like a tensor (non-tensor and fixed for all frames) 
\cite{rbbckul}. These two cases correspond to observer and particle 
Lorentz transforms \cite{carol}. It is to be noted that the 
violation of Lorentz symmetry by extremely tiny 
$\theta^{\mu\nu}$ term is relevant at a very 
short distance or equivalently at a very high energy scale. 
Consequently, these additional $\theta$-dependent 
correction terms can be treated as perturbations. 
As a result, the noncommutative quantum field theory 
is practically considered 
Lorentz invariant in zeroth order in $\theta^{\mu\nu}$, 
with the first order corrections coming from the 
expansion of star product and SW map. 
Various aspects of noncommutative quantum mechanics 
have also been studied, which are usually formulated 
through Schr\"{o}dinger equation written 
in terms of noncommutative wave functions $\hat\psi$. 
Clearly this is in the NR framework. 
The presence of the star product can give rise 
to some new features like, say in presence 
of potential terms, the star product 
expansion gives rise to a Bopp shift 
\cite{bop}, \cite{bop10}, \cite{gamboa} 
in the arguement of the potential. 
Besides, the presence of exotic Galilean symmetry 
have also been found in various noncommutative
quantum mechanical model \cite{dh}.
\vskip 0.003cm
\noindent In this chapter, we study a 
planar noncommutative NR system 
coupled to a $U(1)_{\star}$ gauge field. 
Since the above mentioned condition 
$\theta^{0i} = 0$ is Galilean invariant, 
it is therefore quite interesting to look for 
any violation in Galilean symmetry 
in any NR noncommutative system where matter 
field is coupled to noncommutative gauge fields. 
As in their relativistic counterparts 
(as mentioned above), we shall be 
looking for this violation through 
an effective theory obtained by SW map. 
Interestingly, one can also now carry 
out quantum mechanical analysis in first quantized 
formalism from the Schr\"{o}dinger equation 
derived from this effective theory. 
The main motivation for carrying out this 
investigation in NR framework 
is that here the transition from second 
quantization to first quantization 
is rather quite straightforward, and infact 
first and second quantized formalism 
are completely equivalent as far as Galilean 
invariant models are concerned. 
It may be recalled that there is no particle 
production in a Galilean invariant field theory. 
Also, an $N$-particle state can be constructed by 
superposing in terms of first 
quantized $N$-particle wave functions, 
the states obtained by $N$-fold actions of the 
creation operators on the vacuum. 
Thus, if one restricts the $N$-particle sector, 
while quantizing canonically, one recovers the 
first quantized $N$-particle wave 
functions. So although the noncommutative $\hat \psi$ field 
in Schr\"{o}dinger equation on a plane can have 
an interpretation of probability amplitude, 
it is not clear that this feature will persist 
with the SW field $\psi$ when an effective 
commutative theory is obtained from the original 
noncommutative theory through the use of SW map. We find in this chapter 
that unless the gauge 
field configuration is such that the corresponding magnetic 
field is constant, 
the probabilistic interpretation will not go through. This 
indicates that the nature of the gauge field must 
be of ``background" type, rather than a dynamical one. 
As in the relativistic case, we shall analyse this problem 
by writing down an effective theory of the 
original noncommutative Schr\"{o}dinger action coupled 
to background $U(1)_{\star}$ gauge theory 
in terms of usual commutative variables by using SW map. 
After setting up the formalism, we identify the 
physical variables by proper ``renormalisation" 
of wave function and mass to 
identify the probability current appropriate 
for the first quantized formalism. 
\vskip 0.003cm
\noindent Finally, as an example, we take up the case 
of Hall conductivity in noncommutative plane. 
In this context, we would like to point out an 
important aspect of this noncommutativity which has its deep connection with 
Quantum Hall systems \cite{jb}. Lots of authors have made quite an extensive 
study of this deep connection \cite{su}--\cite{su9}. 
To start with, the simple problem of Landau 
level and Hall conductivity in noncommutative plane 
was addressed by a number of authors 
\cite{dh, du, low, jelal, ko}. However, the results of various authors 
do not seem to be 
convergent on the issue of effect of the 
noncommutative parameter $\theta$ on Hall conductivity; some 
show deviations and others show no deviations 
from usual commutative theory. 
Note that these analysis and their 
subsequent results involve noncommutative electric 
and magnetic fields, which in general are 
not gauge invariant objects even for the 
simplest $U(1)_{\star}$ gauge group; 
they rather transform covariantly. Consequently they 
cannot correspond to any observables 
in a generic case. This limitation 
can be avoided, for example, 
by writing an effective theory in ordinary 
commutative space by making use of 
SW map \cite{switten} and compute Hall conductivity 
in terms of the usual $U(1)$ gauge invariant 
electric and magnetic fields \cite{ja}. 
This will clearly open another avenue 
to compare with the existing results in the literature. 
Here, we would also like to mention that in a recent paper \cite{kod},
the authors also have analysed this problem by 
using a modified  norm--preserving 
unitarised SW map and have studied the effect of 
noncommutativity in Hall systems apart from 
Aharanov--Bohm effect. In contrast, 
in this chapter we apply the usual SW map to 
construct an effective commutative theory and 
identify the probability current after wave 
function and mass renormalisation. This in turn, is used to 
compute the Hall conductivity.

\section{The Seiberg-Witten map}
The SW map has been an important
ingredient in the analysis of noncommutative 
quantum field theories.  The rational behind this 
map derives from the observation that commutative 
and noncommutative field theories result from 
different regularizations of the same gauge theory,
at least in two dimensions.  Thus, 
a map should exist between these 
theories which reflects the fact
that the physical content of the two theories is the same.
In this section we shall present a brief review of this celebrated
map \cite{switten} which has played a very important role
in the study of noncommutative quantum field theory and will also play
a significant role in the rest of this chapter. 
\vskip 0.003cm
\noindent It is an explicit map connecting a given noncommutative gauge
theory with a conventional gauge theory.
Let us consider the case in which the noncommutative gauge
theory is governed by a Yang-Mills (YM) Lagrangian for the gauge potential
$\hat A_\mu$, transforming under gauge transformations according to
\begin{eqnarray}
\hat \delta_{\hat \varepsilon} A_\mu(x) &=&
\hat A'_\mu(x) - \hat A_\mu(x) =   D_\mu[\hat A]\hat \varepsilon(x)~.
\label{seiberg1}
\end{eqnarray}
The SW map
connects the noncommutative YM Lagrangian
to some unconventional Lagrangian on the commutative side.
What is conventional in the latter, apart from the fact  that
fields are multiplied with the ordinary product 
is that the transformation law for the
gauge field $A_\mu$ is governed by the ordinary covariant derivative:
\begin{eqnarray}
\delta_{\varepsilon} A_\mu(x) = A'_\mu(x) - A_\mu(x) =  D_\mu[A]
\varepsilon(x).
\label{seiberg2}
\end{eqnarray}
Note that we are calling $\hat \varepsilon$, the infinitesimal gauge
transformation parameter in the noncommutative theory to distinguish
it from $\varepsilon$, its mapped counterpart in the ordinary theory.
Hence, the mapping
should include, apart from a connection between
$\hat A_\mu$ and $A_\mu$, one for connecting $\hat \varepsilon$ 
and $\varepsilon$.
\vskip 0.003cm
\noindent It turns out that the equivalence holds at the
level of orbit space,
the physical configuration space of gauge
theories. This means that if
two gauge fields $\hat A_\mu$ and $\hat A'_\mu$ 
belonging to the same orbit can be
connected by a noncommutative gauge transformation 
$\exp_{*}(i \hat \varepsilon)$,
then $A'_\mu$
and $A_\mu$, the corresponding mapped gauge fields will also be gauge
equivalent by an ordinary gauge transformation $\exp(i \varepsilon)$.
An important
point is that  the mapping between $\hat \varepsilon$
and $\varepsilon$ necessarily depends on $A_\mu$. 
Indeed, if $\hat \varepsilon$ were
a function of $\varepsilon$ solely, 
the ordinary and the noncommutative gauge
groups would be  identical. That this is not possible can be seen just by
considering the case of a  $U(1)$ gauge theory in which, through
a redefinition of the gauge parameter, one would be establishing an
isomorphism between noncommutative $U_*(1)$ and commutative 
$U(1)$ gauge groups.
\vskip 0.003cm
\noindent Then, the SW mapping consists in finding
\begin{eqnarray}
\hat A &=& \hat A [A;\theta] \nonumber \\
\hat \varepsilon &=& \hat \varepsilon [\varepsilon, A;\theta]
\label{seiberg3}
\end{eqnarray}
so that the   equivalence between orbits holds
\begin{eqnarray}
\hat A [A] + \hat\delta_{\hat\varepsilon}\hat A [A]&=&
\hat A[A + \delta_\varepsilon A].
\label{seiberg4}
\end{eqnarray}
Using the explicit form of gauge transformations and
expanding to first order in $\theta = \delta \theta$, the solution of
eq.(\ref{seiberg4}) reads:
\begin{eqnarray}
&&\hat A_\mu[A] = A_\mu -\frac{1}{4}
 \delta\theta^{\rho \sigma}
 \{A_\rho, \partial_\sigma A_\mu + F_{\sigma \mu}
 \} + O(\delta\theta^2) \nonumber\\
&& \hat\varepsilon [\varepsilon,A] = \varepsilon + \frac{1}{4}
\delta\theta^{\rho \sigma}[\partial_\rho \varepsilon , 
A_\sigma\} + O(\delta\theta^2)
\label{seibergsolsw}
\end{eqnarray}
where the products on the right hand side, such as 
$\{A_\rho, \partial_\sigma A_\mu\}=A_\rho.\partial_\sigma A_\mu
+ \partial_\sigma A_\mu.A_\rho$ are ordinary matrix products.
\vskip 0.003cm
\noindent Concerning the field strength, the connection is given by:
\begin{eqnarray}
\hat F_{\mu\nu}[A] = F_{\mu \nu} +\frac{1}{4}\delta\theta^{\alpha \beta}
\left(
2\{F_{\mu \alpha},F_{\nu \beta}\} - \{A_\alpha, D_\beta F_{\mu\nu}
+ \partial_\beta F_{\mu\nu}\}
\right) + O(\delta\theta^2).
\label{seibergswf}
\end{eqnarray}
One can interpret these equations as differential 
equations describing the passage from
$A_\mu^\theta$ (the gauge field in a theory with parameter $\theta$) to
$A_\mu^{\theta + \delta \theta}$ (the gauge field 
in a theory with parameter $\theta
+ \delta \theta$). Integrating these equations leads to the passage 
from $L_{YM}[\hat A]$ 
(the noncommutative version of YM
Lagrangian), to $L[A,\theta]$ which is a complicated 
but commutative equivalent Lagrangian to all orders in $\theta$.

\section{U(1)$_{\star}$ gauge invariant Schr\"{o}dinger action}
\label{gauge}
We start with the action of a Schr\"{o}dinger field $\psi$ coupled with
$U(1)$ background gauge field $A_{\mu}(x)$ in the  ordinary commutative space
\begin{eqnarray}
S = \int d^{3}x  \psi^{\dag}(iD_{0} + \frac{1}{2m}D_{i}D_{i})\psi 
\label{gal1}
\end{eqnarray}
where, $D_{\mu} = \left(\partial_{\mu} - ig A_{\mu}\right)$ 
is the covariant derivative operator and $g$ is the coupling constant. \\
The corresponding $U(1)_{\star}$ 
gauge invariant action in noncommutative space is
\begin{eqnarray}
\hat S = \int d^{3}x \hat \psi^{\dag}\star(i\hat D_{0} + 
\frac{1}{2m}\hat D_{i}\star \hat D_{i})\star \hat \psi 
\label{gal2}
\end{eqnarray}
where the caret notation indicates noncommutative 
nature of the variables $\hat \psi$ 
(assumed to be Schwartzian \cite{szaborj}) 
which compose through the star product 
(introduced earlier in eqn.(\ref{Weylprodcoord})) defined as
\begin{eqnarray}
\left(\hat f\star \hat g\right)(x) = e^{\frac{i}{2}\theta^{\alpha\beta}
\partial_{\alpha}\partial^{'}_{\beta}} 
\hat f(x)\hat g(x^{'})\big{|}_{x^{'}=x}~. 
\label{gal3}
\end{eqnarray}
Under $ \star $ composition the Moyal bracket between the coordinates is 
\begin{eqnarray}
\left[\hat x^{\mu},\hat x^{\nu}\right]_{\star} = i\theta^{\mu\nu} 
\label{gal4}
\end{eqnarray}
which is isomorphic to the algebra of operator
valued coordinates in noncommutative space 
$\left[x^{\mu}_{op}, x^{\nu}_{op}\right] 
= i\theta^{\mu\nu}$. Also $\left(\hat D_{\mu}\star = 
\partial_{\mu} - ig \hat A_{\mu}\star\right)$ 
is the appropriate covariant derivative operator in noncommutative space.
Under the simultaneous $U(1)_{\star}$ gauge transformation
\begin{equation}
\hat \psi(x) \mapsto \hat \psi^{'}(x) = \hat U(x) \star \hat \psi(x)
\label{gal6}
\end{equation}
\begin{equation}
\hat A_{\mu}(x) \mapsto \hat A^{'}_{\mu}(x) =  
\hat U(x) \star \hat A_{\mu}(x) \star \hat U^{\dag}(x) 
+ \frac{i}{g}\hat U(x) \star \partial_{\mu}\hat U^{\dag}(x)  
\label{gal7}
\end{equation}
where $\hat U(x)$ is the star unitary function satisfying
\begin{equation}
\hat U(x) \star \hat U^{\dag}(x) = \hat  U^{\dag}(x) \star \hat U(x) = 1 
\label{gal8}
\end{equation}
one can show that $\left(\hat D_{\mu}\star \hat \psi \right)\rightarrow
\left(\hat D^{'}_{\mu}\star \hat \psi^{'}\right) = 
\hat U(x)\star \left(\hat D_{\mu}\star \hat \psi \right)$, 
i.e it transforms covariantly. 
Note that $\hat U^{\dag}(x)$ is not equal to 
$\hat U^{-1}(x)$ unless $\hat U(x)\in U(1)_{\star}$-the rank 1 gauge group.\\
\noindent The equation of motion for the fundamental field $\hat \psi(x)$  is
\begin{eqnarray}
(i \hat D_{0} + \frac{1}{2m} \hat D_{i} \star \hat D_{i}) \star \hat \psi = 0.
\label{gal16}
\end{eqnarray}
The usual $\star$--gauge invariant matter or 
probability current density $\hat j_{\mu}$ following 
from eq.(\ref{gal16}) is given by:
\begin{eqnarray}
\hat j_{0} = \hat \rho = \hat \psi^{\dag} \star \hat \psi
\label{gal17}
\end{eqnarray}
\begin{eqnarray}
\hat j_{i} = \frac{1}{2mi} \left[\hat \psi^{\dag} \star \left(\hat D_{i} 
\star \hat \psi \right) -
\left(\hat D_{i} \star \hat \psi \right)^{\dag}\star \hat \psi  \right]
;\quad(i = 1,2)
\label{gal18}
\end{eqnarray}
which satisfy the usual continuity equation
\begin{eqnarray}
\partial_{t}\hat j_{0} + \partial_{i}\hat j_{i} = 0.
\label{gal19}
\end{eqnarray}
Here, we would like to mention that $\hat j_{0}$ is not 
manifestly positive definite. However, it can be made 
so by modifying it by adding a suitable total divergence term, 
so that $\hat j_{0}$ (upto a divergence term) can be 
regarded as a probability density and corresponding 
$\hat j_{i}$'s as probability currents when we 
switch over to first quantized version 
from the second quantized one.
One can at this stage add a $\star$--gauge invariant 
dynamical term $-\frac{1}{4}\int d^{n}x 
\hat F_{\mu\nu}\star \hat F^{\mu\nu}$ to the action 
(\ref{gal2}) where the  field strength $\hat F_{\mu\nu}$ is defined as 
$\hat F_{\mu\nu} = \frac{i}{g}\left[\hat D_{\mu},
\hat D_{\nu}\right]_{\star} = 
 \partial_{\mu}\hat A_{\nu} - 
\partial_{\nu}\hat A_{\mu} - ig 
\left[\hat A_{\mu},\hat A_{\nu}\right]_{\star}$, 
and identify a $U(1)_{\star}$ charge 
current density $\hat J^{\mu}$ through the 
equation of motion for the $\hat A_{\mu}$ field 
$\tilde D_{\nu} \star \hat F^{\mu\nu}  = 
\hat J^{\mu}$ where $\tilde D_{\mu}\star:= 
\partial_{\mu} - ig \left[\hat A_{\mu} , {\quad   }\right]_{\star}$.
The explicit form of $\hat J_{\mu}$ is given by: 
\begin{eqnarray}
\hat J_{0} = g \hat \psi \star \hat \psi^{\dag} 
\label{gal13}
\end{eqnarray}
\begin{eqnarray}
\hat J_{i} = \frac{g}{2mi} \left[ \left(\hat D_{i} 
\star \hat \psi \right) \star
\hat \psi^{\dag} - \hat \psi \star \left(\hat D_{i} 
\star \hat \psi \right)^{\dag} \right] ;\quad(i = 1,2)
\label{gal14}
\end{eqnarray}
Unlike $\hat j_{\mu}$, $\hat J_{\mu}$ 
are not $ U(1)_{\star}$ gauge invariant,
rather they transform covariantly and satisfy a 
covariant version of continuity equation 
$\hat D_{0} \hat J_{0} + \hat D_{i} \hat J_{i} = 0$.
After identifying $\hat J_{\mu}$, we can do away with the dynamical term
and deal with the Galilean invariant action (\ref{gal2}) itself.
Note that similar covariant transformation property holds for
$\hat F_{\mu\nu}$,
i.e. $\hat F_{\mu\nu} \mapsto \hat F^{'}_{\mu\nu} =
\hat U(x) \star \hat F_{\mu\nu} \star \hat U^{\dag}(x)$.
This is reminescent of what happens in Yang-Mills theory.
Consequently, a generic configuration 
for $\hat F_{\mu\nu}$ (except for the special case of $\hat F_{\mu\nu} = $
constant)
does not remain $U(1)_{\star}$ gauge invariant. $\hat F_{\mu\nu}$
therefore, does not correspond to an observable.
A $U(1)_{\star}$ gauge invariant
noncommutative Chern-Simons action
$\hat S_{cs} \sim \int d^{3}x \epsilon^{\mu \nu\lambda}
\left\{\hat A_{\mu}\star\partial_{\nu}\hat A_{\lambda}
+ \frac{2i}{3}\hat A_{\mu}\star\hat A_{\nu}\star\hat A_{\lambda}
\right\}$ could also be added to eq.(\ref{gal2}), 
instead of the noncommutative Maxwell term,
as this dynamical term is not associated with any ``photon"
and can be coupled to  NR matter fields without apparently
spoiling the Galilean symmetry, if there is no spacetime noncommutativity
($\theta^{0i} = 0$). 


\section{Effective Theory constructed in commutative space}
\label{effectivetheory}
We now move on to construct an effective action 
starting from eq.(\ref{gal2}) by 
using the SW map in the lowest order in $\theta^{\mu\nu}$ \cite{sw}: 
\begin{eqnarray}
\hat \psi = \psi - \frac{1}{2}\theta^{mj}A_{m}\partial_{j}\psi
\label{gal40}
\end{eqnarray}
\begin{eqnarray}
\hat A_{i} = A_{i} - \frac{1}{2}\theta^{mj}A_{m}
\left(\partial_{j}A_{i} + F_{ji}\right).
\label{gal41}
\end{eqnarray}
Taking $\theta^{0i} = 0$, we substitute the above form of $\hat \psi$
and $\hat A_{\mu}$ given by
eq.(s) (\ref{gal40}) and (\ref{gal41}) in the action (\ref{gal2}). 
After some algebra
one finds the following usual $U(1)$ gauge invariant expression
for the effective action. 
\begin{eqnarray}
\hat S &\stackrel{\rm{SW \; map}}{=}& \int d^{3}x 
\left[ \left( 1 - \frac{\theta B}{2}\right)
\left( \psi^{\dag}iD_{0}\psi\right) + 
\frac{i}{2}\theta^{mj}\left( \psi^{\dag}D_{j}\psi\right)F_{m0}\right.
\nonumber \\ 
&& \qquad \left.+ \frac{1}{2m}\left( 1 - \frac{\theta B}{2}\right)
\left( \psi^{\dag}D_{i}D_{i}\psi\right)\right.\nonumber\\ 
&& \qquad \left. + \frac{1}{2m}\theta^{mj}
\left( \psi^{\dag}D_{i}D_{j}\psi\right)F_{mi} + \frac{1}{4m} 
\theta^{mj}\left(\psi^{\dag} D_{j} \psi\right) \partial_{i} F_{mi}\right].
\label{gal42}
\end{eqnarray}
The third and fourth terms in the paranthesis can now be combined 
using the relation $F_{mi} = B\epsilon_{mi}$ to get	
\begin{eqnarray}
\hat S &\stackrel{\rm{SW \; map}}{=}&\int d^{3}x 
\left[ \left( 1 - \frac{\theta B}{2}\right)
\left( \psi^{\dag}iD_{0}\psi\right) + 
\frac{i}{2}\theta^{mj}
\left( \psi^{\dag}D_{j}\psi\right)F_{m0}\right.\nonumber \\ 
&&  \qquad +  \left. \frac{1}{2m}
\left( 1 + \frac{\theta B}{2}\right)
\left( \psi^{\dag}D_{i}D_{i}\psi\right)+
\frac{1}{4m} \theta^{mj}\left(\psi^{\dag} D_{j} \psi\right)
\partial_{i} F_{mi}\right].
\label{gal43}
\end{eqnarray}
A hermitian form of this action can easily be written by dropping 
certain boundary terms to get 
\begin{eqnarray}
\hat{S} =&& \int d^{3}x 
\left[\left( 1 - \frac{\theta B}{2}\right)
(\frac{i}{2}\psi^{\dag}\stackrel{\leftrightarrow}{D}_{0}{\psi}) 
- \frac{1}{2m}\left( 1 + \frac{\theta B}{2}\right)(D_{i}{\psi})
^{\dag}(D_{i}{\psi}) + \frac{i}{4}\theta^{mj}({\psi^{\dag}}
\stackrel{\leftrightarrow}{D_{j}}{\psi})F_{m0}\right.\nonumber 
\\  &&+\left.\frac{1}{8m} \theta^{mj}\left({\psi^{\dag}}
\stackrel{\leftrightarrow}{D_{j}}{\psi}\right) \partial_{i} F_{mi} + 
...\right]
\label{gal43A}
\end{eqnarray}
where the dots indicating missing terms, 
involving $\partial_{\mu}F_{\nu\lambda}$, 
have not been written down explicitly, as they play 
no role in the simplectic structure of the theory. 
These terms represent additional possible 
interactions. 
Note that this action is not in the canonical form. 
As a result, the field $\psi$ in second quantized formalism 
does not have a canonical structure for the equal time commutation relation 
between $\psi$ and $\psi^{\dag}$\footnote{In this section, 
we use the same notation $\psi^{\dag}(x)$ to indicate complex (hermitian) 
conjugate of $\psi$ at the classical (quantum) level. 
Also the operator nature of $\psi(x)$ and $\psi^{\dag}(x)$ 
at the quantum level is not displayed explicitly by putting a caret 
on the top; the caret is now reserved to indicate noncommutative variables. 
This, expectedly, will not give rise to any confusion as their 
respective nature should be clear from the context itself.}:
\begin{eqnarray}
\left[\psi(x), \psi^{\dag}(y)\right] = \left(1 + 
\frac{\theta B}{2}\right) \delta^{2}(x - y).
\label{gal4300}
\end{eqnarray}
Note that this commutator is easily obtained by elevating the 
DB between $\psi$ and $\psi^{\dag}$ given as 
\begin{eqnarray}
\{\psi(x), \psi^{\dag}(y)\}_{DB} = - i(1 + \frac{\theta B}{2})
\delta^{2}(x - y) 
\label{gal4300A}
\end{eqnarray}
which in turn is obtained by strong imposition of the following pair 
$(\Lambda_{a};(a = 1,2))$ of second class constraints
\begin{eqnarray}
\Lambda_{1}(x)  = \Lambda_{2}^{*}(x) = \Pi_{\psi}(x) - 
\frac{i}{2}\left(1 - \frac{\theta B}{2}\right)\psi^{\dag}(x) \approx 0
\label{gal4300B}
\end{eqnarray}
where, $\Pi_{{\psi}}$ and $\Pi_{\psi^{\dag}}(= (\Pi_{\psi})^{\dag})$ 
are the canonically conjugate momenta to 
$\psi$ and $\psi^{\dag}$ respectively. 
Since $A_{\mu}$'s are background gauge fields, 
they are not included in the 
configuration space. So we must have $\left[A_{\mu}(x), \psi(y)\right] = 0$. 
This non-standard
form of the commutation relation (\ref{gal4300}) indicates that $\psi$ 
cannot represent the basic field variable or the wave-function 
in the corresponding first
quantized formalism. This is further re-inforced by 
the observation that for the generic case, where $B$ has an $x$-dependence, 
the Euler--Lagrange equation for $\psi^{\dag}$, following 
from eq.(\ref{gal43})
\begin{eqnarray}
\left( 1 - \frac{\theta B}{2}\right) iD_{0}\psi + 
\frac{1}{2m}\left( 1 + \frac{\theta B}{2}\right)D_{i}D_{i}\psi 
+ \frac{i}{2}\theta^{mj}\left(D_{j}\psi\right) F_{m0} \nonumber\\
+ \frac{1}{4m} \theta^{mj}\left(D_{j} \psi\right)\partial_{i} F_{mi} = 0
\label{gal4301}
\end{eqnarray}
can only be brought {\it{almost}} to the form of standard Schr\"{o}dinger 
equation
\begin{eqnarray}
iD_{0}\psi + \frac{1}{2\tilde{m}}D_{i}D_{i}\psi 
+ \frac{i}{2}\theta^{mj} \left(D_{j}\psi \right)F_{m0} +\frac{1}{4m} 
\theta^{mj}\left(D_{j} \psi\right) \partial_{i} F_{mi} = 0
\label{gal4302}
\end{eqnarray}
for the first pair of terms by introducing a non-constant $\tilde{m}$ as 
\begin{eqnarray}
\tilde{m} =  \left( 1 - \theta B\right) m.
\label{gal4303}
\end{eqnarray}
To identify the basic field variable, let us scale $\psi$ as 
\begin{eqnarray}
\psi \mapsto \tilde{\psi} = \sqrt{1 - \frac{\theta B}{2}}\psi
\label{gal4304}
\end{eqnarray}
so that the commutation relation (\ref{gal4300}) can be cast as 
\begin{eqnarray}
\left[\tilde \psi(x), \tilde \psi^{\dag}(y)\right] =  \delta^{2}(x - y)
\label{gal4305}
\end{eqnarray}
and $\tilde \psi$ and $\tilde \psi^{\dag}$ can now be interpreted 
as annihilation and creation operators in second quantized formalism. 
Let us now construct $|x\rangle$  (the state corresponding to a 
single particle located at $x$) by the action of this creation 
operator acting on the normalised vacuum state 
$|0\rangle$ ($\langle 0|0\rangle = 1$) as 
$|x\rangle = \tilde \psi^{\dag}(x)|0\rangle$, 
so that the standard inner product relation 
$\langle y|x\rangle = \delta^{\left(2\right)}\left(x - y\right)$ 
and the resolution of identity ($1 = \int d^{2}x |x\rangle \langle x|$) holds. 
Now writing an arbitrary one-particle state 
$|\tilde \psi\rangle = \int d^{2}x \tilde \psi(x)|x\rangle $ 
in terms of wave function $\tilde \psi(x) = \langle x|\tilde \psi\rangle $ 
corresponding to first quantized formalism, one can easily see 
that the normalisation condition 
\begin{eqnarray}
\int d^{2}x \tilde \psi^{\dag}\tilde \psi = 1 
\label{gal4306}
\end{eqnarray}
follows trivially by demanding 
$\langle \tilde \psi|\tilde \psi \rangle = 1$. 
So this transition from second quantized to first quantized formalism clearly 
shows that it is $\tilde \psi$, rather than $ \psi$ itself, which corresponds 
to the normalised wave-function or the basic field variable in the action. 
It is therefore desirable to re-express the action (\ref{gal43}) 
in terms of $\tilde\psi$ and ensure that  it is 
in the standard form in the first pair of terms 
in both of these expressions. Clearly this can be done only for a constant 
$B$-field\footnote{In addition, if the electric field is also taken
to be constant, then the additional interaction terms in eq.(\ref{gal43A}) 
will vanish thus yielding the simplest possible action incorporating 
noncommutativity.}. Note that with this, 
$\tilde m $ (eq.(\ref{gal4303})) also becomes  constant. 
Such a constant magnetic field can only arise from an 
appropriate background gauge field. In presence of a dynamical term, 
like Chern--Simons action, $B$ cannot be ensured to be a constant 
and consequently the Schr\"{o}dinger equation describing the 
time evolution of the normalised wave-function in terms of the 
``renormalised" 
SW field $\tilde\psi$ cannot be obtained. In rest of the chapter, 
we shall therefore consider a constant background for 
field strength tensor $F_{\mu\nu}$\footnote{Since we are considering a 
constant background magnetic field, it will be advantegeous to 
consider a constant electric field background also. 
Under SW map, a constant configuration of $F_{\mu\nu}$ 
results in a constant $\hat F_{\mu\nu}$ and vice-versa 
($\hat F_{\mu\nu} =  F_{\mu\nu} - \theta^{\alpha \beta} F_{\mu\alpha} 
F_{\nu \beta}$ and $ F_{\mu\nu} = \hat  F_{\mu\nu} + \theta^{\alpha \beta}
\hat F_{\mu\alpha} \hat F_{\nu \beta}$). Also with this constant configuration, all the missing terms in eq.(\ref{gal43A}) vanish.}.
In this case, the above action (\ref{gal43A}) should be written 
in terms of $\tilde m$ (eq.(\ref{gal4303})), 
$\tilde \psi$ (eq.(\ref{gal4304})) to get
a canonical form for the Schr\"{o}dinger action 
\begin{eqnarray}
\hat S \stackrel{\rm{SW \; map}}{=} \int d^{3}x 
\left[ \left( \tilde{\psi}^{\dag}iD_{0}\tilde{\psi}\right) 
+ \frac{1}{2\tilde{m}}\left( \tilde{\psi}^{\dag}D_{i}D_{i}
\tilde{\psi}\right) + \frac{i}{2}\theta^{mj}
\left( \tilde{\psi}^{\dag}D_{j}\tilde{\psi}\right)F_{m0} \right].
\label{gal4307}
\end{eqnarray}
The field $\tilde \psi$ and mass parameter $\tilde m$ can now be regarded 
as renormalised wave-function and mass respectively. 
We shall therefore treat $\tilde{\psi}$ (and not $\psi$) 
as the basic field in our theory. 
This gives the effect of noncommutativity 
in the observed mass $\tilde{m}$. 
A point which is worth mentioning is that the expression for 
$\tilde m$ (eq.(\ref{gal4303})) 
indicates that the external magnetic field $B$ has a critical 
value  $B_{c}= \frac{1}{\theta}$. Clearly, for $B> B_{c}$, $\tilde m$ 
becomes negative which is unphysical. 
We shall see later in chapter 5, how one can define physical
quantities at this critical point $B_{c}$ and beyond it.
Using eq.(\ref{gal4303}), 
one can easily see that the ratio 
of the observed masses $\tilde{m_{1}}$ 
and $\tilde{m_{2}}$ corresponding to two distinct 
magnetic fields $B_{1}$ and $B_{2}$, satisfies 
(upto order $\theta$) 
$\frac{\tilde m_{1}}{\tilde m_{2}} 
= 1 - \theta \left(B_{1} - B_{2} \right)$ 
which in turn, can be used to get 
an estimate for noncommutative 
parameter $\theta$. 
Incidentally, this relation (\ref{gal4303}) 
was also obtained earlier by 
Duval et.al \cite{du}.\\
\noindent The equation of motion for the fundamental field 
$\tilde{\psi}$ (from the action
(\ref{gal4307})) is 
\begin{eqnarray}
 K\tilde{\psi} = 0 
\label{gal48}
\end{eqnarray}
where, $K$ is the operator given by:
\begin{eqnarray}
K = iD_{0} + \frac{1}{2\tilde{m}} D_{i}D_{i} + 
\frac{i}{2}\theta^{mj}F_{m0}D_{j}.
\label{gal480}
\end{eqnarray}
It is easy to verify 
\begin{eqnarray}
-i\left(\tilde{\psi}^{\dag}K\tilde{\psi} - 
(K\tilde{\psi})^{\dag}\tilde{\psi}\right) = \partial_{\mu}j_{\mu}
\label{gal481}
\end{eqnarray}
where, the $3(=1+2)$-currents $j_{\mu}$ are given by:
\begin{eqnarray}
j_{0} = \tilde{\psi}^{\dag}\tilde{\psi} + 
\frac{i}{2}\theta^{mj} \left(D_{m}\tilde{\psi}
\right)^{\dag}\left(D_{j}\tilde{\psi}\right)
\label{gal49cat}
\end{eqnarray}
\begin{eqnarray}
j_{i} &=& \frac{1}{2\tilde{m}i}
\left[\left\{\tilde{\psi}^{\dag}\left(D_{i}
\tilde{\psi}\right) -\left(D_{i}
\tilde{\psi}\right)^{\dag}\tilde{\psi}\right\}\right.
\nonumber\\
&& \left.+\frac{i}{2}\theta^{mj}\left\{\left(D_{m}
\tilde{\psi}\right)^{\dag}\left(D_{i}D_{j}\tilde{\psi}\right) 
+\left(D_{i}D_{j}\tilde{\psi}\right)^{\dag}\left(D_{m}
\tilde{\psi}\right)\right\} \right].
\label{gal50dog}
\end{eqnarray}
Using eq.(s) (\ref{gal48}) and (\ref{gal481}), we find that the 
continuity equation is automatically satisfied by $j_{\mu}$, 
therefore one is tempted to identify $j_{\mu}$ 
(eq.(s) (\ref{gal49cat}, \ref{gal50dog})) 
as the probability density and probability current of the system. 
But as it turns out that the probability density and 
currents have to be determined from 
$\hat j_{\mu}$ (eq.(s) (\ref{gal17},\ref{gal18})) 
as the components of this current played the role of 
gauge invariant probability density and probability current in 
noncommutative formulation (see section 4.2). 
All that we have to do here is to 
apply SW map to rewrite $\hat j_{\mu}$ in terms of field $\psi$ 
and then in terms of the 
renormalised field $\tilde \psi$ (eq.(\ref{gal4304})). 
At this stage, one can note an interesting fact that $\hat j_{\mu}$ 
also has the same form as that of $j_{\mu}$ 
(eq.(s) (\ref{gal49cat}, \ref{gal50dog}) 
except that one has to just replace $\tilde \psi$ by $\psi$ :
\begin{eqnarray}
\hat j_{0} = {\psi}^{\dag}{\psi} + \frac{i}{2}\theta^{mj} 
\left(D_{m}{\psi}\right)^{\dag}\left(D_{j}{\psi}\right)
\label{gal49}
\end{eqnarray}
\begin{eqnarray}
\hat j_{i} &=& \frac{1}{2\tilde{m}i}\left[\left\{{\psi}^{\dag}
\left(D_{i}{\psi}\right) -\left(D_{i}{\psi}\right)^{\dag}{\psi}
\right\}\right.
\nonumber\\
&& \left.+\frac{i}{2}\theta^{mj}\left\{\left(D_{m}\tilde{\psi}
\right)^{\dag}\left(D_{i}D_{j}\tilde{\psi}\right) + 
\left(D_{i}D_{j}\tilde{\psi}\right)^{\dag}\left(D_{m}
\tilde{\psi}\right)\right\} \right]
\label{gal50}
\end{eqnarray}
so that $\hat j_{\mu}$ and $ j_{\mu}$ are related by 
$j_{\mu} = (1 - \frac{\theta B}{2})\hat j_{\mu}$ as follows 
from eq.(\ref{gal4304}) and the fact that the currents are bilinear 
in their respective fields. This is not surprising as $\psi$ 
also satisfies (\ref{gal48}) ($K\psi = 0$) upto order $\theta$. 
However, note that $ \hat j_{0}$ (eq.(\ref{gal49})), does not have 
the standard form because of the presence of the $\theta$-dependent term. 
Not only that, it is not manifestly positive-definite point-wise. 
Consequently, there is a difficulty in identifying $\hat j_0$ 
as the probability density directly
in the ``first quantized" version of single-particle quantum mechanics. 
This problem can be easily seen to be, however, inherited from the original 
noncommutative formulation itself. For that recall, 
this problem was avoided there by 
modifying $\hat j_{0}$ (eq.(\ref{gal17})) 
by a total divergence term to isolate a 
positive definite quantity to be 
identified as the probability density. 
Following the same methodology here, 
we note that $\hat j_{0}$ (eq.(\ref{gal49})) 
can also be brought to almost standard form upto a 
$\left( 1 - \frac{\theta B}{2}\right)$ 
factor (assuming to be positive) 
by dropping a total divergence term, so that we have  
\begin{eqnarray}
\int d^{2}x \hat j_{0} = \left(1 - \frac{\theta B}{2} 
\right)\int d^{2}x \psi^{\dag}\psi
\label{gal51}
\end{eqnarray}
which however takes the canonical form 
\begin{eqnarray}
\int d^{2}x \hat j_{0} = \int d^{2}x \tilde \psi^{\dag}\tilde \psi
\label{gal555}
\end{eqnarray}
when rewritten in terms of renormalised wave-function 
$\tilde \psi$ (eq.(\ref{gal4304})).
With the normalisation condition (\ref{gal4306}), 
it now becomes clear that it is $\tilde{\psi}^{\dag}\tilde{\psi}$ 
(or $\hat j_{0}$ upto a total divergence term) has now to be identified 
as the probability density which is manifestly positive definite at all 
points\footnote{Note that this technique is quite common 
in quantum field theory. 
In this context, it may be recalled that the Noether's expression of 
energy-momentum tensor (say in free Maxwell theory in 3+1-dimension), 
which is nothing but the density and current of conserved energy-momentum 
four-vector, is amended by a four divergence term to render it symmetric 
and gauge invariant (Belinfante method). So here too the original 
$\hat j_{0}$ is modified by dropping a total divergence term at the 
field theoretic level to render it positive definite so that it is 
interpretable as probability density when we switch over to 
``first quantized" version of quantum mechanics. }.
It immediately follows that the spatial components of 
$\hat j_{\mu}$, i.e $\hat j_{i}$ {\it must} correspond to the 
spatial component of the probability current, as $\hat j_{\mu}$ 
satisfies the continuity equation $\partial_{\mu}\hat j_{\mu} = 0$. 
Therefore the particle current (for a single particle) $I_{i}^{(1)}$ 
in the $i$-th direction is obtained by integrating $\hat j_{i}$ 
over the variable in the orthogonal direction, 
i.e $I^{(1)}_{1} = \int dx^{2} \hat j_{1}$ and $I^{(1)}_{2} = 
\int dx^{1} \hat j_{2}$. We shall however be interested 
in the transverse current $I^{(1)}_{2}$ in section $4.5$, 
as the longitudinal current $I^{(1)}_{1}$ will vanish. 

\section{Galilean symmetry generators}
\label{galilean}
In this section we shall try to construct all the Galilean symmetry 
generators for the model defined by the action (\ref{gal4307}) where 
$\tilde \psi$ is taken to be the basic field and $F_{m0}$ representing 
the constant electric field $E_{m}$ in the background. 
The corresponding gauge field $A_{\mu}$ 
is therefore not included in the configuration space variable. 
Before we start carrying out the Hamiltonian analysis, 
we must ensure that the action is in a manifestly hermitian form. 
We therefore rewrite the action (\ref{gal43A}) in terms of 
$\tilde \psi$ (eq.(\ref{gal4304})) as
\begin{eqnarray}
\hat{S} = \int d^{3}x \left[(\frac{i}{2}\tilde{\psi^{\dag}}
\stackrel{\leftrightarrow}{D}_{0}\tilde{\psi}) - 
\frac{1}{2\tilde{m}}(D_{i}\tilde{\psi})^{\dag}(D_{i}\tilde{\psi}) + 
\frac{i}{4}\theta^{mj}\{(\tilde{\psi^{\dag}}
\stackrel{\leftrightarrow}{D_{j}}\tilde{\psi})E_{m}\right].
\label{galG1}
\end{eqnarray}
Coming to the symplectic structure, the conjugate momenta corresponding 
to the configuration space variables are 
\begin{eqnarray}
\Pi_{\tilde{\psi}}  = \frac{i}{2} \tilde \psi^{\dag}\qquad,\qquad 
\Pi_{\tilde{\psi}^{\dag}}  = -\frac{i}{2} \tilde \psi.
\label{galG2}
\end{eqnarray}
The canonical Hamiltonian density can be calculated by a 
Legendre transform which in turn can be integrated to get the Hamiltonian as 
\begin{eqnarray}
H = \int d^{2}x \left[\frac{1}{2\tilde{m}}(D_{i}
\tilde{\psi})^{\dag}(D_{i}\tilde{\psi}) - 
\frac{i}{4}\theta^{mj}\{(\tilde{\psi^{\dag}}
\stackrel{\leftrightarrow}{D_{j}}\tilde{\psi})E_{m} 
- A_{0}(\tilde{\psi^{\dag}}\tilde{\psi})\right].
\label{galG3}
\end{eqnarray}
It is clear from eq.(\ref{galG2}) that the system contains 
second-class constraints which can be strongly implemented 
by Dirac scheme to obtain the following bracket 
\begin{eqnarray}
\left\{\tilde \psi(x), \tilde \psi^{\dag}(y)\right\} = -i \delta^{2}(x - y)
\label{galG3A}
\end{eqnarray}
which in turn can be elevated to obtain the 
quantum commutator (\ref{gal4305}). 
Note that this bracket can also be obtained, in fact more simply 
by using Faddeev--Jackiw (FJ) approach as this Lagrangian (\ref{galG1}) 
is first order in time derivative.
A quick and easy calculation (using eq.(\ref{gal4305})) shows that 
the above Hamiltonian (\ref{galG3}) generates appropriate time translation 
\begin{equation}
\dot {\psi}(x) = \left\{\psi(x), H\right\}.
\label{galG4}
\end{equation}
The generator of spatial translation and 
$SO(2)$ rotation can now be easily constructed 
using Noether's theorem to get 
\begin{equation}
P_{i} = \int d^{2}x \left[\Pi_{\tilde{\psi}}\partial_{i}\tilde{\psi}(x) 
+ \Pi_{\tilde{\psi}^{\dag}}\partial_{i}
\tilde{\psi}^{\dag}(x)\right] =\int d^{2}x\frac{ i}{2}
\tilde{\psi}^{\dag}(x)\stackrel{\leftrightarrow}
{\partial_{i}}\tilde{\psi}(x)
\label{galG5}
\end{equation}
\begin{equation}
J = \frac{i}{2}\int d^{2}x \epsilon_{ij}x_{i} \tilde{\psi}^{\dag}(x)
\stackrel{\leftrightarrow}{\partial_{j}}\tilde{\psi}(x)
\label{galG6}
\end{equation}
which generates appropriate translation and rotation\footnote{The 
adjective ``appropriate'' in this context 
means the brackets $\left\{\Phi (x), \mathcal{G}\right\}$ 
are just equal to the Lie derivative 
$\left[{\mathcal{L}}_{V_{\mathcal{G}}}\left(\Phi\left(x\right)\right)\right]$ 
of a generic field $\Phi(x)$ 
with respect to the vector field $V_{\mathcal{G}}$, 
associated with the symmetry generator $\mathcal{G}$. 
We have not, of course, 
displayed any indices here. The field $\Phi(x)$ may be a 
scalar, spinor, vector or tensor field in general. 
In this case, it corresponds to the field $\tilde \psi(x)$ 
and not $A_{\mu}$ as it is a background field. 
And $\mathcal{G}$ can be, for example, the momentum ($P_{i}$) 
or angular momentum ($J$) operator generating translation 
and spatial rotation, respectively. The associated vector 
fields $V_{\mathcal{G}}$ are thus 
given as $\partial_{i}$  and  $\partial_{\phi}$, respectively ($\phi$ 
being the angle variable in the polar coordinate system in the 
two-dimensional plane). }:
\begin{equation}
\left\{\tilde\psi(x), P_{i}\right\}= \partial_{i} {\tilde\psi}(x)
\label{galG7}
\end{equation}
\begin{equation}
\left\{\tilde\psi(x),J\right\}=
\epsilon_{ij} x_{i}\partial_{j} {\tilde\psi}(x).
\label{galG8}
\end{equation}
Note that $J$ (eq.(\ref{galG6})) consists of only the orbital part of the 
angular momentum as in our simplistic treatment we have ignored the 
spin degree of freedom for the field $\tilde\psi$, 
so that it transforms as an $SO(2)$ scalar. 
Using the DB (\ref{galG3A}), one can verify the following algebra:
\begin{eqnarray}
\left\{P_{i},P_{j}\right\} = \left\{P_{i},H\right\} = \left\{J,H\right\} =0 
\nonumber\\
\left\{P_{k},J\right\} = \epsilon_{kl}P_{l}.
\label{galG81}
\end{eqnarray} 
This shows that $P_{k}$ and $J$ form a closed $E(2)$ (Euclidian) algebra. 
Now coming to the boost, we shall try to analyse the system 
from first principle and shall check the 
covariance of eq.(\ref{gal4302}) 
under Galileo boost. For this, we essentially follow \cite{bcasm}. 
To that end, we consider an infinitesimal 
Galileo boost along the $X$-direction, 
\begin{eqnarray}
t \mapsto t^{\prime} = t,\quad x^{1}{\mapsto} x^{1}{}^{\prime} = 
x^{1} - vt,\quad x^{2} \mapsto x^{2}{}^{\prime} = x^{2}
\label{galG9}
\end{eqnarray}
with an infinitesimal velocity parameter ``$v$''.
Notwithstanding the fact that Galilean spacetime $\mathcal{M}$ 
does not have a metric, one can define tangent space 
$T_{p}(\mathcal{M} )$ or its dual cotangent space 
$T_{p}^{\star}(\mathcal{M} )$ on any point $p \in \mathcal{M}$. 
The canonical basis of $T_{p}(\mathcal{M})$  
corresponding to unprimed and primed frames are thus 
given as $(\partial / \partial t, \partial / \partial x^{i} )$ and 
$(\partial / \partial t^{\prime}, \partial / \partial x^{i}{}^{\prime}) $, 
respectively. They are related as 
\begin{eqnarray}
\frac{\partial }{\partial t^{\prime}} = \frac{\partial }{\partial t} 
+ v \frac{\partial }{\partial x^{1}},\quad 
\frac{\partial }{\partial x^{i}{}^{\prime}} = 
\frac{\partial }{\partial x^{i}}~.
\label{galG10}
\end{eqnarray}
As for the transformation properties of the basic fields are concerned, 
we note that in the first quantized version $\tilde \psi$ 
is going to represent probability amplitude and  
$\tilde\psi^{\dag}\tilde\psi$ represents the probability density. 
Hence in order that $\tilde\psi^{\dag}\tilde\psi$ remains 
invariant under Galileo boost 
($\tilde\psi^{\prime \dag}(x^{\prime},t^{\prime})
\tilde\psi^{\prime}(x^{\prime},t^{\prime}) =
\tilde \psi^{\dag}(x,t)\tilde\psi(x,t)$), we expect 
$\tilde \psi$ to change atmost by a phase factor. 
This motivates us to make the following ansatz :
\begin{eqnarray}
\tilde\psi(x,t) \mapsto \tilde\psi^{\prime}(x^{\prime},t^{\prime}) = 
e^{i v \eta(x,t)}\tilde \psi(x,t) \simeq (1 + i v \eta (x,t) )
\tilde\psi(x,t))
\label{galG11}
\end{eqnarray}
for the transformation of the field $\tilde \psi$ under infinitesimal 
Galileo boost ($v<< 1$). As far as the transformation 
properties of the gauge field $A_{\mu}(x)$ is concerned, 
it should transform like the basis $\frac{\partial }
{\partial x^{\mu}}$ (eq.(\ref{galG10})) of $T_{p}(\mathcal{M} ) $.
This is because $A_{\mu}(x)$'s can be regarded as the 
components of the one-form $A(x) = 
A_{\mu}(x)dx^{\mu} \in T_{p}^{\star}(\mathcal{M} )$. 
It thus follows that
\begin{eqnarray}
A_{0}(x)\mapsto  A_{0}{}^{\prime}(x^{\prime}) &=&A_{0}(x)+ v A_{1}(x)
\nonumber\\A_{i}(x)\mapsto  A_{i}{}^{\prime}(x^{\prime}) &=& A_{i}(x)
\label{galG12}
\end{eqnarray}
under Galileo boost. Now demanding that the action (\ref{gal4307}) 
remains invariant or equivalently the equation of motion 
(\ref{gal48}, \ref{gal480}) 
remains covariant implies that the following pair of equations
\begin{eqnarray}
 iD_{0}\tilde \psi + \frac{1}{2\tilde{m}} D_{i}D_{i}\tilde \psi + 
\frac{i}{2}\theta^{mj}E_{m}D_{j}\tilde \psi = 0
\label{galG13}
\end{eqnarray}
\begin{eqnarray}
 iD_{0}{}^{\prime}\tilde \psi^{\prime} + \frac{1}{2\tilde{m}} 
D_{i}{}^{\prime}D_{i}{}^{\prime}\tilde \psi^{\prime} + 
\frac{i}{2}\theta^{mj}E_{m}{}^{\prime} D_{j}^{\prime}
\tilde \psi^{\prime} = 0
\label{galG14}
\end{eqnarray}
must hold in unprimed and primed frames respectively. 
Now making use of eq.(s) (\ref{galG10},\ref{galG11},\ref{galG12}) 
in eq.(\ref{galG14}) 
and then using eq.(\ref{galG13}) we get the following 
condition involving $\eta$ :
\begin{eqnarray}
D_{1}\tilde \psi + i \partial_{0}\eta \tilde \psi 
= \left[-\frac{1}{\tilde m} \partial_{j} \eta - 
\frac{\theta}{2} \epsilon^{ij} F_{i1}\right] D_{j} \tilde \psi +
\left[- \frac{1}{2\tilde m} \nabla^{2}\eta - \frac{\theta}{2} 
\epsilon^{ij} E_{i}\partial_{j} \eta\right]\tilde \psi~.
\label{galG15}
\end{eqnarray}
Since we have considered the boost along the $x$-axis, the variable $\eta$ 
occuring in the phase factor in eq.(\ref{galG11}) will not have any $x^{2}$ 
dependence. Consequently we can set $\partial_{2}\eta = 0$. 
Also since we have taken the background electric field $E_{i}$ 
to be constant, we have to consider here two independent possibilities : 
$E$ along the direction of the boost and $E$ perpendicular 
to the direction of the boost. Let us consider the former possibility first. 
Clearly in this case the term $\epsilon^{ij} E_{i}\partial_{j} \eta $ 
in the right hand side of eq.(\ref{galG15}) 
vanishes and the above equation becomes 
\begin{eqnarray}
D_{1}\tilde \psi + i \left(\partial_{0}\eta\right)\tilde 
\psi = \left[-\frac{1}{\tilde m} \partial_{1} \eta - 
\frac{\theta B}{2} \right] D_{1} \tilde \psi - 
\frac{1}{2\tilde m}\left(\partial_{1}^{2}\eta\right) 
\tilde \psi~.
\label{galG16}
\end{eqnarray}
Equating the coefficients of $D_{1}\tilde \psi$ and $\psi$ from both 
sides we get the following conditions on $\eta$.
\begin{eqnarray}
\left[\frac{1}{\tilde m} \partial_{1} \eta + \frac{\theta B}{2} \right] = -1
\label{galG17}
\end{eqnarray}
\begin{eqnarray}
i\partial_{0}\eta = - \frac{1}{2\tilde m}\partial_{1}^{2}\eta~.
\label{galG18}
\end{eqnarray}
It is now quite trivial to obtain the following time-independent 
($\partial_{0}\eta = 0$) real solution for $\eta$ :
\begin{eqnarray}
\eta = - \tilde m \left(1 + \frac{\theta B}{2}\right) x^{1}~.
\label{galG19}
\end{eqnarray}
This shows that boost in the direction of the electric field is a 
symmetry for the system. This is, however, not true 
when electric field is perpendicular to the direction of the boost. 
This can be easily seen by re-running the above analysis for this case, 
when instead of eq.(\ref{galG18}) one gets 
\begin{eqnarray}
 i\partial_{0}\eta = - \frac{1}{2\tilde m}\partial_{1}^{2}\eta + 
\frac{\theta E}{2}\partial_{1}\eta
\label{galG20}
\end{eqnarray} 
along with eq.(\ref{galG17}) which, however, remains unchanged. 
Clearly this pair (eq.(s) (\ref{galG17}, \ref{galG20})) 
does not admit any real solution. 
In fact, the solution can just be read off as 
\begin{eqnarray}
\eta = - \tilde m \left(1 + \frac{\theta B}{2}\right) x^{1} + 
\frac{i}{2} \theta E \tilde m t~.
\label{galG21}
\end{eqnarray}
This complex solution of $\eta$ implies the wave-function (\ref{galG11}) 
does not preserve its norm under this boost transformation 
as this transformation is no longer unitary. This demonstrates 
that the boost in the perpendicular direction of the 
applied electric field is 
not a symmetry of the system. This is clearly a noncommutative effect 
as it involves the noncommutative parameter $\theta$. 
This violation of boost symmetry rules out the possibility 
of Galilean symmetry, let alone any exotic Galilean symmetry 
obtained by \cite{dh} in their model. We shall however see in
chapter 7 that a twisted version of Galileo group can be made
compatible with noncommutativity. Indeed, it turns out that the
Galilean boost symmetry is taken care of rather trivially there,
despite the appearance of mass as a central charge in Galilean algebra.

\section{Hall Conductivity in commutative variables}
In this section, we are going to compute the effect of 
noncommutativity on Hall conductivity, if any, 
using the formalism we have developed in section $4.3$, 
in particular eq.(s) (\ref{gal48}), (\ref{gal480}). 
The violation of Galilean boost symmetry observed
in the preceding section is not expected to interfere with this computation.
Admittedly, the value of $\theta$ is very small, if it has its origin in the
fundamental noncommutativity of nature, if any. On the other hand, 
the presence of electric field is known to 
lift the degeneracy of the Landau level, 
but the basic noncommutativity of the 
coordinates of the particle confined in the first 
Landau level (given in terms  of the reciprocal of the magnetic field) 
is expected to persist even in the presence 
of a very small electric field. 
If this is really true then the value of the 
noncommutative parameter may be appreciable even for any 
condensed matter experiment that one can think of. 
We are however not going to discuss about this issue any further.
The sole objective of the following exercise is to just illuminate 
the formalism we have developed so far.\\
\noindent Hence we now take up the problem of Hall effect in terms of 
commutative variables and
attempt to solve the equation of motion (\ref{gal48}) in Landau gauge.
\begin{eqnarray}
A_{0} = E x^{1} , A_{1} = 0 , A_{2} = B x^{1}~.
\label{gal53}
\end{eqnarray}
Taking the trial solution of standard Landau gauge problem, 
appropriate for the gauge fixing condition (\ref{gal53})
\begin{eqnarray}
\tilde \psi(t, x^{1}, x^{2}) = e^{-i\omega t}e^{ip_{2} x^{2}}\phi(x^{1})
\label{gal530}
\end{eqnarray}
we obtain
\begin{eqnarray}
\left[\omega + Ex^{1} - \frac{1}{2\tilde m}\left\{- \partial_{1}^{2} + 
\left(p_{2} - Bx^{1}\right)^{2}\right\} - \frac{\theta E}{2}
\left(p_{2} - Bx^{1}\right)\right]\phi(x^{1}) = 0.
\label{gal54}
\end{eqnarray}
Now using the following change of variables 
\begin{eqnarray}
x^{1}\rightarrow X = x^{1} - \frac{p_{2} + \tilde{m}E/B}{B}
\label{gal540}
\end{eqnarray}
we get the equation 
\begin{eqnarray}
\left[- \frac{1}{2\tilde{m}}\partial_{X}^{2} + \frac{\tilde{m} 
\tilde{\omega}_{c}^{2}}{2}\left( X -\frac{\tilde{m}E 
\theta}{2B}\right)^{2} \right]\phi^{'}\left(X\right) = 
\xi \phi^{'}\left(X\right)  
\label{gal55}
\end{eqnarray}
where, 
$\phi^{'}(X)=\phi(x^{1})$ 
and 
\begin{eqnarray}
\xi &=&\left[ \left(\omega + p_{2}E/B + 
\frac{\tilde{m}}{2}\left(E/B\right)^{2}\right) 
+ \frac{\tilde{m}}{2}\theta\left(E^{2}/B\right)\right].
\label{evgal55}
\end{eqnarray}
A further change of variables 
\begin{eqnarray}
\bar{X}  = (X -\frac{\tilde{m}E \theta}{2B})
\label{gal5500}
\end{eqnarray} 
yields the standard harmonic oscillator equation with an enhanced frequency 
$\tilde \omega_{c} = (1+ \theta B)\omega_{c}$:
\begin{eqnarray}
\left[- \frac{1}{2\tilde{m}}\partial_{\bar X}^{2} + \frac{\tilde{m} 
\tilde{\omega}_{c}^{2}}{2} \bar {X} ^{2} \right]\phi^{''}
\left(\bar X\right) = \xi \phi^{''}\left(\bar X\right)  
\label{gal550}
\end{eqnarray}
where, $\phi^{''}(\bar{X}) = \phi^{'}(X) = \phi(x^{1})$ and $\xi $ 
is the harmonic oscillator energy eigen-value. 
The admissible eigen-functions are given in terms of Hermite polynomials as
\begin{eqnarray}
\phi^{''}_{n}(\bar{X}) = C_{n}\exp ({-\frac{\tilde{m} 
\tilde{\omega}_{c}}{2}\bar{X}^{2}})H_{n}
\left(\sqrt{\tilde{m}\tilde{\omega_{c}}}\bar{X}\right)
\label{gal56}
\end{eqnarray}
and the eigen-values are
\begin{eqnarray}
\xi_{n} = (n + \frac{1}{2})\tilde\omega_{c}.
\label{gal34}
\end{eqnarray}
Also note that the $\theta$-dependent term appearing in the 
harmonic oscillator energy eigen-value $\xi$ (\ref{evgal55}) 
is due to electric field term in 
eq.(\ref{gal4307}). 
This will imply a quantization 
condition for $\omega$ 
\begin{equation}
\omega_{n} = (n + \frac{1}{2})\tilde\omega_{c} -\left[\left( p_{2}E/B + 
\frac{\tilde{m}}{2}\left(E/B\right)^{2}\right) + 
\frac{\tilde{m}}{2}\theta\left(E^{2}/B\right)\right].
\label{gal34Z}
\end{equation}
This indicates that the degeneracy of the Landau level has now been 
lifted by the external electric field as states with different $p_{2}$ 
values will have different energy eigen-values $\omega_{n}$. 
Now the normalisation condition (\ref{gal4306}) becomes
\begin{eqnarray}
1 = \int d\bar{X}dx^{2} |\phi^{''}(\bar{X})|^{2}
\label{gal57}
\end{eqnarray}
which for a sample width $L_{y}$ yields the condition 
\begin{eqnarray}
\int d\bar{X}|\phi^{''}(\bar{X})|^{2} = \frac{1}{L_{y}}~.
\label{gal58}
\end{eqnarray}
Since $\hat j_{1} = 0$, corresponding to the wave-function (\ref{gal530}), 
the longitudinal current vanishes. Now coming to the transverse current, 
we note that $I_{2}^{(1)}$ contains only $x^{1}$ integration. 
This indicates that only integration by parts over $x^{1}$ 
variable can be performed, so that one can write, for example 
$\int dx^{1}\tilde \psi^{\dag}\left( D_{1}\tilde \psi \right) 
= -\int dx^{1}\left( D_{1}\tilde \psi \right)^{\dag}\tilde \psi$.
However such an expression with $D_{1}\rightarrow D_{2}$ in the 
above equation can also be written for the {\it particular form} 
of the wave-function (\ref{gal530}) we have chosen to work with. 
In fact, this equality holds between the integrands themselves as one gets
\begin{eqnarray}
\tilde \psi^{\dag}D_{2}\tilde \psi = -(D_{2}\tilde \psi)^{\dag}\tilde 
\psi = i(p_{2} - A_{2})(\phi(x^{1}))^{2}~.
\label{gal580}
\end{eqnarray}
One can therefore write 
$\int dx^{1}\tilde \psi^{\dag}\left( D_{2}\tilde \psi \right) 
= -\int dx^{1}\left( D_{2}\tilde \psi \right)^{\dag}\tilde \psi$ 
and the same thing also holds for higher order covariant 
derivatives appearing in $\hat j_{2}$, as one can verify. 
We can therefore write $I^{(1)}_{2}$ more compactly as
\begin{eqnarray}
I^{(1)}_{2} = \int dx^{1} \frac{1}{2\tilde{m}i}\left(1 - 
\frac{\theta B}{2}\right) \left\{\tilde{\psi}^{\dag}
\left(D_{2}\tilde{\psi}\right) -\left(D_{2}\tilde{\psi}
\right)^{\dag}\tilde{\psi} \right\}.
\label{gal582}
\end{eqnarray}
Here the $\left(1 - \frac{\theta B}{2}\right)$-factor\footnote{As we are 
expressing everything in terms of the 
renormalised mass (i.e the observed mass) 
so there is no point in absorbing the factor 
$(1 - \theta B)$ in $\tilde m$ to give $m$.}   
stems from the presence of the electric field term 
in the action (\ref{gal4307}). 
Now using eq.(\ref{gal580}), the pair of co-ordinate 
transformations (\ref{gal540}, \ref{gal5500}) and eq.(\ref{gal58}) 
one can cast the expression for transverse current for a 
single particle as
\begin{eqnarray}
I^{(1)}_{2} = -\int d\bar{X} E \left(\frac{1}{B} + 
\frac{\theta }{2} \right)\left(1 - \frac{\theta B}{2}\right)
|\phi^{''}(\bar{X})|^{2} = - \frac{1}{L_{y}}\left(\frac{E}{B}\right).
\label{gal59}
\end{eqnarray}
Observe that $I^{(1)}_{2}$ is independent of both the indices $n$ 
and $p_{2}$, so that all the electronic states $|n, p_{2}\rangle$ 
carry the same Hall current just as happens usually. 
Therefore to obtain the total current $I_{2}$, (following \cite{st})
we just multiply $I^{(1)}_{2}$ (eq.(\ref{gal59})) 
by the number of available states ($\rho L_{x}L_{y}$) 
within an arbitrarily chosen rectangular area $L_{x}L_{y}$, 
where $\rho$ is the density of such states. 
We therefore have the total current (upto order $\theta$) as   
\begin{eqnarray}
I = -\rho L_{x}\frac{E}{B} = -\frac{\rho}{B}V
\label{gal5822}
\end{eqnarray}
where, $V = E L_{x}$. Hence the Hall-conductivity 
$\sigma_{H} = \frac{I_{2}}{V} = -\frac{\rho}{B}$ 
has no explicit $\theta$-dependence. 
At this stage one can easily see that the usual expression 
for degeneracy per unit area (for $E=0$)
holds, enabling one to define the filling fraction\footnote{Note 
that the filling fraction $\nu$ can be defined 
because the expression of $\hat j_{0}$ in eq.(\ref{gal555}) 
suggests (using eq.(s) (\ref{gal530}),(\ref{gal540}),(\ref{gal5500})) that 
the centre of the harmonic oscillator, i.e. the centre of the charge 
distribution now will be located at $x^{1} = p_{2}/(B)$. 
Now a range $\Delta x^{1} = L_{x}$ for $x^{1}$ 
implies a range $\Delta p_{2} = B L_{x}$ for $p_{2}$ 
which clearly can accomodate 
$\frac{\Delta p_{2}}{2\pi/L_{y}} = \frac{ B}{2\pi}L_{x}L_{y}$ 
number of charged states within an area $L_{x}L_{y}$, 
if periodic boundary condition is imposed in the $x^{2}$-direction. 
One thus recovers the usual expression 
for degeneracy per unit area to be $ B/(2\pi)$ 
with no accompanying noncommutative corrections.}
in the conventional way, using which one can write down an 
alternative expression
for Hall conductivity as $\sigma_{H}=-\frac{\nu}{2\pi}$.

\section{Summary}
In this chapter we have obtained an effective $U(1)$ 
gauge invariant action and correspondingly $U(1)$ 
gauge covariant Schr\"{o}dinger equation starting from $U(1)_{\star}$ 
gauge invariant action, describing noncommutative 
Schr\"{o}dinger field coupled 
to a background noncommutative $U(1)_{\star}$ gauge field, by using 
SW map followed by wave-function and mass renormalisation. 
The effect of noncommutativity on the mass parameter 
appears naturally in our analysis. Interestingly, 
we observe that the external magnetic field has to be 
static and uniform in order to get a canonical form of 
Schr\"{o}dinger equation upto $\theta$-corrected terms, 
so that a natural probabilistic interpretation emerges. 
The Galilean symmetry of the model is next 
investigated where the translation and the rotation 
generators are seen to form a closed Euclidean sub-algebra 
of Galilean algebra. However, the boost is not found to be a 
symmetry of the system. This shows that even though the 
condition $\theta^{0i} = 0$ is Galilean invariant, 
a violation in the Galilean symmetry is exhibited for boost perpendicular 
to the electric field. Further, as a quantum mechanical 
application of our model, we take up the problem of Hall effect, 
where we compute the  Hall conductivity (considering a set of free particles) 
and find no noncommutative correction upto first order in $\theta$. 
Thus, in our formalism we reproduce the standard result of  
Hall conductivity  with the 
filling fraction $\nu$ taking all possible values. 
The presence of impurities/disorder are essential 
for any quantization of $\nu$, appropriate for 
Quantum Hall effect (integer/fractional), 
which may be the topic of future investigation. 


\chapter{Dual families of noncommutative quantum systems}
We have seen in the previous chapter that the 
SW map provides a correspondence from the
noncommutative to the commutative space which preserves
the gauge invariance and the physics \cite{switten}. 
However, it should be noted that this map is classical in nature, 
and therefore it is not clear whether this map will
hold at the quantum level or not \cite{jurco}, \cite{jurco1}, 
\cite{jurco2}, \cite{jurco3}. It is therefore natural to
enquire about the status of this map in noncommutative
quantum mechanics where, apart from a few works
\cite{bcsgas, kod} which consider the
SW map only to lowest order in the 
noncommutative parameter, very little has been done.    
\vskip 0.003cm
\noindent A second motivation for the present work comes from 
the by now well known noncommutative paradigm associated 
with the quantum Hall effect \cite{bop, np, pahorvathy}.  
In particular, \cite{jelal} explores the possibility 
of tuning the noncommutative parameter $\theta$ such
that the electrons moving in two dimensional noncommutative space
(in presence of both uniform external magnetic and electric fields)
can be interpreted as either leading to the fractional quantum Hall
effect or composite fermions in the usual coordinates. 
On the other hand, the discovery of the fractional quantum Hall effect
led to the immediate realization that the Coulomb interaction 
plays an essential role in
the understanding of this phenomenon \cite{laughlin}.
This raises the question whether the noncommutative
Hamiltonian introduced by \cite{jelal} in a somewhat
ad hoc way can be reinterpreted as an effective
noncommutative Hamiltonian which describes the same
physics as the interacting commutative theory, at
least in some approximation.  Clearly, this equivalence
cannot be exact as it is well known 
\cite{mezincescu, bcsgas} that a
 noninteracting commutative Hamiltonian with
 constant magnetic field maps onto a noninteracting
 noncommutative Hamiltonian with constant magnetic
 field. However, one might think about the possibility
 that there is some preferred value of the noncommutative
 parameter which minimizes the interaction on the
 noncommutative level.  If this is the case the 
corresponding noninteracting noncommutative Hamiltonian
 might be a good starting point for a computation which
 treats the residual interaction as a perturbation.
  This might seem problematic due to the degeneracy 
of the Landau levels.  However, under the assumption of
 a central potential this construction can be carried out
 in each angular momentum sector, which effectively lifts
 this degeneracy and allows for a perturbative treatment
 in each sector (see section 5.5).  
\vskip 0.003cm
\noindent With the above remarks in mind,
 i.e. the physical equivalence of different noncommutative
 descriptions, the following question arises quite
 naturally: how should a family of noncommutative
 Hamiltonians be parameterized as a function of the
 noncommutative parameter to ensure that they are
 physically equivalent?  This is the central issue
 addressed here. The relation to the SW
 map and the possible use to construct dualities
 are natural secondary issues that arise which has
 also been addressed here, although not in complete generality.    
\vskip 0.003cm
\noindent This chapter is organized as follows. In section
5.1, the general construction
of a one parameter family of noncommutative, physically
equivalent Hamiltonians is considered. In section 5.2 and
5.3, application of this general 
construction is carried out to
a particle in two dimensions moving in a constant
magnetic field without interactions and in the presence
of a harmonic potential, respectively.  The construction
is carried out to all orders in the noncommutative parameter.
The relation between this construction 
and the SW map is discussed in section 5.4.
In section 5.5, an approximate duality between 
the interacting commutative
Hamiltonian and a noninteracting 
noncommutative Hamiltonian is constructed for 
an harmonic oscillator potential. Section 5.6 contains our
discussion and conclusions. An appendix summarizes
notational issues at the end.

\section{General considerations}
\label{dual}
We consider a NR particle moving in a plane 
under a potential $V$ and coupled minimally 
to a $U(1)$ gauge field $A$.  
In commutative space the Hamiltonian reads ($\hbar=c=e=1$)
\begin{eqnarray}
H=\frac{\left({\bf{p}} - \bf{A}\right)^2}{2m}+V(x). 
\label{dual1aa}
\end{eqnarray}
The prescription to go over to the noncommutative space is to replace 
the commutative quantities by noncommutative ones, 
denoted by a hat, and introduce the star product, 
defined in the usual way (\ref{gal3}).
The spacetime noncommutativity 
is assumed to vanish $(\theta^{0i}=0)$ and, for a planar system, 
the spatial part of the $\theta$-matrix can 
be written as $\theta^{ij} = \theta \epsilon^{ij}$.  
The Schr\"{o}dinger equation in noncommutative space therefore reads
\begin{eqnarray}
i\frac{\partial \hat\psi({\bf{x}}, t)}{\partial t}
&=&\left[\frac{\left({\bf{p}} - \bf{\hat A}\right)\star \left({\bf{p}} - 
\bf{\hat A}\right)}{2\hat m}+\hat V(x)\right]\star
\hat\psi({\bf{x}}, t)\nonumber\\
&=&\hat H\star\hat\psi({\bf x}, t)\equiv\hat H_{BS}(\theta)\hat\psi(\theta).
\label{dualg1}
\end{eqnarray}
Here, $\hat H_{BS}(\theta)$ denotes the Hamiltonian after the star product 
has been replaced by a Bopp-shift, defined by 
\cite{bop, gamboa, mezincescu}
\begin{eqnarray}
\left(\hat f\star \hat g\right)(x) =\hat f\left(x-\frac{\theta}
{2}\epsilon^{ij}p_j\right)\hat g(x).
\label{dualBoppshift}
\end{eqnarray}
Note that the quantities appearing in $\hat H_{BS}(\theta)$ 
are still the noncommutative ones.
\vskip 0.003cm
\noindent The condition that the physics remains invariant 
under a change in $\theta$ requires that
$\hat H_{BS}(\theta)$ and $\hat H_{BS}(0)$ 
are related by a unitary transformation
\begin{eqnarray}
\hat H_{BS}(\theta)=U(\theta)\hat H_{BS}(0)U^\dagger(\theta)\;
\label{dualg2}
\end{eqnarray}
and that 
\begin{eqnarray}
\hat\psi(\theta)=U(\theta)\hat\psi(0)\;.
\label{dualg3}
\end{eqnarray}
Differentiating eq.(\ref{dualg2}) with respect to $\theta$, we obtain
\begin{eqnarray}
\frac{d\hat H_{BS}(\theta)}{d\theta}=[\eta(\theta), \hat H_{BS}(\theta)]\; 
\label{dualg4}
\end{eqnarray}
where,
\begin{eqnarray}
\eta(\theta)=\frac{dU(\theta)}{d\theta}U^\dagger(\theta)
\label{dualg41}
\end{eqnarray}
is the generator of the unitary transformation 
relating the noncommutative Bopp-shifted Hamiltonian with the 
commutative Hamiltonian. 
\vskip 0.003cm
\noindent We now consider under what conditions 
eq.(\ref{dualg4}) admits a solution for $\eta$. 
These conditions will, of course, 
provide us with the constraints on the parameterization of the 
noncommutative Hamiltonian necessary to ensure unitary equivalence, 
i.e., the existence of $\eta$. It is a simple matter 
to verify that eq.(\ref{dualg4}) admits a solution for $\eta$ if and only if
\begin{eqnarray}
\langle n , \theta \vert \frac{d\hat{H}_{BS}(\theta)}{d\theta} \vert n , 
\theta\rangle = 0\quad ,\quad \forall n
\label{dualgp5} 
\end{eqnarray}
where, $\vert n , \theta\rangle$ are eigenstates of 
$\hat{{H}}_{BS}(\theta)$, i.e., 
\begin{eqnarray}
\hat{H}_{BS}(\theta) \vert n , \theta\rangle = E_n \vert n , \theta\rangle .
\label{dualgp4} 
\end{eqnarray}
If eq.(\ref{dualgp5}) holds, the off-diagonal part of $\eta$ 
is uniquely determined by 
\begin{eqnarray}
\eta=\sum_{n\neq m}\frac{\langle n,\theta|
\frac{d {\hat H}_{BS}}{d\theta}|m,\theta\rangle}{E_m-E_n}
|n,\theta\rangle\langle m,\theta|
\label{dualeta}
\end{eqnarray}       
while the diagonal part is arbitrary, 
reflecting the arbitrariness in the phase of the eigenstates. 
Here we have assumed no degeneracy in the spectrum of 
$\hat H_{BS}(\theta)$.  The generalization to the case 
of degeneracies is straightforward.
\vskip 0.003cm
\noindent The set of conditions (\ref{dualgp5}) 
should be viewed as the set of conditions 
which determines the $\theta$-dependency of the 
matrix elements of the noncommutative potential 
$\hat V$ and gauge field $\hat A$. 
Expectedly these matrix elements are 
under-determined, i.e., that not both $\hat V$ and $\hat A$ 
are uniquely determined by them. 
Instead one can fix one of these and compute the other.  
For comparison with the SW map, it is therefore 
natural to take for $\hat A$ 
the noncommutative gauge field as determined from the SW map. 
Note that this procedure implies that $\hat V$ will be gauge dependent. 
\vskip 0.003cm
\noindent Consider the SW map for the 
noncommutative wave-function (\ref{gal40}).
Below we consider two dimensional systems in a constant magnetic field.  
Taking the symmetric gauge, the SW map reduces to a 
$\theta$ dependent scaling transformation. 
Clearly this is not a unitary transformation 
and a unitary SW map can be constructed as in 
\cite{kod}. However, a more convenient point of view, closer 
in spirit to the SW map, would be to relax the 
condition of unitarity above.  It therefore seems worthwhile, 
in particular to relate to the SW map, 
to generalize the above considerations by relaxing 
the condition of unitarity.   
\vskip 0.003cm
\noindent This generalization is straightforward.  
The unitary transformation in 
eq.(s) (\ref{dualg2}) and (\ref{dualg3}) needs to be replaced
by a general similarity transformation
\begin{eqnarray}
\hat H_{BS}(\theta)=S(\theta)\hat H_{BS}(0)S^{-1}(\theta)
\label{dualg2a}
\end{eqnarray}
while
\begin{eqnarray}
\hat\psi(\theta)=S(\theta)\hat\psi(0)
\label{dualg3a}
\end{eqnarray}
and note that a new inner product 
$\langle \psi|\phi\rangle_T=\langle \psi|T|\phi\rangle$ 
can be defined such that $\hat H_{BS}(\theta)$ 
is hermitian with respect to it. 
In particular $T$ is given by $T=(S^{-1})^{\dagger} 
S^{-1}$ and has the property 
$T\hat H_{BS}(\theta)=\hat H_{BS}^\dagger(\theta)T$. 
Under this prescription the same physics results.  
A detailed exposition of these issues can be found in \cite{scholtzgeyer}. 
\vskip 0.003cm
\noindent Differentiating eq.(\ref{dualg2a}) with respect to $\theta$, 
we obtain
\begin{eqnarray}
\frac{d\hat H_{BS}(\theta)}{d\theta}=[\eta(\theta), \hat H_{BS}(\theta)]\;
\label{dualg4a}
\end{eqnarray}
where, 
\begin{eqnarray}
\eta(\theta)=\frac{dS(\theta)}{d\theta}S^{-1}(\theta)
\label{dualg44a}
\end{eqnarray}
is now the generator of the similarity transformation 
relating the noncommutative Bopp-shifted Hamiltonian 
with the commutative Hamiltonian. 
\vskip 0.003cm
\noindent It can now be easily verified that 
eq.(\ref{dualgp5}) gets replaced by
\begin{eqnarray}
\langle n , \theta \vert T\frac{d\hat{H}_{BS}(\theta)}{d\theta} 
\vert n , \theta\rangle = 0\quad ,\quad \forall n
\label{dualgp5a} 
\end{eqnarray}
where, $\vert n , \theta\rangle$ are eigenstates of 
$\hat {H}_{BS}(\theta)$ (note that the eigenvalues 
will be real as $\hat {H}_{BS}(0)$ is assumed to be
hermitian and thus has real eigenvalues). 
As before, if eq.(\ref{dualgp5a}) holds, the 
off-diagonal part of $\eta$ is uniquely determined by 
\begin{eqnarray}
\eta=\sum_{n\neq m}\frac{\langle n,
\theta|T\frac{d \hat H_{BS}}{d\theta}|m,
\theta\rangle}{E_m-E_n}|n,\theta\rangle\langle m,\theta|T
\label{dualeta10}
\end{eqnarray}       
while the diagonal part is arbitrary, reflecting the 
arbitrariness in the phase and now also the 
normalization of the eigenstates. 
\vskip 0.003cm
\noindent Under the above description, 
the Hamiltonians $\hat {H}_{BS}(\theta)$ and 
$\hat {H}_{BS}(0)$ are physically equivalent. 
There is, however, one situation in which this 
equivalence may break down and of which careful 
note should be taken. This happens when the 
similarity transformation $S(\theta)$ becomes 
singular for some value of $\theta$, 
which will be reflected in the appearance of zero norm or 
unnormalizable states in the new inner product. 
Only values of $\theta$ which can be reached by integrating 
eq.(\ref{dualg44a}) from $\theta=0$ without passing through a singularity, 
can be considered physically equivalent to the commutative system. 
\vskip 0.003cm  
\noindent To solve eq.(s) (\ref{dualgp5}) or (\ref{dualgp5a}) 
in general is of course impossible. 
Therefore we take a slightly different approach in what follows. 
An ansatz for $\eta$ motivated by the SW map is taken to 
solve eq.(\ref{dualg4}) or eq.(\ref{dualg4a}) 
directly. We have already noted above 
that in the cases of interest to us, i.e., 
two dimensional systems in constant magnetic fields, 
the SW map for the noncommutative wave-function 
corresponds to a scaling transformation in the symmetric gauge.  
This motivates us to make the following ansatz             
\begin{eqnarray}
\eta(\theta)= f(\theta)r\partial_{r}=if(\theta)x.p
\label{dualg5}
\end{eqnarray}
with $f$ being an arbitrary function to be determined.  
The finite form of this scaling transformation 
can be readily obtained by integrating eq.(\ref{dualg44a}) to yield
\begin{eqnarray}
S(\theta)= e^{i\left(\int_0^\theta 
f(\theta^{\prime})d\theta^{\prime}\right)x.p}.
\label{dualscaletrans}
\end{eqnarray}
Clearly this is not a unitary transformation and therefore 
falls in the class of more general 
transformations described above eq.(\ref{dualg2a}). 
Furthermore we note that the non-singularity of $S(\theta)$ 
requires that the integral $\int_0^\theta f(\theta^{\prime})
d\theta^{\prime}$ exists.

\section{Free particle in a constant magnetic field}
\label{magnetic}
In this section, we apply the considerations discussed 
above to the case of a free particle ($\hat V=0$) 
moving in a noncommutative plane in the presence of a 
constant noncommutative magnetic field. 
The Schr\"{o}dinger equation is given by eq.(\ref{dualg1}) 
with $\hat V$ set to zero. 
\vskip 0.003cm
\noindent In the symmetric gauge 
$\hat A_{i}=-\frac{\bar B(\theta)}{2}
\epsilon_{ij}x_{j}$ \footnote{We use $\bar B(\theta)$ 
to denote the noncommutative counterpart of $B$
in eq.(\ref{symgauge}). It should not be confused with the 
noncommutative magnetic field $\hat B$ as 
determined from the field strength (see eq.(\ref{s2})).  
In the limit $\theta=0$, $\bar B(\theta)=B$.}, 
the Bopp-shifted Hamiltonian (\ref{dualg1}) is easily found to be 
\begin{eqnarray}
\hat H_{BS}(\theta) &=& \frac{\left(1 + 
\frac{\bar B\theta}{4}\right)^2}{2\hat 
m(\theta)}\left({\bf{p}} - \frac{1}{1 + 
\frac{\bar B\theta}{4}}{\bf A}\right)^2 \nonumber\\
&=& \frac{1}{2M(\theta)}\left(p^{2}_{x} + p^{2}_{y}\right) + 
\frac{1}{2}M(\theta){\Omega}(\theta)^{2}\left( x^2 + y^2 
\right)\nonumber\\ &&- \Omega(\theta) L_{z}\; 
\label{duality5} 
\end{eqnarray}
where,
\begin{eqnarray} 
\frac{1}{2M(\theta)} = \frac{\left(1 
+ \frac{\bar B\theta}{4}\right)^2}
{2\hat m(\theta)}\quad,\quad \frac{1}{2}M(\theta)
{\Omega(\theta)}^{2} = \frac{\bar B^2}{8\hat m(\theta)}\;.
\label{dualfreepar}
\end{eqnarray}
Substitution of the above form of the Hamiltonian in eq.(\ref{dualg4}) 
with $\eta$ as in eq.(\ref{dualg5}), 
leads to the following set of differential equations :
\begin{eqnarray}
\frac{d M^{-1}(\theta)}{d\theta} &=& -2f(\theta)M^{-1}(\theta)\; \\  
\frac{d \left(M(\theta){\Omega(\theta)}^{2}\right)}{d\theta} 
&=& 2M(\theta){\Omega(\theta)}^{2}f(\theta)\;
\label{dual6a} 
\end{eqnarray}
\begin{eqnarray}
\frac{d\Omega(\theta)}{d\theta} &=& 0.
\label{dual5a} 
\end{eqnarray}
Eq.(\ref{dual5a}) ensures the stability of the energy spectrum, i.e
the cyclotron frequency $\Omega(\theta)=\Omega(\theta=0)=B/2m$, where
$m=\hat m(\theta=0)$.
This is the physical input in our analysis and will play a very
important role as we shall see later. 
The above equations (\ref{dualfreepar}, \ref{dual6a}, \ref{dual5a}) 
immediately lead to
\begin{eqnarray}
f(\theta) = \frac{1}{2M(\theta)}
\frac{dM(\theta)}{d\theta} = 
\frac{\partial_{\theta}\bar{B}(\theta) - 
\frac{\bar{B}(\theta)^2}{4}}{2\bar{B}(\theta)
\left(1 + \frac{\theta\bar{B}(\theta)}{4}\right)}\;
\label{dual7a} 
\end{eqnarray}
which fixes $f$ once $\bar B$ has been determined. As indicated 
before, we take $\hat A$
as the noncommutative gauge field determined from the SW map.
With this in mind we now proceed to determine $\bar B$.\\
\noindent It is easy to see that a symmetric gauge configuration 
\begin{equation}
\label{symgauge}
A_i=-\frac{B}{2}\epsilon_{ij}x^j 
\end{equation}
with magnetic field  $B=F_{12}=(\partial_1A_2-\partial_2A_1)$, 
transforms to a symmetric gauge field configuration 
at the noncommutative level under the SW transformation (\ref{gal41}). 
Using the same 
notation as in eq.(\ref{symgauge}), we write
\begin{equation}
\hat{A}_i=-\frac{\bar{B}}{2}\epsilon_{ij}x^j
\label{symgauge1}
\end{equation}
where, $\bar B$ is determined to leading order in $\theta$ 
from eq.(\ref{gal41}) to be
\begin{equation}
\bar{B}=B\left(1+\frac{3\theta B}{4}\right).
\label{dualsolution}
\end{equation}
Note that $\bar{B}(\theta)$ should 
not be identified with the 
noncommutative magnetic field $\hat{B}$, 
which has an additional Moyal bracket term 
$[\hat{A}_1, \hat{A}_2 ]_{\star}$:
\begin{equation}
\hat{B}=\hat{F}_{12} = \partial_1\hat{A}_2-\partial_2
\hat{A}_1 - i(\hat{A}_1{\star}
\hat{A}_2-\hat{A}_2{\star}\hat{A}_1)=\bar B(1+\frac{\theta\bar B}{4}).
\label{s2}
\end{equation}
This is precisely the same expression one gets if 
one applies the SW map directly at the level 
of the field strength tensor, which is given by \cite{sw}:
\begin{equation} 
\hat{F}_{\mu\nu}=F_{\mu\nu}+\theta\epsilon^{ij}F_{\mu i}F_{\nu j}\;.
\label{duals3}
\end{equation}
Note that the expression (\ref{s2}) relating $\hat B$ 
with $\bar B$ is an exact one in contrast with eq.(\ref{dualsolution}) 
which relates $\bar B$ to $B$ only up to leading order in $\theta$. 
For a constant field configuration, the SW equation 
for the field strength tensor can be integrated exactly 
to give the result \cite{sw}
\begin{equation}
\hat{B}=\frac{1}{1 - \theta B}B.
\label{duals4}
\end{equation}
From eq.(s) (\ref{s2}) and (\ref{duals4}), 
we obtain a quadratic equation in $\bar B(\theta)$ 
that can be solved exactly to give
\begin{equation}
\bar B(\theta)=\frac{2}{\theta }\left[(1 - \theta B)^{-1/2} - 1\right].
\label{duals5}
\end{equation}
The above expression for $\bar B(\theta)$ 
is exact up to all orders in $\theta$. 
When substituted in eq.(\ref{symgauge1}) an expression, 
correct to all orders in $\theta$, 
for the noncommutative gauge field $\hat A_{i}$ result 
\begin{equation}
\hat A_{i} = -\frac{1}{\theta }\left[(1 - \theta B)^{-1/2} - 1\right]
\epsilon_{ij}x^{j}.
\label{duals6}
\end{equation}
Substituting $\bar{B}(\theta)$ from eq.(\ref{duals5}) 
into eq.(\ref{dual7a}) yields
\begin{equation}
f(\theta)=\frac{\bar B(\theta)}{4}\;.
\label{dualf}
\end{equation}
Upon differentiating eq.(\ref{dualg3a}) with respect to $\theta$ 
and using $f$ from eq.(\ref{dualf}), 
we find that $\hat \psi(\theta)$ 
must satisfy the following equation:
\begin{eqnarray}
\frac{d{\hat\psi(\theta)}}{d\theta} = 
\frac{\bar B(\theta)}{4}r\frac{d\hat \psi(\theta)}{dr}\;. 
\label{dualswmap20} 
\end{eqnarray}
This result can now be compared to the corresponding 
SW transformation rule for $\hat\psi$.  
The SW equation (\ref{gal40}) 
for an arbitrary $\theta + \delta\theta$ reads 
\begin{eqnarray}
\hat\psi(\theta + \delta\theta)- \hat\psi(\theta) = 
-\frac{1}{2}\theta\epsilon^{ij}\hat A_i\star\partial_j\hat\psi(\theta).
\label{dualswmap10} 
\end{eqnarray}
Upon substituting $\hat A_i$ from eq.(\ref{symgauge1}), 
eq.(\ref{dualswmap20}) indeed results.  
Thus the transformation rule as obtained from the 
requirement of physical equivalence agrees 
with that of the SW map.
\vskip 0.003cm
\noindent Finally, substituting $\bar B(\theta)$ in the condition 
$\Omega=B/2m$ yields the following expression for $\hat m(\theta)$:
\begin{eqnarray}
\hat m(\theta)=\frac{m}{1-\theta B}\;.
\label{dual1000d} 
\end{eqnarray}
The above equation relates the noncommutative mass 
$\hat m(\theta)$ with the commutative mass $m$. 
This generalizes the result obtained in eq.(\ref{gal4303})  
to all orders in $\theta$.
\vskip 0.003cm
\noindent The Schr\"odinger equation can of course be solved exactly
in a simple case such as this. 
It is useful to see what the above procedure entails 
from this point of view.  
To solve for the eigenvalues and eigenfunctions of eq.(\ref{duality5}) 
is a standard procedure and for notational 
completeness we summarize the essential steps in appendix. 
This results in the degenerate eigenvalue spectrum
\begin{eqnarray}
E_{n_-,\ell}=2\Omega\left( n_{-} + 
\frac{1}{2}\right)\nonumber\\ n_-=0,1,\ldots\quad;\;
\ell=-n_-,-n_-+1\ldots\;
\label{dual10.1} 
\end{eqnarray}
where, $\ell$ denotes the eigenvalues of the 
angular momentum operator $L_3$. 
The corresponding eigenstates are 
obtained by acting with the creation operators 
$b_{\pm}^\dagger$ defined in eq.(\ref{10bapp}) on the ground-state
\begin{eqnarray}
\hat\psi(z, \bar z;\theta) &=& N\exp\left[-\frac{M\Omega}{2}
\bar{z}z\right]\nonumber\\
&=&  N\exp\left[- \frac{\bar B(\theta)}
{4\left(1 + \frac{\bar B(\theta)\theta}{4}\right)}\bar{z}z\right]\;.
\label{dual10c.1} 
\end{eqnarray}
Comparing with our previous results, we note that 
eq.(\ref{dual5a}) ensures invariance of the 
spectrum under a change of $\theta$. 
Furthermore direct inspection shows that the 
{\it unnormalised} ground-state and, subsequently, 
also all excited states satisfy the transformation rule 
(\ref{dualswmap20}). 
The fact that the unnormalised wave-functions satisfy the 
transformation rule (\ref{dualswmap20}) 
is consistent with our earlier remarks on the 
non-unitary nature of the scaling transformation.
\vskip 0.003cm
\noindent Finally, note that although the 
noncommutative parameters $\bar B(\theta)$ 
and $\hat m(\theta)$ have singularities 
at $\theta=1/B$ \footnote{This singularity was also encountered
in the previous chapter.}, these singularities cancel 
in the parameter $\Omega$, which is by construction 
free of any singularities, i.e., the spectrum is 
not affected by this singularity. 
This is also reflected by the fact that the 
integral of $f$, as determined in eq.(\ref{dualf}), 
is free of this singularity. Thus, 
despite the appearance of this singularity in the parameters of the 
noncommutative Hamiltonian, 
there is no breakdown of the physical equivalence 
(see the discusion in section $5.1$).

\section{Harmonic oscillator in a constant magnetic field}
\label{bf}
In this section, we include a harmonic oscillator potential 
$V=\lambda r^2$ in the commutative Hamiltonian (\ref{dual1aa}). 
If the physical equivalence between the noncommutative 
and commutative Hamiltonians is indeed implementable 
through a scale transformation, we expect the potential 
to be form preserving (this is certainly not true for arbitrary potentials). 
We therefore take for the noncommutative potential in eq.(\ref{dualg1}) 
$\hat V=\hat\lambda(\theta)r^2$, 
where the oscillator strength $\hat\lambda(\theta)$ has to be determined. 
Obviously we must also require that 
$\hat{\lambda}(\theta) = \lambda $ 
in the limit $\theta = 0$. 
The Bopp-shifted Hamiltonian with this form for the 
noncommutative Hamiltonian (\ref{dualg1}), is easily found to be
\begin{eqnarray}
\hat H_{BS}(\theta) &=& \frac{\left(1 + 
\frac{\bar B\theta}{4}\right)^2}{2\hat m}
\left({\bf{p}} - \frac{1}{1 + 
\frac{\bar B\theta}{4}}{\bf A}\right)^2 \nonumber\\
 &&+\hat{\lambda}(\theta)\left[\frac{{\theta}^2}{4}\left(p^{2}_{x} 
+ p^{2}_{y}
\right) + \left( x^2 + y^2\right) - \theta L_z \right]
\nonumber\\
&=& \frac{1}{2M}\left(p^{2}_{x} + p^{2}_{y}\right) + \frac{1}{2}M{\Omega}^{2}
\left( x^2 + y^2 \right)\nonumber\\
&&- \Lambda(\theta)L_{z}\;
\label{dual101d} 
\end{eqnarray}
where,
\begin{eqnarray}
\label{dualharpar}
\frac{1}{2M} &=& \frac{\left(1 + \frac{\bar B\theta}{4}\right)^2}{2\hat m} 
+ \frac{\hat{\lambda}{\theta}^2}{4}\; \nonumber\\
\frac{1}{2}M{\Omega}^{2} &=& \frac{\bar B(\theta)^2}{8\hat m(\theta)} 
+ \hat{\lambda}(\theta)\;\\
\Lambda(\theta) &=& \left[\frac{M{\Omega}^2\theta}{2} + 
\frac{\bar B\left[1 - \left({\frac{M\Omega \theta}{2}}\right)^2 
\right]}{2\left(1 + \frac{\bar B\theta}{2}\right)M} \right]\nonumber\;.
\end{eqnarray}
Here $\bar B(\theta)$ is again taken from the SW map 
(\ref{duals5}). 
Substituting the above form of the Hamiltonian in eq.(\ref{dualg4}) 
with $\eta$ as in eq.(\ref{dualg5}), we obtain
the following set of differential equations:
\begin{eqnarray}
\frac{dM^{-1}(\theta)}{d\theta} &=& -2f(\theta)M^{-1}(\theta)\;\\     
\frac{d \left(M(\theta){\Omega(\theta)}^{2}\right)}{d\theta} 
&=& 2M(\theta){\Omega(\theta)}^{2}f(\theta)\;
\label{dual61a} 
\end{eqnarray}
\begin{eqnarray}
\frac{d\Lambda(\theta)}{d\theta} &=& 0\;.
\label{dual51a} 
\end{eqnarray}
Eq.(\ref{dual51a}) requires that $\Lambda(\theta)$ 
is independent of $\theta$ and hence we 
have the condition $\Lambda(\theta)=\Lambda(0)=B/2m$. 
Substituting the form of $M(\theta)$ in terms of 
$\hat m(\theta)$ and $\hat \lambda(\theta)$, 
we obtain the following solution for $\hat m(\theta)$ 
in terms of $\hat \lambda(\theta)$:
\begin{eqnarray}
\hat{m}(\theta) = \frac{m}{\left(1 - \theta B\right)}\frac{B}
{\left(B - 2m\theta\hat{\lambda}(\theta)\right)}\;.
\label{dual600a} 
\end{eqnarray}
The set of differential equations in (\ref{dual61a}) 
can also be combined to obtain
\begin{eqnarray}
\frac{d\Omega^2}{d\theta} = 0\;.
\label{dual601a} 
\end{eqnarray}
This shows that $\Omega$ is a constant and therefore we have 
\begin{eqnarray}
\Omega^2(\theta) = \Omega^2(\theta = 0) = \frac{B^2}{4m^2} + 
\frac{2\lambda}{m}\;.
\label{dual602a} 
\end{eqnarray}
Substituting $\Omega^2(\theta)$ in eq.(\ref{dualharpar}) and using
eq.(\ref{dual600a}), a quadratic equation for $\hat \lambda(\theta)$
is obtained:
\begin{eqnarray}
\left[ B^3 + 8(1 - \theta B)mB\hat{\lambda}(\theta) - 
16(1 - \theta B)m^2\theta \hat{\lambda}(\theta)^2\right]
\nonumber\\= B^3 + 8\lambda mB\;.
\label{dual603a} 
\end{eqnarray}
The solution for $\hat\lambda(\theta)$ yields
\begin{eqnarray}
\hat{\lambda}(\theta) = \frac{B}{4m\theta}\left[1 - \left(1 - 
\frac{8\lambda m\theta}{B(1 - \theta B)}\right)^{\frac{1}{2}}\right]\;
\label{dual604a} 
\end{eqnarray}
where we have taken the negative sign before the square root 
since with this choice we have $\hat\lambda(\theta=0)=\lambda$. 
\vskip 0.003cm
\noindent With the value of $\bar B(\theta)$ 
fixed from the SW map and 
$\hat m(\theta)$ and $\hat\lambda(\theta)$ determined as above, 
we can compute the value of $M(\theta)$ from eq.(\ref{dualharpar}) 
and subsequently the value of $f(\theta)$ from eq.(\ref{dual61a}) 
as $f(\theta)=\frac{1}{2M(\theta)}\frac{dM(\theta)}{d\theta}$. 
The expression is a lenghthy one and 
we do not need to list it here. 
What is important to note, however, is that once $f(\theta)$ is fixed, 
the transformation rule satisfied by $\hat\psi(\theta)$ is determined 
from eq.(\ref{dualg3a}) and that this transformation rule is 
not the same as the one derived from the SW map (\ref{dualswmap20}). 
In fact, it turns out that the transformation rule for $\hat\lambda(\theta)$ 
is also different from the SW map. 
We discuss these points in more detail in the next section. 
\vskip 0.003cm
\noindent As a consistency check, one can once 
again solve for the eigenvalues and eigenstates. 
The procedure is the same as in appendix and one 
finds for the eigenvalues
\begin{eqnarray}
E_{n_-,\ell}= 2\Omega \left(n_{-} + \frac{1}{2}\right) +
(\Omega- \Lambda)\ell\;\nonumber \\
 n_-=0,1,\ldots;\;\ell=-n_-,-n_-+1,\ldots\;.
\label{dualz2} 
\end{eqnarray}
From the above expression of the energy eigenvalues, it is easy to see that 
the degeneracy in $\ell$ has been lifted. However, 
in the limit $\lambda = 0$, the energy spectrum given by 
eq.(\ref{dual10.1}) is recovered.  
The corresponding eigenstates are again 
obtained by acting with the creation operators 
$b_{\pm}^\dagger$ defined in eq.(\ref{10bapp}) on the ground-state
\begin{eqnarray}
\hat\psi(z, \bar z; \theta) = N\exp\left[-\frac{1}{4}
\sqrt{\frac{2\bar B(\theta)^2 + 16 \hat
\lambda \hat m}{2\left(1+\frac{\theta \bar B(\theta)}{4}\right)^2+ 
\theta^2 \hat\lambda \hat m}}\bar{z}z\right]\;.\nonumber\\
\label{dualz3} 
\end{eqnarray}
Once again we note that eq.(s) (\ref{dual61a}) and (\ref{dual51a}) 
ensures invariance of the spectrum under a change in $\theta$. 
Using the values of $\bar B(\theta)$, $\hat m(\theta)$ and 
$\hat\lambda(\theta)$as determined above, one finds that the 
{\it unnormalised} wave-functions indeed satisfy the 
transformation rule as determined by eq.(\ref{dualg3a}) and not the 
SW transformation rule (\ref{dualswmap20}).
Also, in the $\theta=0$ limit, eq.(\ref{dualz3}) 
smoothly goes over to the standard 
commutative result
\begin{eqnarray}
\hat\psi(z, \bar z, \theta=0) = \psi(z, \bar z)= N
\exp\left[-\frac{1}{4}\sqrt{B^2+ 8\lambda m }\bar{z}z\right]\;.
\nonumber\\
\label{dualz300} 
\end{eqnarray}
\vskip 0.003cm
\noindent Finally we remark on the 
non-singularity of the scaling 
transformation $S(\theta)$. As already pointed out in section 
$5.1$, this requires the existence of 
the integral of $f$, which in the present case is 
simply given by $\log(M(\theta)/m)/2$.  
This turns out to be free of singularities, 
although the noncommutative parameters again 
exhibit singularities at $\theta=1/B$. 
As in the free case these singularities cancel in the 
parameters $\Omega$ and $\Lambda$ which determine the physical spectrum.

\section{Connection with Seiberg-Witten map}
\label{relation}
In this section, we are going to discuss the relationship of the flow
equations for $\hat{m}(\theta)$ and $\hat{\lambda}(\theta)$ 
obtained from the stability
analysis of the previous section to the flow equation obtained from
the SW map. To that end, let us write down 
the $U(1)_{\star}$ gauge
invariant action from which the ${\star}$ gauge covariant 
one-particle Schr\"{o}dinger
equation (\ref{dualg1}) follows as Euler-Lagrangian equation:
\begin{eqnarray}
\hat {S}=\int d^3{x} \hat{\psi}^{\dagger}\star(i\hat{D}_0+\frac{1}
{2\hat {m}}\hat{D}_i\star\hat{D}_i+\hat{V})\star\hat{\psi}\;.
\label{dualp420} 
\end{eqnarray}
\vskip 0.003cm
\noindent The preservation of $U(1)_\star$ 
gauge invariance of the action requires 
that the potential $\hat{V}$ must transform adjointly under 
$\star$ gauge transformation
\begin{eqnarray}
\hat{V}(x)\longrightarrow\hat{V}^\prime(x)=\hat{U}(x){\star}
\hat{V}(x){\star}\hat{U}^{\dagger}(x)
\label{dualp4} 
\end{eqnarray}
for $\hat U(x)\in U(1)_{\star}$.
The reason for this is quite simple to see. 
If it were to remain invariant, this would have implied that the
Moyal bracket between $\hat{V}$ and $\hat{U}$, 
$\forall$ $\hat U \in  U(1)_{\star}$ vanishes 
$([\hat{V}, \hat{U}]_{\star}=0)$. 
Through Wigner-Weyl correspondence (\ref{Weylstar}),
this in turn implies that $V_{\rm op}$ commutes 
with $U_{\rm op}$ at the operator level:  
$[V_{\rm op}, U_{\rm op}]=0\; \forall U_{\rm op}$.  
Applying Schur's lemma, assuming that $U_{\rm op}$ acts irreducibly, 
this indicates $V_{\rm op}=$constant.
Clearly this does not have the desired property. Now the SW 
transformation property of $\hat{V}(x)$ can be easily obtained as 
\begin{eqnarray}
\hat{V}^{\prime}(x) = \hat{V}(x) - 
\delta\theta\epsilon^{ij}\hat{A}_i\star\partial_j\hat{V}(x)\;
\label{dualp5} 
\end{eqnarray}
which relates the noncommutative potential 
$\hat{V}(x;\theta) \equiv \hat{V}(x)$ for 
noncommutative parameter $\theta$ to the corresponding 
noncommutative potential $\hat{V}(x;\theta + \delta\theta) 
\equiv \hat{V}^{\prime}(x)$ for noncommutative parameter 
$(\theta + \delta\theta)$. For the noncommutative 
gauge potential (\ref{symgauge1}), 
this leads to the following differential equation 
\begin{eqnarray}
\frac{d{\hat V(\theta)}}{d\theta} = 
\frac{\bar B(\theta)}{2}r\frac{d\hat V(\theta)}{dr} 
\label{dualp6} 
\end{eqnarray}
which can be solved by the method of seperation of variables\footnote{Such 
a seperation of variables can be made as one can expect that
a commutative central potential goes over to another central potential
of the same form but of different coupling constant 
at the noncommutative level. With this only the coupling constant
is subjected to SW flow.}, i.e. by taking $\hat V(r,\theta) = 
V(r)\hat\lambda_{sw}(\theta)$. We also have the boundary condition 
$\hat\lambda_{sw}(\theta=0)=\lambda$.
Using this, eq.(\ref{dualp6}) simplifies to 
\begin{eqnarray}
\frac{2}{\bar B(\theta)\hat\lambda_{sw}(\theta)}
\frac{d\hat\lambda_{sw}(\theta)}{d\theta} = 
\frac{r}{V(r)}\frac{dV(r)}{dr} = k(=\mathrm{constant}).\nonumber\\
\label{dualz9} 
\end{eqnarray}
Solving we get 
\begin{eqnarray}
V(r) &=& \lambda r^k\; \nonumber \\
\hat\lambda_{sw}(\theta) &=& \lambda\exp\left[\frac{k}{2}
\int^{\theta}_{0}d\theta^{'}\bar B(\theta^{'})\right]\nonumber \\
&=& \lambda\left(\frac{1 + \left(1 - \theta B\right)^{\frac{1}{2}}}{2}
\right)^{-2k}\;.
\label{dualz10} 
\end{eqnarray}
For $k = 2$, we get the usual harmonic oscillator, i.e.
\begin{eqnarray}
\hat V(r,\theta) &=& \hat\lambda_{sw}(\theta){r}^{2}\nonumber \\
                 &=& \lambda\left(\frac{1 + \left(1 - \theta B\right)
^{\frac{1}{2}}}{2}\right)^{-4}{r}^{2}\;. 
\label{dualz11}
\end{eqnarray}
If we now demand as in the free case that eq.(\ref{dualz3}) 
satisfies eq.(\ref{dualswmap20})
then the solution of eq.(\ref{dualswmap20}) 
can also be found by taking the trial solution 
$\hat\psi(z,\bar{z};\theta) = N\exp\left(- \frac{\bar{z}z}{4} 
g(\theta)\right)$ subject to the boundary condition (\ref{dualz300}) 
at $\theta = 0 $. This leads to the solution
\begin{eqnarray}
\hat\psi_{sw}(z, \bar z; \theta) = N
\exp\left[-\bar{z}z\frac{\sqrt{\left(B^2 + 8m\lambda \right)}}
{\left((1 - \theta B)^{\frac{1}{2}} + 1\right)^2}\right]\;.
\label{dualz5} 
\end{eqnarray}
Comparing eq.(s) (\ref{dualz3}) and (\ref{dualz5}), 
we get an algebraic equation
\begin{eqnarray}
\frac{4\left(B^2 + 8m\lambda \right)^{\frac{1}{2}}}
{\left[\left(1 - \theta B\right)^{\frac{1}{2}} +1\right]^2} 
= \left[\frac{2\bar{B}^{2}(\theta) + 16\hat\lambda_{sw}
\hat m_{sw} }{2\left(1 + \frac{\theta\bar{B}(\theta)}{4}
\right)^2 + \theta^2 \hat\lambda_{sw}\hat m_{sw}}\right]^{\frac{1}{2}}\;
\label{dualz6} 
\end{eqnarray}
which leads to
\begin{eqnarray}
&&\hat\lambda_{sw}\hat m_{sw} =\nonumber\\
&&\frac{4m\lambda\left[1 + \left(1 - \theta B\right)^{\frac{1}{2}}
\right]^2}{\left(1 - \theta B\right)\left(\{
\left(1 - \theta B\right)^{\frac{1}{2}} + 1\}^4 - 
\theta^{2}\left(B^2 + 8m\lambda \right)\right)}\;.\nonumber\\
\label{dualz7} 
\end{eqnarray}
Substituting the value of $\hat\lambda_{sw}(\theta)$ from eq.(\ref{dualz10}), 
we obtain the value of $\hat m_{sw}(\theta)$ as
\begin{eqnarray}
\hat m_{sw}(\theta) = \frac{m}{4\left(1 - \theta B\right)}
\frac{\left[1 + \left(1 - \theta B\right)^
{\frac{1}{2}}\right]^6}
{\left[\{\left(1 - \theta B\right)^
{\frac{1}{2}} + 1\}^4 - \theta^{2}
\left(B^2 + 8m\lambda\right)\right]}\;.\nonumber\\
\label{dualz12} 
\end{eqnarray}
The flow structure of $\hat\lambda_{sw}$ (eq.(\ref{dualz10})) and 
$\hat m_{sw}$ (eq.(\ref{dualz12})) in $\theta$ 
shows that the SW-flow is different
(in the presence of interactions) from the flows obtained
in the previous section eq.(s) (\ref{dual600a}) and (\ref{dual604a}) from
the consideration of the stability of the spectrum, although the formal
structure of the wave-functions $\hat\psi_{sw}$ (eq.(\ref{dualz5})) 
and $\hat\psi$ (eq.(\ref{dualz3})) are the same.  Indeed, it can be checked 
easily and explicitly that the flow obtained here (eq.(s) (\ref{dualz10}) 
and (\ref{dualz12})) from the SW map is not spectrum preserving, 
as is the case with the flow of the previous section. 
This indicates that these flows are not equivalent 
or related in some simple way. 
\vskip 0.003cm
\noindent We have already seen that in absence of 
interaction $(\hat\lambda=0)$ the noncommutative 
wave-function $\hat\psi_{sw}$ satisfies the SW map, 
subject to the boundary condition (\ref{10dapp}) at $\theta = 0$, when 
$\hat{\psi}_{sw}$ becomes identifiable with the commutative 
wave-function $\psi$. Also, unlike its 
noncommutative counterpart $\hat{\psi}$,
the commutative wave-function $\psi$ does not have a 
flow of its own in $\theta$. However, the situation 
changes drastically in the presence of interactions. 
To see this more clearly, let us consider the Schr\"odinger equation
\begin{eqnarray}
iD_0\psi = -\frac{1}{2m}D_iD_i\psi - \frac{i\theta}{2}\epsilon^{ij}
F_{i0}D_j\psi\;
\label{dualp3} 
\end{eqnarray}
obtained from the $U(1)$ gauge invariant effective action 
in the presence of a background gauge field, derived in 
the previous chapter to leading order in the noncommutative 
parameter $\theta$. Note that the temporal component 
$A_0$ of the background gauge field can be regarded as 
$(-V)$, where $V$ is the potential since this background 
gauge field is time independent. Indeed the SW transformation 
property of both $A_0$ and $V$ become identical, as can be seen 
from  eq.(s) (\ref{dualp5}) and (\ref{gal41}). This helps us to identify, 
again to leading order in $\theta$, the corresponding Hamiltonian as
\begin{eqnarray}
H = \frac{\left({\bf{p}} - \bf{A}\right)^2}{2m} + V - \frac{\theta}{2}
\epsilon^{ij}\partial_iV\left({p_j} -  A_j\right).
\label{dualp7} 
\end{eqnarray}
For a central potential $V(r)$, this simplifies in the symmetric gauge 
(\ref{symgauge}) to
\begin{eqnarray}
H = \frac{\left({\bf{p}} - \bf{A}\right)^2}{2m} + V - \frac{\theta}{2r}
\frac{\partial V}{\partial r}\left(L_z - \frac{B}{2}r^2\right).
\label{dualp8} 
\end{eqnarray}
Again for a harmonic potential $V(r) = \lambda r^2$, this takes the form 
\begin{eqnarray}
H = \frac{{\bf{p}^2}}{2m} + \frac{B^{\prime2}}{8m}r^2 - \tilde{\Lambda}L_z
\label{dualp9} 
\end{eqnarray}
where, ${B^{\prime}} = B\sqrt{1 + \frac{8m\lambda}{B^2}
(1 + \frac{\theta B}{2})}$ and $\tilde{\Lambda} = 
\frac{B}{2m} + \theta\lambda$.
Recognising that the structure of eq.(\ref{dualp9}) is the same 
as that of eq.(\ref{duality5}),
we can readily write down the ground state wave-function as
\begin{eqnarray}
\psi_{0}(z,\bar z; \theta) &=& \exp\left(-\frac{B^{\prime}(\theta)}
{4}\bar{z} z
\right)\nonumber\\
 &=& \exp\left(-\frac{1}{4}
\bar{z}z\sqrt{B^2 + 8m\lambda
\left(1+\frac{\theta B}{2}\right)}\right)\;;\nonumber\\
 &&|\theta|<<1\;.
\label{dualp90} 
\end{eqnarray}
This expression clearly reveals the fact that the commutative
wave-function has a non-trivial flow in $\theta$ of its own, only in the 
presence of interaction ($\lambda\neq0$) and the values of both
noncommutative wave-functions $\hat \psi$, $\hat \psi_{sw}$ and the 
commutative one $\psi$ 
coincide at $\theta=0$. One can, in principle, determine the exact
expression of this wave-function, valid upto all orders in $\theta$, 
but we shall not require this here. In fact the
wave-function (\ref{dualp90}) or higher angular
momentum states $z^l\psi_{0}(z,\bar z;\theta)$ can
be alternatively determined from perturbation theory
applied to each angular momentum sector $l$ for small
$\theta$ and $\lambda$. A point that we would
like to emphasise is that the SW map does not map the 
noncommutative field 
$\hat\psi_{sw}(z,\bar z;\theta)$ at value $\theta$ 
to the corresponding one at the
commutative level $\psi(z, \bar z;\theta)$; 
the SW map or equivalently the
SW equation (\ref{dualswmap20}) only relates 
$\hat\psi(z, \bar z;\theta)$ to
$\hat\psi(z, \bar z;\theta=0)=\psi(z, \bar z;\theta=0)$.
Furthermore, the fact that the
parameter $\hat m_{sw}(\theta)$ (eq.(\ref{dualz12}))
does not reproduce the expression to leading order in $\theta$, 
derived in the previous chapter eq.(\ref{gal4303}) 
can be seen to follow from the observation 
that the parameter $m$ was basically fixed by demanding the
form invariance of the Schr\"{o}dinger action
which is equivalent to the stability analysis (in absence of interaction)
we have carried out in the previous sections. Also observe that in
eq.(\ref{gal4303}) the ``renormalised" mass parameter $m$
does not get modified by the interaction term in any way,
in contrast to both $\hat m_{sw}$ (eq.(\ref{dualz12})) and $\hat m$ 
(eq.(\ref{dual600a})). 
On the other hand, the commutative wave-function $\psi$ in eq.(\ref{dualp90}) 
gets modified in presence of interaction, as we mentioned 
above, in such a way that it has a non-trivial flow in $\theta$. 
This is in contrast to the noncommutative wave-functions $\hat\psi$ 
(eq.(\ref{dualz3}))
and $\hat\psi_{sw}$ (eq.(\ref{dualz5}))
which have flows in $\theta$ even in absence of interactions.
Finally, note that we have three versions of the Hamiltonians
here with distinct transformations properties : (i) $\hat{H}$ occuring
in eq.(\ref{dualg1}) transforms adjointly under $U(1)_\star$ 
gauge transformation, (ii) $H$ 
occuring in  eq.(\ref{dualp7}) transforms adjointly under ordinary $U(1)$ 
gauge transformation and (iii) the Bopp-shifted Hamiltonian 
$\hat{H}_{BS}$ occuring in eq.(\ref{dualg1}) which, 
however, does not have any of these transformation 
properties under either type of gauge transformation as it was
constructed just by disentangling the $\star$ product but retaining 
the noncommutative variables. In this context, it will be worthwhile
to remember that in order to have the symmetry under 
$\star$ gauge transformation we must have noncommutative variables
composed through $\star$ product and to have the corresponding 
symmetry under ordinary gauge transformation, we must replace the 
noncommutative variables by commutative ones by making use of 
the SW map apart from disentangling the $\star$ product as was 
done in the previous chapter \cite{bcsgas}. Consequently, the issue of 
maintaining the gauge invariance/covariance is not relevant here, 
since we are dealing with $\hat{H}_{BS}$ in this chapter.

\section{Constructing dualities}
In the earlier sections, we have seen how physically equivalent families
of noncommuting Hamiltonians can be constructed.
In this construction $\theta$ simply plays the role
of a parameter and subsequently, as the physics does not change,
physical quantities can be computed with any value of this parameter. 
A natural question to pose, therefore,
 is whether there is any advantage in choosing
 a specific value of $\theta$, i.e., is there any advantage
 in introducing noncommutativity in the first place. The motivation
 for asking this question was already outlined
 in section $5.1$, where it was pointed
 out that in some existing literature \cite{jelal},
 the noncommutative quantum Hall system is considered
 a paradigm for the fractional quantum Hall effect which,
 however, requires the presence of interactions.
 If this interpretation is to be taken seriously
 a natural possibility that presents itself is that
 interacting commuting systems may in some approximation
 be equivalent to a particular non-interacting noncommutative system. 
If this turns out to be true, it would provide a
 new rational for the introduction of noncommutativity
 in quantum Hall systems.
 In this section we explore this possibility within
 a very simple setting.
\vskip 0.003cm
\noindent We consider the noncommutative harmonic oscillator moving
 in a constant magnetic field discussed in section $5.3$.
 After undoing the star product through a Bopp-shift
 we find the Hamiltonian
\begin{eqnarray}
\hat H_{BS}(\theta) &=& \frac{{\bf p}^{2}}{2M_0} + 
\frac{{\bf x}^2}{2}M_0{\Omega_0}^{2} - 
\Omega_0(\theta)L_{z}\nonumber\\&&+\hat\lambda 
\left(\frac{\theta^2}{4}{\bf p}^{2}+{\bf x}^{2}-\theta L_z\right)\nonumber\\
&=& \hat H_0+\hat V
\label{dual1} 
\end{eqnarray}
where,
\begin{eqnarray}
&&\frac{1}{2M_0} = 
\frac{\left(1 + \frac{\bar B\theta}{4}\right)^2}{2\hat m}
\nonumber \\
&&\frac{1}{2}M_0{\Omega_0}^{2} = \frac{\bar B(\theta)^2}{8\hat m(\theta)}\;.
\label{dual2}
\end{eqnarray}
To represent equivalent systems, the parameters $\bar B$, $\hat m$ and 
$\hat\lambda$ are parameterized as in eq.(s) (\ref{duals5}), (\ref{dual600a}) 
and (\ref{dual604a}), respectively.  
\vskip 0.003cm
\noindent Naively one might argue that when the 
noncommutative coupling constant $\hat\lambda$ becomes small, 
the interaction term can be neglected on the noncommutative level. 
However, as this happens when $\theta$ becomes 
large ($\hat\lambda\sim 1/\theta$), 
one sees from the Bopp-shifted equivalent of 
the Hamiltonian that this is not true due to the 
$\theta$ dependency that is generated by the Bopp-shift. 
One therefore has to use a different criterion to decide 
when the interaction term $\hat V$ is small and can be neglected. 
One way is to introduce a norm on the space of operators and check 
that $\hat V$ is small in this norm. The trace norm 
${\rm tr}(\hat V^\dagger \hat V)$ is divergent and cannot 
be used; a regularization is required. 
An obvious alternative candidate to use is the following
\begin{eqnarray}
Z(\theta)=\frac{{\rm tr}(\hat V^\dagger e^{-\beta \hat H_0} 
\hat V)}{{\rm tr} e^{-\beta \hat H_0}}\;.
\label{dual3}
\end{eqnarray}
\vskip 0.003cm
\noindent Here $\beta$ plays the role of an energy cutt-off. 
It is clear that $Z(\theta)$ has all the properties of a norm, 
in particular $Z(\theta)=0$ if and only if $\hat V=0$. 
As remarked before, it is impossible to eliminate 
$\hat V$ completely, however, we can minimize 
$Z(\theta)$ with respect to $\theta$ and in doing 
this find the value of $\theta$ for which the  
noncommutative non-interacting Hamiltonian $\hat H_0$ 
gives the best approximation to the interacting Hamiltonian. 
Since the low energy spectrum of $\hat H_0$ is 
biased in the norm 
(\ref{dual3}), one can expect that the low energy spectrum of 
$\hat H_0$ would give good agreement 
with the interacting spectrum, while 
the agreement will become worse as one moves 
up in the spectrum of $\hat H_0$. Before 
implementing this program, there is one further 
complication to take care of. 
Due to the degeneracy of $\hat H_0$ in the angular momentum, 
the norm (\ref{dual3}) is still 
divergent when summing over angular momenta in the trace. 
However, since $\hat V$ is a central potential and 
subsequently different angular momentum sectors decouple, 
it is quite sufficient to implement the program above 
in each angular momentum sector seperately. 
Under minimization this will give rise to an 
angular momentum dependent value of $\theta$, 
giving rise to a lifting in the degeneracy in 
angular momentum, which is what one would expect 
in the presence of interactions.  To proceed 
we therefore replace eq.(\ref{dual3}) by           
\begin{eqnarray}
Z(\theta,\ell)&=&\frac{{\rm tr}_\ell(\hat V^\dagger e^{-\beta 
\hat H_0} \hat V)}{{\rm tr}_\ell e^{-\beta \hat H_0}}\nonumber\\
&=&\sum_{n_-=0}^{\infty}|\langle n_-,\ell|V|n_-,\ell\rangle
|^2e^{-\beta \Omega_0 (2n_-+1)}\;
\label{dual3a}
\end{eqnarray}
where ${\rm tr}_\ell$ denotes that the trace is 
taken over a fixed angular momentum sector, 
eq.(\ref{dual10.1}) was used and $|n_-,\ell\rangle$ 
denote the eigenstates of $\hat H_0$. 
This expression can be evaluated straightforwardly to yield
\begin{eqnarray}
Z(\theta,\ell)&=&\hat\lambda^2(\theta)\left[\Gamma(\theta)^2\left(1+\frac{2}
{\sinh^2(\beta\Omega_0)}\right)\right.\nonumber\\
&+&2\ell\coth(\beta\Omega_0)\Gamma(\theta)\left(\Gamma(\theta)-
\theta\right)\nonumber\\
&+&\left.\ell^2\left(\Gamma(\theta)-\theta\right)^2\right]\;
\label{dual4}
\end{eqnarray}
where,
\begin{eqnarray}
\Gamma(\theta)=\frac{M_0\Omega_0\theta^2}{4}+\frac{1}{M_0\Omega_0}\;.
\label{dual4a}
\end{eqnarray}
For $\beta>>1/B$, one finds the value of $\theta$ 
that minimizes this expression to be 
\begin{eqnarray}
\theta(\ell)=\frac{2(1+\ell)}{B(1+2\ell)}
\label{dual5}
\end{eqnarray}
at which value $Z(\theta,\ell)\sim\frac{1}{B^2}$, which means 
that the potential at these values of $\theta$ 
can be treated as a correction of order $1/B$. 
The eigenvalues of $\hat H_0$ at these values 
of $\theta$ are easily evaluated to be 
\begin{eqnarray}
E_{n_-}(\ell)&=&2\Omega_0(\ell)(n_-+1/2)\;\nonumber\\
\Omega_0(\ell)&=&\frac{B}{4m}\left(1+\sqrt{1+\frac{16\lambda 
m(\ell+1)}{B^2}}\right)\;.
\label{dual6}
\end{eqnarray}
From the above considerations it is clear that the approximation 
is controlled by $1/B$.  One therefore expects 
eq.(\ref{dual6}) to agree with the exact result eq.(\ref{dualz2}), 
at least for the lowest eigenvalues, to order $1/B$. 
This indeed turns out to be the case. 
Expanding the lowest eigenvalues of eq.(s) (\ref{dual6}) 
and (\ref{dualz2}) to leading order in $1/B$, one finds in both cases
\begin{eqnarray}
E_0(\ell)=\frac{B}{2m}+\frac{2(\ell+1)\lambda}{B}\;.
\label{dual7}
\end{eqnarray}
This result suggests that it is indeed possible to trade 
the interactions for noncommutativity, 
at least in the lowest Landau level and 
for weak Landau level mixing (large $B$). 
It would, of course, be exceedingly naive 
to immediately extrapolate from the above 
to realistic quantum Hall systems. 
However, the above result does suggest 
a new paradigm for noncommutative quantum 
Hall systems worthwhile to explore. Within this 
paradigm interactions get traded, at least in the 
lowest Landau level, for noncommutativity, 
explaining the fractional filling fractions 
and emergence of composite fermions from a new perspective.

\section{Summary}
We have demonstrated how physically equivalent 
families of noncommutative Hamiltonians can be constructed. 
This program was explicitly implemented to all orders 
in the noncommutative parameter in the case of a 
free particle and harmonic oscillator moving in a 
constant magnetic field in two dimensions. 
It was found that this spectrum preserving 
map coincides with the SW map 
in the case of no interactions, but not in the presence 
of interactions. A new possible paradigm for 
noncommutative quantum Hall systems was demonstrated 
in a simple setting.  In this paradigm an interacting 
commutative system is traded for a weakly interacting 
noncommutative system, resulting in the same physics 
for the low energy sector. 
This provides a new rational for the introduction 
of noncommutativity in quantum Hall systems.  
\section*{Appendix: Eigenvalues and eigenstates of the free
and harmonic oscillator Hamiltonians }
To solve for the eigenvalues and eigenstates of eq.(\ref{duality5}), 
creation and annihilation operators are introduced through the equations
\begin{eqnarray}
b_x = \sqrt{\frac{M\Omega}{2}}\left(x + \frac{ip_x}{M\Omega}\right)\quad,\quad  
b_{x}^{\dagger} = \sqrt{\frac{M\Omega}{2}}\left(x - \frac{ip_x}{M\Omega}
\right)\;\nonumber\\
b_y = \sqrt{\frac{M\Omega}{2}}\left(y + \frac{ip_y}{M\Omega}\right)\quad,
\quad  b_{y}^{\dagger} = \sqrt{\frac{M\Omega}{2}}\left(y - 
\frac{ip_y}{M\Omega}\right)\;.\nonumber\\
\label{6app} 
\end{eqnarray}
In terms of these operators the Hamiltonian (\ref{duality5}) takes the form:
\begin{eqnarray}
 H = \Omega \left(b_{x}^{\dagger}b_x + b_{y}^{\dagger}b_{y} + 1\right) - 
i\Omega\left(b_{x}b_{y}^{\dagger} - b_{x}^{\dagger}b_{y}\right)\;.
\label{7app} 
\end{eqnarray}
\vskip 0.003cm
\noindent Now making
use of the following set of transformations
\begin{eqnarray}
b_{+} = \frac{1}{\sqrt{2}}\left(b_{x} - ib_{y}\right)\; , \quad 
b_{+}^{\dagger} = \frac{1}{\sqrt{2}}\left(b_{x}^{\dagger} + ib_{y}
^{\dagger} \right) \;\nonumber\\
b_{-} = \frac{1}{\sqrt{2}}\left(b_{x} + ib_{y}\right)\;, 
\quad  b_{-}^{\dagger} = \frac{1}{\sqrt{2}}\left(b_{x}^{\dagger} 
- ib_{y}^{\dagger}\right)\;
\label{9app} 
\end{eqnarray}
the Hamiltonian (\ref{7app}) reads
\begin{eqnarray}
 H &=& \Omega \left(b_{+}^{\dagger}b_{+} + b_{-}^{\dagger}b_{-} + 1\right) - 
\Omega\left(b_{+}^{\dagger}b_{+} - b_{-}^{\dagger}b_{-}\right) \nonumber \\
 &=& \Omega \left(n_{+} + n_{-} + 1\right) - \Omega\left(n_{+} - n_{-}\right)
\nonumber \\
&=& 2\Omega\left( n_{-} + \frac{1}{2}\right)\;.
\label{10app} 
\end{eqnarray}
Note that the energy spectrum depends only on $n_{-}$ and is independent 
of $n_{+}$. Therefore, it results in an infinite degeneracy 
in the energy spectrum.
 The above cancellation of the terms involving $n_{+}$ has taken place since
 the coefficients of $n_{+}$ are equal. This is also true
 in the limit $\theta = 0$. This feature does not persist in presence
of interactions (see section $5.3$). \\ 
\noindent Introducing complex coordinates $z = x + iy$ 
and $\bar z = x - iy$, 
eq.(\ref{9app}) takes the form
\begin{eqnarray}
b_{+} = \frac{1}{2}\sqrt{M\Omega}\left[\bar z + \frac{2}{M\Omega}
\partial_{z}\right]\;, \   b_{+}^{\dagger} = 
\frac{1}{2}\sqrt{M\Omega}\left[z - \frac{2}{M\Omega}
\partial_{\bar z}\right]\;\nonumber\\
b_{-} = \frac{1}{2}\sqrt{M\Omega}\left[z + 
\frac{2}{M\Omega}\partial_{\bar z}\right]\;, 
\   b_{-}^{\dagger} = \frac{1}{2}\sqrt{M\Omega}
\left[\bar z - \frac{2}{M\Omega}\partial_{z}\right]\;.\nonumber\\
\label{10bapp} 
\end{eqnarray}
The ground state wave-function is annihilated by $b_{-}$, i.e. 
$b_{-}\hat\psi(z, \bar z; \theta) = 0$. 
This immediately leads to the solution
\begin{eqnarray}
\hat\psi_{0}(z, \bar z;\theta) = N\exp\left[-\frac{M\Omega}{2}
\bar{z}z\right]
 =  N\exp\left[- \frac{\bar B}{4\left(1 + \frac{\bar B\theta}{4}\right)}
\bar{z}z\right].\nonumber\\
\label{10capp} 
\end{eqnarray}
Since $\bar B(\theta = 0) = B$,  the above solution goes smoothly
to the commutative result
\begin{eqnarray}
\psi(z, \bar z) = N\exp\left[- \frac{B}{4}\bar{z}z\right]\;.
\label{10dapp} 
\end{eqnarray}
This state is also annihilated by $b_{+}$ and
therefore corresponds to zero angular momentum state, as the angular momentum 
operator $L_{3}=(xp_{y}-yp_{x})$ takes the following form 
\begin{eqnarray}
L_{3}=i\left(b_{x}b_{y}^{\dag}-b_{x}^{\dag}b_{y}\right)
=\left(b_{+}^{\dag}b_{+}-b_{-}^{\dag}b_{-}\right).
\label{1000aapp} 
\end{eqnarray}
If this $xy$-plane is thought to be embedded in $3-d$ Euclidean space
${\mathcal{R}}^{3}$, then the other rotational generators $L_{1}$ and  $L_{2}$
obtained by cyclic permutation would result in the standard angular
momentum $SU(2)$ algebra 
\begin{eqnarray}
\left[L_{i}, L_{j}\right]=i\epsilon_{ijk}L_{k}.
\label{1000bapp} 
\end{eqnarray}
One can, however, define the $SU(2)$ algebra using 
the creation and annihilation operators alone, 
which in the cartesian basis (\ref{6app}), is given by:
\begin{eqnarray}
J_{1} &=& \frac{1}{2}\left(b^{\dagger}_{x}b_{x} - 
b^{\dagger}_{y}b_{y}\right)\; \nonumber \\
J_{2} &=& \frac{1}{2}\left(b^{\dagger}_{x}
b_{y} + b^{\dagger}_{y}b_{x}\right)\; \nonumber \\
J_{3} &=& \frac{1}{2i}\left(b^{\dagger}_{x}b_{y} - 
b^{\dagger}_{y}b_{x}\right)\;
\label{8app} 
\end{eqnarray}
satisfying $\left[J_{i}, J_{j}\right]=i\epsilon_{ijk}J_{k}$. As one can
easily verify, by computing the PB(s) of the generators
with phase-space variables that $J_{1}$ generates
rotation in $(x, p_{x})$ and $(y, p_{y})$ planes, $J_{y}$ in $(x, p_{y})$
and $(y, p_{x})$ planes and $J_{z}$ in $(x, y)$ and $(p_{x}, p_{y})$ planes.
Also note that $L_{3}$ is not identical to $J_{3}$ but differs by a factor
of $2$: $L_{3}=2J_{3}$.
\vskip 0.003cm
\noindent The Casimir operator in terms of $J_{i}$ 
representation now becomes
\begin{eqnarray}
\vec{J}^2= \frac{1}{4}\left(b_{+}^\dag b_{+}+b_{-}^\dag b_{-}\right)
\left(b_{+}^\dag b_{+}+b_{-}^\dag b_{-}+2\right)
\label{840app} 
\end{eqnarray}
with eigenvalues
 $\vec{J}^2= \frac{1}{4}\left(n_{+}+n_{-}\right)\left(n_{+}+n_{-}+2\right).$
Defining $n_{+}+n_{-}=2j$, the Casimir becomes $\vec{J}^2=j(j+1)$. Also,
if the eigenvalues of $J_{3}$ is given by $l^{\prime}$, then the
eigenvalues of $L_{3}$ will be given by $n_{+}-n_{-}=2l^{\prime}=l~\epsilon 
~\mathcal{Z}$. Note that, like $l^{\prime}$, $j$ also admits half-integral
values. Finally, one can write down the eigenvalues (\ref{10app}) as
\begin{eqnarray}
E_{n_{-}}=\Omega\left(2j-2l^{\prime}+1\right)=\Omega\left(2j-l+1\right)
\label{841app} 
\end{eqnarray}
which agrees with \cite{gamboa}. Any arbitrary state can now be
obtained by repeated application of 
$b_{\pm}^{\dag}$ on eq.(\ref{10capp}) as
\begin{eqnarray}
\vert n_{-}, l\rangle \sim \left(b_{-}^{\dag}\right)^{n_{-}}
\left(b_{+}^{\dag}\right)^{l}\hat\psi_{0}(z, \bar z;\theta)\;.
\label{842app} 
\end{eqnarray}
\chapter{Noncommutativity and quantum Hall systems}
\section{Introduction}
\label{intro}
After investigating noncommutative quantum mechanics in the 
previous chapter, where we tried to provide a new rationale for
introducing noncommutativity in quantum Hall systems in the
sense that interactions can be traded with noncommutativity within
certain approximation, we now try to present a ``complementary"
point of view on the impact of noncommutativity stemming from the 
inter-particle interactions in quantum Hall systems. This issue has 
recently attracted considerable attention from
the point of view of noncommutative quantum mechanics and quantum
field theory \cite{np}, \cite{bop}, \cite{dh}, \cite{su3}, 
\cite{du}, \cite{jelal}, \cite{myung} 
as it is probably the simplest
physical realization of a noncommutative spatial geometry. 

\noindent Some time ago Dunne, Jackiw and Trugenberger 
\cite{dunne}, \cite{dunne1} already observed this
noncommutativity by noting that in the limit $m\rightarrow 0$
the $y$-coordinate is effectively 
constrained to the momentum canonical
conjugate to the $x$-coordinate. This result can also be
obtained \cite{myung1,ouvry1} by keeping the mass fixed and taking
the limit $B\rightarrow\infty$. An alternative point of view is to keep
the magnetic field and mass finite, but to project the position coordinates
onto the lowest (or higher) Landau level. These projected operators
indeed satisfy the commutation relation ($\hbar=e=m=c=1$) \cite{ouvry2}
\begin{equation}
\label{commutator}
\left[P_0xP_0,P_0yP_0\right]=\frac{1}{iB}P_0=\frac{1}{iB}
\end{equation}
where $P_0$ denotes the projector onto the lowest Landau level, which is
also just the identity operator on the projected subspace, as reflected
in the last step.
\vskip 0.003cm
\noindent                                                                
This result allows a simple heuristic understanding of quantum Hall fluids.
Recall the elementary uncertainty relation (see {\it e.g.} \cite{gaz})
for two noncommuting operators $A$ and $B$
\begin{eqnarray}
\label{uncertain}
(\Delta A)^2(\Delta B)^2&\geq& \frac{1}{4}\langle i[A,B]\rangle^2\,\\
(\Delta A)^2=\langle A^2\rangle-\langle A\rangle^2\,\quad,
&&(\Delta B)^2=\langle B^2\rangle-\langle B\rangle^2\,\nonumber
\end{eqnarray}
where, $\langle\cdot\rangle$ denotes the normalized expectation value in
some state. Using this we note that the noncommutativity of the coordinates
implies a lower bound to the area a particle in the lowest Landau
level occupies. This bound follows easily from eq.(\ref{uncertain}) to be
\begin{eqnarray}
\Delta A= 4|\Delta (P_0xP_0)||\Delta (P_0yP_0)|\geq\frac{2}{B}
\equiv \Delta A_0~.
\label{bound707}
\end{eqnarray}
Therefore the number of states available in
a Landau level is given by:
\begin{eqnarray}
M=\frac{A}{\Delta A_0}=\frac{AB}{2}~. 
\label{bound7077}
\end{eqnarray}
The filling fraction is defined in the usual 
way as 
\begin{eqnarray}
\nu=\frac{N}{M}=\frac{2N}{AB}
\label{bound77077}
\end{eqnarray}
where, $N$ is the number of electrons. 
For fermions it then follows that at
maximum filling of $p$ Landau levels the particles must occupy the minimal
allowed area, i.e. $N\Delta A_0=pA$ and $\nu=p$ with $p$ 
being an integer.
As the next available states are in the higher Landau level, separated
in energy by the cyclotron frequency, one expects that the quantum fluid
will be incompressible at these values of the filling fraction.
\vskip 0.003cm
\noindent                                                             
The literature mentioned above does not take into account 
the effect that interactions between
electrons might have on the noncommutativity.
In \cite{bop}, an harmonic potential between two interacting
particles was considered, but there the noncommutativity of the center
of mass coordinates was investigated, 
which is again a Landau problem effectively. 
In particular the analysis of \cite{ouvry2} has
been done in the absence of any interactions between particles. 
The conjectured equivalence between a noncommutative $U(1)$ Chern-Simons
theory \cite{su1, su3} and the composite fermion description for
the fractional Hall effect, which is an effective non-interacting theory
for the interacting quantum Hall system, urges one to have a
better understanding of the relationship between 
noncommutativity and interactions.
A similar picture arises in the much simpler setting of noncommutative
quantum Hall systems where it seems as if the fractional quantum Hall
effect, associated with an interacting quantum Hall system, can effectively
be described by a non-interacting noncommutative quantum Hall
system \cite{jelal}, again suggesting an interplay between
noncommutativity and interactions. Indeed, keeping the picture of
the composite fermion in mind, which replaces electrons
interacting through a short ranged repulsive interaction by
non-interacting composite fermions moving in a reduced magnetic field,
one would expect that the interactions must modify the commutation
relation (\ref{commutator}) as the magnetic field is reduced. A similar
conclusion was reached from a completely different point of view in
the previous chapter \cite{fgsbcsgagh}. Here we want
to investigate this question in more detail using the approach
of \cite{ouvry2}.
\vskip 0.003cm
\noindent                                                                 
To set the scene, let us consider two interacting 
particles with the same masses
and charges moving in a plane with constant 
magnetic field perpendicular
to the plane. In the symmetric gauge the Hamiltonian is given
by ($\hbar=e=m=c=1$)
\begin{eqnarray}
\label{ham1}
H &=&\frac{1}{2}\left(\vec{p}_1-\vec{A}(\vec{x}_1)\right)^2+
\frac{1}{2}\left(\vec{p}_2-\vec{A}(\vec{x}_2)\right)^2+
V(|\vec{x}_1-\vec{x}_2|)\;\\
A_{i}(\vec{y})&=&-\frac{B}{2}\epsilon_{ij}y^{j},\;B\geq 0\;.\nonumber
\end{eqnarray}
Introducing relative and center of mass coordinates through
 \begin{equation}
\vec{R}=\frac{1}{2}(\vec{x}_1+\vec{x}_2)\quad
,\quad \vec{r}=\vec{x}_1-\vec{x}_2
\end{equation}
the Hamiltonian reduces to
\begin{eqnarray}
\label{ham2}
H &=&\frac{1}{4}\left(\vec{P}-2\vec{A}(\vec{R})\right)^2+
\left(\vec{p}-\frac{1}{2}\vec{A}(\vec{r})\right)^2+V(|\vec{r}|)\\
\vec{P}&=&(\vec{p}_1+\vec{p}_2),\quad \vec{p}=
\frac{1}{2}\left(\vec{p}_1-\vec{p}_2\right)\;.\nonumber
\end{eqnarray}
We find that the original problem have got splitted into 
two decoupled problems. The center of mass motion
corresponds to that of a particle with mass $M=2$ and charge $q=2e=2$
moving in a magnetic field $B$, while the relative motion is that of
a particle with reduced mass $\mu=\frac{m}{2}=\frac{1}{2}$ and
charge $q=\frac{e}{2}=\frac{1}{2}$ moving in the same magnetic field $B$ and
radial potential $V(|\vec{r}|)$. Clearly the 
cyclotron frequency for both problems
is $B$. The center of mass motion can clearly be analysed as
in \cite{ouvry2}; projection onto the lowest Landau level will lead
to noncommutative coordinates $[P_0XP_0,P_0YP_0]=\frac{1}{2iB}$.
Our analysis here concerns the relative motion. This might seem
problematic as the potential $V(|\vec{r}|)$ lifts the degeneracy of
the Landau levels so that one can apparently no longer think of projection
onto Landau levels, and particularly the lowest Landau level. Closer
inspection of the argument in \cite{ouvry2} reveals, however, that
the degeneracy is not essential. Indeed, the only requirement is that
the subspace on which is to be projected is infinite dimensional as
the noncommutative coordinates can only be realized in this case.
A natural generalization of the analysis in \cite{ouvry2} would
therefore be to identify a low energy infinite dimensional subspace on
which to perform the projection. In the case of short range interactions
$V(|\vec{r}|)$, for which the interaction energy scale is much less than
the cyclotron frequency, which is the situation normally assumed, this
can still be done. The reason is that the spectrum for the relative
motion will clearly be close to that of the Landau problem for large
values of the relative angular momentum as the particles are then well
separated. For small values of the angular momentum the potential will
have its main effect. However, if the interaction energy scale is much
less than the cyclotron frequency, one will still have well separated
bands of eigenstates, with the cyclotron frequency being the energy scale
determining the separation between bands and the interaction energy scale
determining the separation within bands. We can therefore identify an infinite
dimensional low energy subspace as the lowest Landau level perturbed
by the interaction and proceed to study the commutation relations of
the relative coordinates projected onto this subspace.
\vskip 0.003cm
\noindent
This chapter is organized in the following way. 
The general procedure of projection onto the low energy subspace
is described in section \ref{general}.
We then apply this procedure 
to a number of exactly
soluble interacting models to obtain insight 
into the underlying physics in section \ref{soluble}.
Finally, we conclude with a summary in section \ref{qhesummary}.
                                                                              
\section{General projection on the low energy sector}
\label{general}
We start by recalling a few basic facts about the Landau problem
discussed in the previous chapter.
A particle moving on a plane, subjected to a perpendicular constant
magnetic field $B$, has a discrete set of energy eigenstates, known as
Landau levels, and are labelled as $|n, \ell\rangle$, where $n$ and
$\ell$ are integers labelling the various Landau levels $(n)$ and
the degenerate angular momentum eigenstates with integer eigenvalues
$\ell(\geq-n)$ within the same Landau level $n$.
\noindent                                                                      
We focus on the relative motion of the two particles described by
the second part of the Hamiltonian (\ref{ham2})
\begin{equation}
\label{hamrel}
H =\left(\vec{p}-\frac{1}{2}\vec{A}(\vec{r})\right)^2+\tilde V(|\vec{r}|).
\end{equation}
From the rotational symmetry this problem can be solved as usual through
the separation of variables and the wave functions have the generic form
\begin{equation}
\psi_{n,\ell}(\vec{r})=\langle \vec{r}|n,\ell\rangle=
R_{n,\ell}(r)e^{i\ell\phi}
\end{equation}
where, $R_{n,\ell}$ solves the radial equation
\begin{equation}
\label{radial}
\left[-\frac{\partial^2}{\partial r^2}-\frac{1}{r}
\frac{\partial}{\partial r}+\frac{\ell^2}{r^2}-\omega_{c}\ell+
\frac{1}{4}\omega_{c}^{2}r^2 + V(|\vec{r}|)\right]R_{n,\ell}=
E_{n,\ell}R_{n,\ell}
\end{equation}
$n$ is the principle quantum number and $\omega_{c}=B/2$ is half of
the cyclotron frequency. Under the conditions discussed in
section \ref{intro} the separated bands of eigenstates will be labelled
by the principle quantum number, $n$, while the states within a band will
be labelled by the angular mometum $\ell$. In particular we assume that
the lowest energy states are described by $n=n_0$ (say) and
$\ell=0,1,2\ldots$ where we noted from eq.(\ref{radial}) that a change in
sign of the angular momentum will require an energy of the order of
the cyclotron frequency, so that negative angular momenta will not occur
in the low energy sector, {\it i.e.}, we are restricting to the lowest
Landau level, perturbed by interactions.
We can now construct the projection operator on the low energy sector as
\begin{equation}
\label{proj}
P_0=\sum_{\ell=0}^\infty |n_0,\ell\rangle\langle n_0\ell|.
\end{equation}
We now compute the projected relative coordinates
\begin{eqnarray}
\label{projop}
P_{0}xP_{0}&=&\sum_{l,l^{\prime}=0}^{\infty}
\langle n_0,\ell^{\prime}|x|n_0,\ell\rangle|n_0, \ell^{\prime}\rangle
\langle n_0, \ell|\,\nonumber\\
P_{0}yP_{0}&=&\sum_{l,l^{\prime}=0}^{\infty}
\langle n_0,\ell^{\prime}|y|n_0,\ell\rangle|n_0, \ell^{\prime}\rangle
\langle n_0, \ell|\,
\end{eqnarray}
with
\begin{eqnarray}
\label{me}
\langle n_0,\ell^{\prime}|x|n_0,\ell\rangle&=&\Omega_{\ell^\prime,\ell}
\left(\delta_{\ell^{\prime},\ell+1}+\delta_{\ell^{\prime},\ell-1}\right)
\,\nonumber\\
\langle n_0,\ell^{\prime}|y|n_0,\ell\rangle&=&-i\Omega_{\ell^\prime,\ell}
\left(\delta_{\ell^{\prime},\ell+1}-\delta_{\ell^{\prime},\ell-1}\right)
\,\nonumber\\
\Omega_{\ell^\prime,\ell}&=&\pi\int_0^\infty\; drr^2R^*_{n_0,\ell^{\prime}}
R_{n_0,\ell}\,.
\end{eqnarray}
The commutator of the relative coordinates then yields
\begin{equation}
\label{commrel}
\left[P_0xP_0,P_0yP_0\right]=2i\sum_{\ell=0}^{\ell=\infty}
|\Omega_{\ell,\ell+1}|^2\left[|n_0,\ell+1\rangle\langle n_0,\ell+1|-
|n_0,\ell\rangle\langle n_0,\ell|\right].
\end{equation}
\noindent                                                                      
We now simply have to compute the matrix elements 
$\Omega_{\ell^{\prime},\ell}$
to determine the commutator. For some potentials this can be done
analytically and exactly, but in most cases one has to resort to
approximations. In this regard we note that since the potential has
radial symmetry, it will not mix different angular momentum sectors.
Within a particular angular momentum sector there is of course no
degeneracy of the Landau states, so that one can safely apply
perturbation theory to compute the radial wave-functions $R_{n_0,\ell}$
and therefore matrix elements $\Omega_{\ell^{\prime},\ell}$. Indeed,
this corresponds to a $1/B$ expansion.
When the interaction is switched off ($V(r)=0$) the radial wave-functions
are those of the Landau problem and this result is easily seen
to reduce to eq.(\ref{commutator}), except for a factor of two. The reason
for this is simply that since we are working with the relative coordinates
between two particles this commutator should yield in the non-interacting
case the minimal area occupied by two particles, which is consistent
with eq.(\ref{commutator}).
\noindent 
In contrast to the non-interacting case eq.(\ref{commutator}), 
this commutator is in general no longer proportional to $P_0$. 
However, since we are
dealing with a central potential, the different angular momentum
($\ell$) sectors decouple and one can interpret this result as
a noncommutative theory with an effective $\ell$ dependent noncommutative
parameter in the same spirit as was done in 
the previous chapter \cite{fgsbcsgagh}.
\noindent 
As the area occupied by the two particles will increase with increasing
relative angular momentum, one can deduce from eq.(s) (\ref{commrel}) and
(\ref{uncertain}) an absolute lower bound to the average area that
a particle in the low energy sector may occupy
\begin{equation}
2\Delta A=4|\Delta (P_0xP_0)||\Delta (P_0yP_0)|\geq 4|\Omega_{0,1}|^2.
\end{equation}
The factor of two on the left is required as the right hand side is
the average area occupied by two particles, as pointed out earlier.
                                                                                
\section{Noncommutativity in some soluble models}
\label{soluble}
                                                                              
In this section, we study the noncommutative structure that arises in
a number of soluble interacting models to gain deeper insight into
the underlying physics.
                                                                                
\subsection{Harmonic oscillator}
\label{harosc}
                                                                                
\begin{figure}[t]
\begin{center}
       \epsfig{file=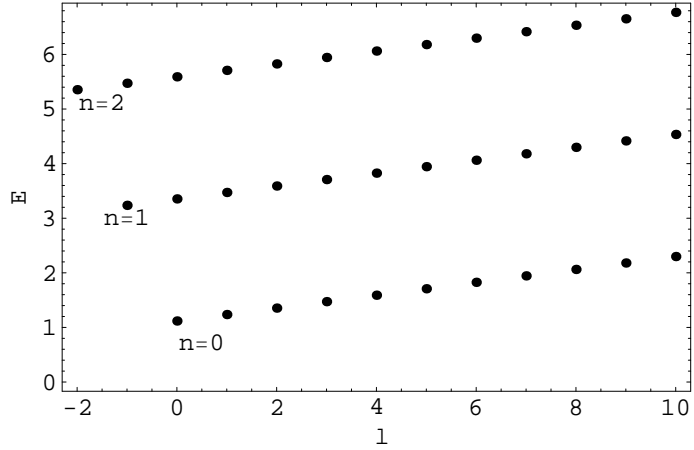,height=6.0cm,clip=,angle=0}
       \caption{The spectrum for the harmonic oscillator
potential $\omega_c=1$ and $\lambda=0.5$.}
        \label{Fig1}
\end{center}
\end{figure}

We take $V(|\vec{r}|)=\frac{\lambda^2}{4} r^2$.  This is not a
short range potential and the spectrum will not approach that of the
Landau problem for large values of $\ell$. Indeed, here one gets a
spectrum linearly growing in $\ell$ (see Fig.\ref{Fig1}) so that
one cannot claim that projection onto the lowest principle quantum number
will correspond to the lowest energy sector. However, as was pointed out
in the introduction, one can in principle project onto any
infinite dimensional subspace, not necessarily just the lowest energy,
and that is the spirit in which the current calculation is done.
\noindent                                                                 
The radial equation for the
lowest principle quantum number is easy to solve in this case 
and one obtains:
\begin{eqnarray}
\label{harmonic}
R_{0,\ell}&=&N_{\ell}r^\ell
\exp(-\frac{1}{4}\sqrt{\omega_{c}^2+\lambda^2} r^2)\quad;\quad \ell\geq 0\\
N_{\ell}&=&\frac{\left(\omega_{c}^2+\lambda^2\right)^{(\ell+1)/2}}
{\sqrt{\pi 2^{\ell+1}\Gamma(\ell+1)}}.
\end{eqnarray}
The spectrum is given by:
\begin{equation}
E_{0,\ell}=\ell(\sqrt{\omega_{c}^2+\lambda^2}-\omega_c)+
\sqrt{\omega_{c}^2+\lambda^2}
\end{equation}
and is linearly growing with increasing $\ell$. The spectrum and
eigenfunctions for higher quantum numbers can of course also be solved
easily and projection onto those subspaces can also be done. The full
spectrum is given by $E_{n,\ell}=\sqrt{\omega_c^2+\lambda^2}(2n+1)+
\ell(\sqrt{\omega_c^2+\lambda^2}-\omega_c)$, $n=0,1,2\ldots$, $\ell\geq -n$
and is shown in Fig.\ref{Fig1} for $\omega_c=1$ and $\lambda=0.5$.
As no new features appear we restrict ourselves here to the solutions
with the lowest principle quantum number.
                                                                                
\noindent The commutator of the relative coordinates can now be evaluated
from eq.(\ref{commrel}) and yields
\begin{eqnarray}
\label{harcomm}
\left[P_0xP_0,P_0yP_0\right]&=&\frac{i}{\sqrt{\omega_{c}^2+\lambda^2}}
\sum_{\ell=0}^{\ell=\infty}(\ell+1)\left[|n_0,\ell+1\rangle\langle n_0,
\ell+1|-
|n_0,\ell\rangle\langle n_0,\ell|\right]\nonumber\\
&=&\frac{1}{i\sqrt{\omega_{c}^2+\lambda^2}}P_0\nonumber\\
&=&\frac{2}{iB}(1+\frac{4\lambda^2}{B^2})^{-1/2}\,.
\end{eqnarray}
In the last line we have noticed that $P_0$ is just the identity on
the projected subspace. As was discussed in general, we note that when
the interaction is switched off ($\lambda=0$), this result differs
by a factor of two from eq.(\ref{commutator}).
The first important point to note from this computation is that,
generically, the noncommutative parameter is renormalized by
the interactions.
                                                                                
\noindent We can follow the same heuristic line of 
reasoning as for the free case
to compute the filling fractions at which the interacting quantum Hall
fluid behaves incompressibly. The filling factor is
$\nu=\frac{N}{M}=\frac{2N}{AB}$. Arguing as in section 6.2,
it follows from eq.(\ref{harcomm}) that the average area occupied by
a particle is strictly bounded from below by
$2\Delta A=4|\Delta (P_0xP_0)||\Delta (P_0yP_0)|\geq 4/B
\sqrt{1+\frac{4\lambda^2}{B^2}}\equiv 2\Delta A_0$. At maximum filling
of the $p$ lowest Landau levels (bands) one expects the particles to occupy
the minimum allowed area, {\it i.e.}, $N\Delta A_0=pA$ and
$\nu=p\sqrt{1+\frac{4\lambda^2}{B^2}}$, $p$ integer. As the next available
states are in the next Landau level, which are still separated on an energy
scale of the cyclotron frequency under the assumption that
the interaction energy scale is much less than the cyclotron frequency,
one expects that the quantum fluid will be incompressible at these values
of the filling. Note that these filling fractions are larger than
the non-interacting values. This is easily understood from the attractive
nature of the interaction which, effectively, enhances the magnetic field.
\subsection{Inverse square potential}
\label{inverse}
                                                                                
Here we take $V(|\vec{r}|)=\frac{2\lambda^2}{r^2}$. This Hamiltonian is
very similar in structure to the Hamiltonian of a charged particle moving
in a plane and coupled to the gauge potential
$A_{i}=-\frac{\alpha}{r^2}\epsilon_{ij}x^j$ corresponding to a singular
flux tube located at the origin, augmented by a harmonic potential.
We investigate this case in detail in the next section as it is of
particular importance in quantum Hall systems. Taking a cue from
the wave function of this Hamiltonian \cite{ouvry3},
the lowest energy wave functions ($n=0, \ell\geq0$) are obtained by
making the following ansatz:
\begin{eqnarray}
\psi_{n=0, \ell}(r, \phi)= N_\ell r^{\Lambda(\ell)}e^{i\ell\phi}
\exp\left(-\frac{\omega_{c}}{4}r^2\right)
\label{19proj}
\end{eqnarray}
where, $\Lambda(\ell)$ is some unknown quantity which will get fixed
by eq.(\ref{radial}). The solution for $\Lambda(\ell)$, the exact low energy
eigenvalues $E_{n=0,\ell}$ and the normalisation constant $N(\ell)$ are
given by:
\begin{eqnarray}
\Lambda(\ell) = \left(\ell^2+2\lambda^2\right)^{1/2}\;
\label{20proj}
\end{eqnarray}
\begin{eqnarray}
E_{n=0, \ell}=\left[\left(\ell^2+2\lambda^2\right)^{1/2}-\ell+1\right]
\omega_c\;
\label{21proj}
\end{eqnarray}
\begin{eqnarray}
N_\ell=\left[\frac{\omega_{c}^{\Lambda(\ell) +1}}{\pi 2^{\Lambda(\ell)+1}
\Gamma(\Lambda(\ell)+1)}\right]^{1/2}\;.
\label{21aproj}
\end{eqnarray}
                                                          
\begin{figure}[t]
\begin{center}
\epsfig{file=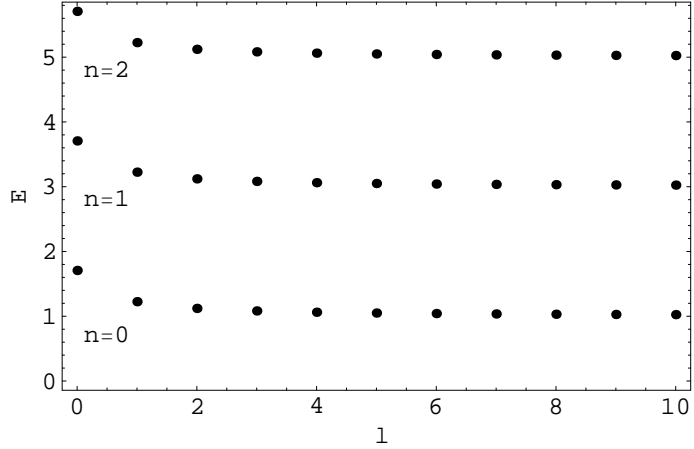,height=6.0cm,clip=,angle=0}
        \caption{The spectrum for the inverse square potential
with $\omega_c=1$ and $\lambda=0.5$.}
        \label{Fig2}
\end{center}
\end{figure}

\noindent In this case the eigenvalues and eigenfunctions for higher Landau
levels can also be solved as in \cite{ouvry3}. The full spectrum is
given by $E_{n, \ell}=\left[2n+\left(\ell^2+2\lambda^2 \right)^{1/2}-
\ell+1\right]\omega_c$, $n=0,1,2\ldots$, $\ell\geq 0$ and is shown in
Fig.\ref{Fig2} for $\omega_c=1$ and $\lambda=0.5$. The expected features
for a short range repulsive interaction can clearly be seen from this graph.

\noindent The commutator of the relative coordinates can now be evaluated
from eq.(\ref{commrel}) and yields:
\begin{eqnarray}
\left[P_{0} xP_{0}, P_{0} yP_{0}\right]
=\frac{2i}{B}\sum_{\ell=0}^{\infty}F(\ell)^2
\left[|0, \ell+1\rangle\langle0, \ell+1|-
|0, \ell\rangle\langle0, \ell|\right]\;
\label{23proj}
\end{eqnarray}
where, $F(\ell)$ is given by
\begin{eqnarray}
F(\ell)= \frac{\Gamma\left(\frac{\Lambda(\ell)+\Lambda(\ell+1)+3}{2}\right)}
{[\Gamma(\Lambda(\ell)+1)\Gamma(\Lambda(\ell+1)+1)]^{1/2}}\;.
\label{24proj}
\end{eqnarray}
Note that in this case the right hand side of eq.(\ref{23proj}) 
is not proportional
to the projection operator $P_{0}$ and, as pointed out earlier, one
should interpret this as an effective noncommutative theory with
an $\ell$-dependent renormalized noncommutative parameter.
                                                                                
\noindent Note that contrary to what one might 
naively expect, the lower bound of
the area of the particle in angular momentum sector $l$, given in terms
of the quantities $|F(\ell-1)^2-F(\ell)^2|$, are not monotonically
increasing functions of $\ell$. To understand this, one must note that
this lower bound is only achieved for minimum uncertainty states.
The actual area is to be computed from $\langle r^2\rangle$ in the
appropriate eigenstate, which is indeed a monotonically
increasing function of $\ell$. One therefore concludes that
the corresponding expression, evaluated at $\ell=0$ gives an absolute
lower bound. Arguing as before, it follows from eq.(\ref{23proj}) that the
average area occupied by a particle is strictly bounded from below
by $2\Delta A=4|\Delta (P_0xP_0)||\Delta (P_0yP_0)|\geq \frac{4F(0)^2}{B}
\equiv 2\Delta A_0$. As before the filling fractions at which the fluid
is incompressible are $\nu=\frac{p}{F(0)^2}$, $p$ integer.
As $F(0)^2\geq 1$ this yields a fractional filling factor.
                                                                                
\subsection{Singular magnetic fields}
\label{anyon}
In this section we consider the relative motion of the two particles
without any interaction, but with a singular flux tube located at
the position of the particles. To obtain the appropriate
Hamiltonian \cite{ouvry3}, we perform a singular gauge transformation
in the relative coordinate on the Hamiltonian (\ref{ham2}). To be precise
we perform the gauge transformation $e^{i\alpha\phi} H e^{-i\alpha\phi}$
with $\phi=\tan^{-1}\left(\frac{y}{x}\right)$, with $y$ and $x$
the components of the relative coordinates. Dropping the center of mass
part of (\ref{ham2}), which is not affected by the gauge transformation,
the gauge transformed Hamiltonian for the relative coordinate,
which corresponds to a singular flux tube inserted at the position of
the particles, reads \cite{ouvry3}
\begin{equation}
\label{anyonham}
H =\left(\vec{p}-\frac{1}{2}\vec{A}\right)^2\,
\end{equation}
where,
\begin{equation}
A_{i}=-(\frac{B}{2}+\frac{2\alpha}{r^2})\epsilon_{ij}x^{j},\; B\geq 0\;.
\end{equation}
Written out explicitly the Hamiltonian reads:
\begin{equation}
H=-\frac{\partial^2}{\partial r^2}-\frac{1}{r}\frac{\partial}{\partial r}+
\frac{1}{r^2}\left(i\frac{\partial}{\partial \phi}+\alpha\right)^2+
i\omega_{c}\frac{\partial}{\partial\phi}+\frac{1}{4}\omega_{c}^{2}r^2+
\alpha\omega_c\;.
\end{equation}
As in the composite fermion paradigm, we choose $\alpha\leq 0$ so that
it leads to an effective reduction of the magnetic flux seen by
the particles. The low energy eigenfunctions and spectrum are easily
found to be \cite{ouvry3}:
\begin{eqnarray}
\psi_{0,\ell}&=&N_\ell r^{|\ell-\alpha|}e^{i\ell\phi}
e^{-\frac{\omega_c r^2}{4}},\;\ell\geq 0\,\nonumber\\
N_\ell&=&\left[\frac{\omega_c^{|\ell-\alpha|+1}}
{\pi 2^{|\ell-\alpha|+1}\Gamma\left(|\ell-\alpha|+1\right)}\right]^{1/2}\,\\
E_{0,\ell}&=&\omega_c\,.\nonumber
\end{eqnarray}
                                                                                
\noindent The commutator of the relative coordinates 
yields from eq.(\ref{commrel})
\begin{eqnarray}
\label{sing}
\left[P_{0} xP_{0}, P_{0} yP_{0}\right]
&=&\frac{2i}{B}\sum_{\ell=0}^{\infty}\left(l+|\alpha|+1\right)
\left[|0, \ell+1\rangle\langle0, \ell+1|-|0, \ell\rangle\langle0, \ell|\right]
\nonumber\\
&=&\frac{2}{iB}\sum_{\ell=0}^{\infty}\left(|\alpha|\delta_{0,\ell}+1\right)
|0, \ell\rangle\langle 0, \ell|.
\end{eqnarray}
                                                                                
\noindent As before, it follows from eq.(\ref{sing}) that the 
average area occupied
by a particle is strictly bounded from below by $2\Delta A=
4|\Delta (P_0xP_0)||\Delta (P_0yP_0)|\geq \frac{4(1+|\alpha|)}{B}\equiv
2\Delta A_0$ and the filling fractions at which the fluid is incompressible
are $\nu=\frac{p}{1+|\alpha|}$, $p$ integer. Keeping in mind the phase
factor associated with the singular gauge transformation, unchanged
statistics requires, as usual, that one must choose $\alpha=2k$ with
$k$ a negative integer. This choice indeed yields the 
fractional fillings as obtained
from the composite fermion picture \cite{jain} when appropriate
choices of $p$ and $k$ are made.

\section{Summary}
\label{qhesummary}                                                                                
In this chapter, we have discussed the role that 
interactions play in the noncommutative
structure that arises when the relative coordinates of two interacting
particles are projected onto the lowest Landau level. 
The fact that the interactions in general renormalize 
the noncommutative parameter
away from the non-interacting value $\frac{1}{B}$ is transparent
from our analysis. The effective
noncommutative parameter also depends on the angular momentum in general,
as was also found from other considerations in the previous
chapter\cite{fgsbcsgagh}.
An heuristic argument, based on the noncommutative coordinates, was
given to find the filling fractions at incompressibility and the results
are consistent with known results in the case of singular magnetic fields.
It should be kept in mind however that this argument was very simplistic
as all possible many-body correlations were ignored. Probably due to
this oversimplification, this argument cannot explain, for a general
short range repulsive interaction, the quantized values of the filling
fraction at incompressibility observed in the fractional quantum Hall
effect. Indeed, from naive perturbative considerations in the above
setting one would expect that the (screened) Coulomb interaction will
have only a perturbative effect on the noncommutative parameter and
filling fraction, which is certainly not the case. As in other treatments,
it is only when one already assumes the existence of composite fermions,
as was done in section \ref{anyon}, that the quantized filling fraction
can be explained. The apparently non-perturbative microscopic origin
of composite fermions as effective 
non-interacting degrees of freedom to describe
the Coulomb interacting quantum Hall fluid is indeed still an illusive
and controversial issue \cite{dyakonov}.

\chapter{Twisted Non-Relativistic Quantum Field Theory}
In this chapter, let us take up the issue of Galilean symmetry
again in the light of an observation made in the literature recently
that the Lorentz symmetry in noncommutative quantum field theory
can be restored under a twisted 
implementation of the 
Lorentz group\footnote{So far we have investigated the untwisted 
formulation of noncommutative 
quantum field theory in the earlier chapters.}
\cite{chaichian, masud, rjszabo1, zahn1, zahn2}. The twist approach
was proposed as a way to circumvent the breaking 
of Lorentz invariance that follows from the 
choice of a particular noncommutative matrix $\theta$.
It has been shown that by invoking the concept of twisted Poincar\' e
symmetry of the algebra of functions on a Minkowski spacetime, 
the noncommutative spacetime with the
commutation relations (\ref{commrelation}), with
$\theta_{\mu\nu}$ being a {\it constant} real antisymmetric matrix,
can be interpreted in a Lorentz-invariant way. This is interesting
because unlike the earlier scheme studied in chapter 4, this does
not make use of SW map, apart from undoing the star product and
then terminating the series upto certain order in $\theta$. This
is not a very satisfactory standpoint, as, apart from the approximation
involved, SW map is not spectrum preserving in an interacting theory
as we have seen in chapter 5. For simplicity however, we shall discuss
a free theory only in this chapter and see that it can have certain
non-trivial consequences in the form of violation of Pauli principle.

\noindent To discuss the basic set up, note that the Poincar\'{e} group 
$\mathcal{P}$ or the diffeomorphism group $\mathcal{D}$ which 
acts on the noncommutative spacetime ${\mathcal{R}}^{d+1}$ defines a 
natural action on smooth functions 
$\alpha \in C^{\infty}({\mathcal{R}}^{d+1})$ as
\begin{equation}
\label{gaction}
( g\alpha) (x) = \alpha (g^{-1} x)
\end{equation}
for $g \in \mathcal{P}$ or $\in \mathcal{D}$.
However, in general 
\begin{equation}
(g \alpha) \ast_\theta (g \beta) \neq g (\alpha \ast_\theta \beta)
\label{gastarb} 
\end{equation}
showing that the action of the group 
$\mathcal{P}$ or $\mathcal{D}$ is not
an automorphism of the algebra 
${\cal A}_\theta({\mathcal{R}}^{d+1})$, unless
one considers the translational sub-group.
This violation of Poincar\'e symmetry in particular is accompanied
by the violation of microcausality, 
spin statistics and CPT theorem
in general \cite{szaborj, greenberg}. These results, which 
follow from the basic axioms
in the canonical (commutative) quantum field theory, are no longer
satisfied in presence of noncommutativity in the manner discussed
above. Besides, noncommutative field theories are afflicted with 
infra-red/ultra-violet (IR/UV) mixing. It is however possible for  some of 
these results to still go through
even after postulating weaker versions of the axioms used in 
standard quantum field theory. For example one can consider the proof
of CPT theorem given by Alvarez-Gaume et.al \cite{alvarez} where
they consider the breaking of Lorentz symmetry down to the subgroup
$O(1,1) \times SO(2)$, and replace the usual causal structure,
given by the light cone, by the light-wedge associated with the
$O(1,1)$ factor of kinematical symmetry group. One can also consider
the derivation of CPT and Spin-Statistics theorems by Franco et.al
\cite{franco} where they invoke only ``asymptotic commutativity''
i.e. assuming that the fields to be commuting at sufficiently large
spatial separations.

\noindent All the above problems 
basically stemmed from the non-invariance (\ref{gastarb}), 
and therefore it is desirable
to look for some way to restore the invariance. Indeed, 
the invariance
can be restored by introducing a deformed coproduct, 
thereby modifying the corresponding Hopf algebra \cite{chaichian, 
aschieri, Dimitrijevic:2004rf} (see
also the prior work of \cite{Oeckl:2000eg}). 
Since then,  this deformed or twisted coproduct has been used 
extensively in the framework of relativistic quantum field theory,
as this approach seems to be  quite promising.

\noindent The twisted 
implementation of the Poincar\'{e} group leads to 
two interesting consequences. 
The first is that there is apparently no longer any IR/UV mixing 
\cite{bal1}, which indicates that the high and low energy 
sectors decouple, in contrast to 
the untwisted formulation.  The second striking consequence
is an apparent violation of Pauli's principle \cite{bal}. 
This seems to be unavoidable if one wants to restore 
Poincar\'{e} invariance through the twisted coproduct. 
If there is no IR/UV mixing, 
one would expect that any violation of Pauli's 
principle would impact in either the high 
or low energy sector. Experimental observation 
at present energies seems to rule out any 
effect at low energies, therefore if this picture is a 
true description of nature, we expect 
that any violation of the Pauli principle can only appear at 
high energies. It does, 
therefore, seem worthwhile as a consistency check 
to investigate this question in more 
detail and to establish precisely what the possible 
impact it may have at low energies and why 
it may not be observable. 

\noindent One of the quantities where 
spin statistics manifests itself very 
explicitly is the two particle correlation function.  
A way of addressing this issue would 
therefore be to study the low temperature limit of the 
two particle correlation function in 
a twisted implementation of the Poincar\'{e} group. Since we 
are at low energies it would, however,
be sufficient to study the NR limit, i.e. 
the Galilean symmetry. 
The other motivation for studying the Galilean symmetry
is that the second and first quantized formulation in the 
NR set up is completely equivalent enabling us to extract the
probabilistic interpretation quite easily. This is necessary to
relate the above mentioned two particle correlation function to
joint probability. We therefore need
to consider the question whether the Galilean symmetry can 
also be restored by a suitable twist 
of the coproduct. This is a non-trivial point 
and should be looked at carefully, 
as the Galilean algebra admits a central extension, 
in the form of mass,
unlike the Poincar\'{e} case and the 
boost generator does not have the form of a vector field in spacetime.
It may be recalled, in this context, 
that the presence of spacetime
noncommutativity spoils the noncommutative structure under Galileo
boost. This question is  all the more important because of  
the observation made by \cite{doplicher} that 
the presence of spacetime noncommutativity  
does not spoil the unitarity
of the noncommutative theory.
However, we have shown that the presence of spacetime noncommutativity
in the relativistic case does not have a well defined NR
($c \to \infty$) limit. Furthermore, spacetime noncommutativity
gives rise to certain operator ordering ambiguities rendering
the extraction of a NR field in the 
$c \to \infty$ limit non-trivial.

\noindent This chapter is organised as follows. The mathematical
preliminaries are discussed in section 7.1 where we 
introduce the concept of Hopf algebra 
and the deformed or twisted coproduct.
Sub-section (7.2.1) of section 7.2 deals with a 
brief review of the twisted Lorentz
transformation properties of quantum spacetime, 
as was discussed by \cite{chaichian, wess}. This is then extended
to the NR case in sub-section (7.2.2).
In section 7.3, we then discuss briefly the NR reduction
of the Klein-Gordon field to the Schr\"{o}dinger field in
(2 + 1) dimensions in commutative space,
which is then used to obtain the action of the twisted 
Galilean transformation on the Fourier coefficients in section 7.4.
We eventually obtain the action of twisted 
Galilean transformation on NR Schr\"{o}dinger fields
in section 7.5. In section 7.6, we discuss the 
implications of the subsequent deformed 
commutation relations on the two particle correlation 
function of a free gas in two spatial dimensions. 
We conclude in section 7.7. Finally, we have added
an Appendix where we have included some important aspects
of Wigner-In$\ddot{o}$nu group contraction in this context (i.e. 
Poincar\'{e} $\to$ Galileo), which we have made use of in the main text.

\section{Mathematical preliminaries}
In this section we give a brief review of the essential 
results in \cite{bal} for the purpose of application in later sections.

\noindent
Consider a group $G$ that acts on a complex vector space $V$ by a
representation $\rho$. This action is denoted by
\begin{equation}
v \rightarrow \rho(g) v 
\label{rhov}
\end{equation}
for $g \in G$ and $v \in V$. Then the group algebra $G^*$ also acts on
$V$.  A typical element of $G^*$ is
\begin{equation}
\int dg \,\alpha(g)\, g \quad,\quad \alpha(g)\in\mathcal{C}  
\label{llll}
\end{equation}
where $dg$ is an invariant measure on $G$. Its action is
\begin{equation}
v \rightarrow \int dg \,\alpha(g) \, \rho (g) \, v~.
\label{actgstar}
\end{equation}
Both $G$ and $G^*$ act on $V \otimes V$, the tensor product of
$V$'s, as well. These actions are usually taken to be
\begin{equation}
v_1 \otimes v_2 \rightarrow \left[ \rho(g) \otimes \rho(g) \right]
(v_1 \otimes v_2 ) = \rho(g) v_1 \otimes \rho(g) v_2 
\label{acttens}
\end{equation}
and
\begin{equation}
v_1 \otimes v_2 \rightarrow
\int dg \, \alpha(g) \, \rho(g) v_1 \otimes \rho(g) v_2 
\label{acens}
\end{equation}
respectively, for $v_1, v_2 \in V$.
\noindent
In Hopf algebra theory \cite{majid, chai}, the action of 
$G$ and $G^*$ on tensor products
is defined by the coproduct $\Delta_{0}$, a homomorphism from $G^*$
to $G^* \otimes G^*$, which on restriction to $G$ gives a homomorphism
from $G$ to $G^* \otimes G^*$. 
This restriction specifies $\Delta_{0}$ on
all of $G^*$ by linearity. Hence, if
\begin{eqnarray}  
&& \Delta_{0}: \, g \rightarrow \Delta_{0}(g)  \\ \nonumber
&& \Delta_{0}(g_1)\Delta_{0}(g_2)=\Delta_{0}(g_1g_2)  
\label{xev}
\end{eqnarray} 
we have 
\begin{eqnarray}  
\Delta_{0}\left(\int dg~\alpha(g)g\right)=\int~dg~\alpha(g)\Delta_{0}(g). 
\label{twisted1}
\end{eqnarray} 
We now make an elevation.
Suppose that $V$ is an algebra ${\cal A}$.
As ${\cal A}$ is an algebra, we have a rule for taking products of
elements of ${\cal A}$, which means that there exists a multiplication map
\begin{eqnarray}
m: {\cal A} \otimes {\cal A} \rightarrow {\cal A}  \\
\alpha \otimes \beta
\rightarrow m (\alpha \otimes \beta)  \nonumber 
\label{multmap}
\end{eqnarray}
for $\alpha,\beta \in {\cal A}$, the product $\alpha \beta$ being $m
(\alpha \otimes \beta)$.

\noindent The compatibility of $\Delta_{0}$ 
with $m$ is now essential, so that: 
\begin{equation}
m\left( (\rho \otimes \rho) \Delta_{0}(g) \left(\alpha \otimes \beta
\right)\right)=
\rho(g) m ( \alpha \otimes \beta)~.
\label{compatib}
\end{equation}
In the Moyal plane, the multiplication denoted by the
map $m_{\theta}$ is noncommutative and depends on
$\theta^{\mu \nu}$. It is defined by\footnote{The signature
we are using is $(+ , - , - , ...)$.}
\begin{eqnarray}
m_\theta ( \alpha \otimes \beta) &=& m_0 
\left( e^{-\frac{i}{2} (i
\partial_\mu) \theta^{\mu \nu} \otimes (i \partial_\nu)} \alpha
\otimes \beta \right)\nonumber \\
&=& m_0 \left( F_\theta \alpha \otimes \beta
\right)\,
\label{multmoyal}
\end{eqnarray}
where, $m_{0}$ is the usual point-wise multiplication of two functions.
Note that here we have introduced a new twist element
$F_\theta$ given by
\begin{eqnarray}
F_\theta &=& e^{-\frac{i}{2}\theta^{\mu \nu} P_{\mu} \otimes P_{\nu}}
\nonumber \\
&=& e^{-\frac{i}{2} (i \partial_\mu) \theta^{\mu \nu}
\otimes (i \partial_\nu) } \quad ;\quad P_{\mu} = i\partial_\mu.
\label{ftheta}
\end{eqnarray}
The twist element $F_\theta$ changes the coproduct to
\begin{eqnarray}
\Delta_{0} (g) \rightarrow \Delta_\theta (g) = 
\hat{F}^{-1}_\theta \Delta_{0} (g) \hat{F}_\theta
\label{newcoprod}
\end{eqnarray}
in order to maintain compatibility with $m_{\theta}$, as
can be easily checked.
\noindent
In the case of the Poincar\'{e} group, if $\exp( i P {\cdot} a) $ is a
translation, we have:
\begin{eqnarray}
(\rho \otimes \rho)\Delta_\theta \left( e^{i P {\cdot} a} \right) 
e_p \otimes e_q 
&=&(\rho \otimes \rho)\left[\hat{F}^{-1}_\theta(e^{i P {\cdot} a}\otimes 
e^{i P {\cdot} a})\hat{F}_\theta\right]\nonumber\\
&=& e^{i (p+q) {\cdot} a}  e_p \otimes
e_q  \quad;\quad (e_p(x) = e^{-ip\cdot x})  
\label{planetrans}
\end{eqnarray}
while if $\Lambda$ is a Lorentz transformation
\begin{equation}
(\rho \otimes \rho)\Delta_\theta(\Lambda) e_p \otimes e_q
= \left[e^{\frac{i}{2}(\Lambda p)_\mu \theta^{\mu \nu}
    (\Lambda q)_\nu } e^{-\frac{i}{2} p_\mu \theta^{\mu \nu}
    q_\nu } \right] e_{\Lambda p} \otimes e_{\Lambda q}~.  
\label{lorentz}
\end{equation}
These relations are derived in \cite{bal}.
Finally, we mention the action of the coproduct $\Delta_{0}$ on 
the elements of a Lie-algebra $\cal{A}$ . 
The coproduct is defined on $\cal{A}$ by
\begin{equation}
\Delta_{0}(X) = X \otimes 1 + 1 \otimes X.
\label{cop}
\end{equation}
Its action on the elements of the corresponding 
universal covering algebra $\cal{U}(\cal{P})$
can be calculated through the homomorphism \cite{varilly} :
\begin{equation}
\Delta_{0}(XY) = \Delta_{0}(X) \Delta_{0}(Y) = XY \otimes 1 + X \otimes Y 
+ Y \otimes X +  1 \otimes XY.
\label{co}
\end{equation}
One can also easily check that this action of the coproduct on
the Lie-algebra is consistent with the action on the group element
defined by
\begin{equation}
\Delta_{0}(g) = g \otimes g.
\label{copg}
\end{equation}

\section{Transformation properties of tensors under 
spacetime transformation}
\subsection{Lorentz transformation}
\label{p-mcs}
In this sub-section, we give a brief review of the
Lorentz transformation properties in the commutative case 
to set the scene for the rest of the chapter.
This turns out to be essential in understanding
the action of the Lorentz generators on any vector or tensor field.

\noindent Let us consider an infinitesimal Lorentz transformation
\begin{eqnarray}
x^{\mu}\rightarrow x^{\prime\mu}=x^{\mu}+\omega^{\mu\nu}x_{\nu}
\label{1}
\end{eqnarray}
where, $\omega^{\mu\nu}$ is an infinitesimal constant 
($\omega^{\mu\nu}=-\omega^{\nu\mu}$).
Any vector field $A_{\mu}$ under this transformation transforms as
\begin{eqnarray}
A_{\mu}\rightarrow A_{\mu}^{\prime}(x^\prime)=A_{\mu}(x)
+{\omega_{\mu}}^{\lambda}A_{\lambda}(x)~.
\label{2}
\end{eqnarray}
The functional change in $A_{\mu}(x)$ therefore reads
\begin{eqnarray}
\delta_{0}A_{\mu}(x)&=&A_{\mu}^{\prime}(x)-A_{\mu}(x)\nonumber\\
&=&\omega^{\nu\lambda}x_{\nu}\partial_{\lambda}A_{\mu}(x)+
\omega_{\mu\nu}A^{\nu}
\nonumber\\
&=& - \frac{i}{2}\omega^{\nu\lambda}J_{\nu\lambda}A_{\mu}
\label{3}
\end{eqnarray}
where, $J_{\nu\lambda}=  M_{\nu\lambda}+S_{\nu\lambda}$ are the total
Lorentz generators with $M_{\mu\nu}$ and  $S_{\mu\nu}$ 
identified with orbital and spin parts, respectively. This immediately
leads to the representation of $M_{\nu\lambda}$
\begin{eqnarray}
M_{\nu\lambda}= i(x_{\nu}\partial_{\lambda}-x_{\lambda}\partial_{\nu})
=  \left(x_{\nu}P_{\lambda}-x_{\lambda}P_{\nu}\right); 
\quad P_{\lambda} = i\partial_{\lambda}\,.
\label{4}
\end{eqnarray}
The representation of $S_{\nu\lambda}$ can be found by making 
use of the relation
$\frac{i}{2}\omega^{\rho\lambda}(S_{\rho\lambda}A)_{\mu} 
= \omega_{\mu\nu}A^{\nu}$
obtained by comparing both sides of eq.(\ref{3}). This leads to
\begin{eqnarray}
(S_{\alpha\beta})_{\mu\nu} = i(\eta_{\mu\alpha}\eta_{\nu\beta}-
\eta_{\mu\beta}\eta_{\nu\alpha})\,.
\label{5}
\end{eqnarray}
It can now be easily checked that $M_{\mu\nu},\ S_{\mu\nu}$ and
$J_{\mu\nu}$ all satisfy the same homogeneous Lorentz algebra
$SO(1 , 3)$:
\begin{eqnarray}
\left[M_{\mu\nu} , M_{\lambda\rho}\right] = i\left(
\eta_{\mu\lambda}M_{\nu\rho} - \eta_{\mu\rho}M_{\nu\lambda}
- \eta_{\nu\lambda}M_{\mu\rho} + \eta_{\nu\rho}M_{\mu\lambda}\right).
\label{5a}
\end{eqnarray}

\noindent
Setting $A_{\mu} = x_{\mu}$, where $x_{\mu}$ represents a
position coordinate of a spacetime point, yields
\begin{eqnarray}
\delta_{0}x_{\mu} = -\frac{i}{2}w^{\nu\lambda}(M_{\nu\lambda}
+S_{\nu\lambda})x_{\mu}=0
\label{6}
\end{eqnarray}
as expected, since the Lie derivative of the ``radial'' vector field
$\vec{X} = x^{\mu}\partial_{\mu}$ w.r.t. the ``rotation''
generators (\ref{4}) $M_{\mu\nu}$ vanishes i.e. 
${\cal{L}}_{M_{\mu\nu}}\vec{X} = 0 $.

\noindent 
Now we observe that the change in $x_{\mu}$ (not the functional change
$\delta_{0}x_{\mu}$ as in eq.(\ref{3}))
defined by
\begin{eqnarray}
\delta x_{\mu} = x^{\prime}_{\mu}-x_{\mu} = {\omega_{\mu}}^{\nu}x_{\nu}
\label{6a}
\end{eqnarray}
can be identified as the action of $S_{\nu\lambda}$
on $x_{\mu}$
\begin{eqnarray}
\delta x_{\mu} = {\omega_{\mu}}^{\nu}x_{\nu}
= - \frac{i}{2}\omega^{\nu\lambda}\left(S_{\nu\lambda}x\right)_{\mu}
\label{7q}
\end{eqnarray}
with the representation of $S_{\nu\lambda}$ given in eq.(\ref{5}).
Using eq.(\ref{6}), one can also obtain the action
of $M_{\nu\lambda}$ on $x_{\mu}$\footnote{Note 
that $\delta A_{\mu}=A_{\mu}^{\prime}(x^\prime)-A_{\mu}(x)
={\omega_{\mu}}^{\lambda}A_{\lambda}(x)$ is not the functional change
and $\delta x_{\mu}$ in eq.(\ref{7}) is
obtained by setting $A_{\mu}=x_{\mu}$.}
\begin{eqnarray}
\delta x_{\mu} = -\frac{i}{2}\omega^{\nu\lambda}M_{\nu\lambda}x_{\mu}.
\label{7}
\end{eqnarray}
The generalization of this to higher second rank tensors 
$f_{\rho\sigma}(x)=x_{\rho}x_{\sigma}$ is straightforward as
\begin{eqnarray}
\delta \left(x_{\lambda} x_{\sigma}\right) =
\left(- \frac{i}{2} w^{\mu\nu} M_{\mu\nu}\right) 
\left(x_{\lambda} x_{\sigma}\right)
\label{13}
\end{eqnarray}
since we can write
\begin{eqnarray}
M_{\mu\nu}f_{\rho\sigma} &=& i(x_{\mu}
\partial_{\nu}-x_{\nu}\partial_{\mu})
f_{\rho\sigma}\nonumber\\
&=&i(f_{\mu\sigma}\eta_{\nu\rho}-f_{\nu\sigma}\eta_{\mu\rho}+
f_{\rho\nu}\eta_{\mu\sigma}-f_{\rho\mu}\eta_{\nu\sigma})
\label{1b}
\end{eqnarray}
where we have made use of eq.(\ref{4}). This indeed shows the 
covariant nature of the transformation properties
of $f_{\rho\sigma}$.

\noindent
We now review the corresponding covariance property in the 
noncommutative case under the twisted coproduct of Lorentz generators
\cite{chaichian}, \cite{wess}.
The issue of violation of Lorentz symmetry in noncommutative
quantum field theories has been known for a long time,
since field theories defined on a noncommutative
spacetime obeying the commutation
relation (\ref{commrelation}) between the coordinate operators,
where $\theta_{\mu\nu}$ is treated as a 
constant antisymmetric matrix, are obviously
not Lorentz invariant. However, a new kind of symmetry known as 
twisted Poincar\'{e} symmetry has been found in \cite{chaichian} 
under which quantum field theories defined on noncommutative
spacetime are still Poincar\'{e}  invariant.

\noindent
To generalise to the noncommutative case, first note that 
the star product between two vectors
$x_{\mu}$ and $x_{\nu}$  given as $x_{\mu} \star x_{\nu}$ is
not symmetric, unlike in the commutative case. One can,
however, write this as
\begin{eqnarray}
x_{\mu} \star x_{\nu} = x_{\{\mu} \star x_{\nu\}} + \frac{i}{2}
\theta_{\mu \nu}
\label{14}
\end{eqnarray}
where the curly brackets $\{ \}$ denotes symmetrization in the indices
$\mu$ and $\nu$. 
This can be easily generalised to higher ranks, showing that every 
tensorial object of the form $(x_{\mu} \star x_{\nu} \star .....\star
x_{\sigma})$ can be written as a sum of symmetric tensors of equal
or lower rank, so that the basis representation is symmetric.
Consequently $f_{\rho \sigma}$ should be replaced by 
the symmetrized expression
$f^{\theta}_{\rho \sigma} = x_{\{\rho} \star x_{\sigma \}} = 
\frac{1}{2} (x_{\rho} \star x_{\sigma } + x_{\sigma } \star x_{\rho})$,
and correspondingly the action of the Lorentz generator should be 
applied through the twisted coproduct (\ref{newcoprod})
\begin{eqnarray}
M^{\theta}_{\mu \nu} f^{\theta}_{\rho \sigma} &=& M^{\theta}_{\mu \nu} 
m_{\theta} \left(x_{\rho} 
\otimes x_{\sigma} \right) = m_{\theta}\left(\Delta_{\theta}\left(
M_{\mu \nu}\right)\left(x_{\rho} 
\otimes x_{\sigma} \right)\right) \nonumber \\
&=& i(f^{\theta}_{\mu\sigma}\eta_{\nu\rho} - 
f^{\theta}_{\nu\sigma}\eta_{\mu\rho} +
f^{\theta}_{\rho\nu}\eta_{\mu\sigma} - 
f^{\theta}_{\rho\mu}\eta_{\nu\sigma}).
\label{17}
\end{eqnarray}
In the above equation, $M^{\theta}_{\mu \nu}$ denotes the usual 
Lorentz generator, but with the action of a twisted coproduct.
In \cite{chaichian}, it was shown that $M^{\theta}_{\mu \nu}
(\theta^{\rho \sigma}) = 0$, and 
\begin{eqnarray}
 M^{\theta}_{\mu \nu}\left(S^{2}_{t}\right) = 0 \ ;
\quad (S^{2}_{t} = x_{\sigma} \star x_{\sigma})
\label{1744}
\end{eqnarray}
i.e. the antisymmetric tensor $\theta^{\rho \sigma}$ 
is twisted-Poincar\'{e} invariant.
\subsection{Twisted Galilean Invariance}
We extend the results of the earlier sub-section on twisted Poincar\'{e}
invariance to the corresponding NR case in this sub-section. 
To demonstrate the need for this, 
consider the Galilean boost transformation
\begin{eqnarray}
t \rightarrow t^{\prime} &=& t \nonumber \\
x^{i} \rightarrow x^{\prime i} &=&  x^{i} - v^{i}t
\label{18}
\end{eqnarray}
applied in the noncommutative Galilean spacetime having the
following noncommutative structure 
\begin{eqnarray}
\left[t , x^i\right] = i\theta^{0 i}
\ \ \ ; \ \ \  \left[ x^i , x^j\right] = i \theta^{ij}.
\label{18kj}
\end{eqnarray}
In the boosted frame, the corresponding expression is given by
\begin{eqnarray}
\left[t^{\prime} , x^{\prime i}\right] &=& \left[t , x^{i}\right]
= i \theta^{0 i}\nonumber \\
\left[x^{\prime i} , x^{\prime j}\right] &=& i \theta^{i j}
+  i  \left( \theta^{0 i} v^j - \theta^{0 j}v^i\right).
\label{18a}
\end{eqnarray}
This shows that the noncommutative structure in the primed frame
does not preserve its structure unless spacetime noncommutativity
disappears i.e. $\theta^{0 i} = 0$.
\noindent
Here we show that even in the presence of spacetime 
noncommutativity the Galilean symmetry can be restored 
through an appropriate twist.
To do this we consider a tangent vector field $\vec{A}(x) = 
A^{\mu}(x)\partial_{\mu}$, in  Galilean spacetime. 
Under Galilean transformations (\ref{18}), we have
\begin{eqnarray}
\label{19}
A^{i}(x) \rightarrow A^{\prime i}(x^{\prime}) &=&
\frac{\partial x^{\prime i}}{\partial x^{\mu}} A^{\mu}(x)
= A^{i}(x) - v^{i}A^{0}(x)  \\
A^{0}(x) \rightarrow A^{\prime 0}(x^{\prime}) &=& A^{0}(x). \nonumber
\end{eqnarray}
From eq.(\ref{19}), it follows that
\begin{eqnarray}
\delta_{0} A^{\mu}(x) &=& A^{\prime \mu}(x) - A^{\mu}(x) \nonumber \\
&=& i v^{j} \left(- i t \partial_{j} A^{\mu}(x) + 
i \delta^{\mu}_{j} A^{0}(x)\right) \nonumber \\
&=& i v^{j} K_{j} A^{\mu}(x)
\label{20}
\end{eqnarray}
where,
\begin{eqnarray}
K_{j} A^{\mu}(x) &=& \left(- i t \partial_{j} A^{\mu}(x) + 
i \delta^{\mu}_{j} A^{0}(x)\right) \nonumber \\
&=& - tP_{j} A^{\mu}(x) +  i \delta^{\mu}_{j} A^{0}(x).
\label{21}
\end{eqnarray}
Setting $A^{\mu}(x) = x^{\mu}$\footnote{Here we identify $x^0$
to be just the time $t$, rather than $ct$.} we  easily see that 
$ K_{j} x^{\mu} = 0$, from which we get
\begin{eqnarray}
\delta x^{\mu} = i v^{j} t P_{j} x^{\mu} = i v^{j} K^{(0)}_{j} x^{\mu}
\label{22}
\end{eqnarray}
where, $K^{(0)}_{j} =  t P_{j}$. This is the counterpart of eq.(\ref{7}) 
in the Galilean case. 
In other words, here $K^{(0)}_{j}$ plays the same role as
$M_{\mu \nu}$ in the relativistic case.
Indeed, it can be easily checked that at the commutative level it has 
its own coproduct action
\begin{eqnarray}
K^{(0)}_{j} m\left( x^{\mu} \otimes x^{\nu} \right) = m \left(\Delta_{0}
\left(K^{(0)}_{j} \right)\left(x^{\mu} \otimes x^{\nu} \right)\right).
\label{23}
\end{eqnarray}
Here $K^{(0)}_{j}$ is clearly the boost generator 
$K^{(M)}_{j}$ (see eq.(\ref{12c}) in Appendix) with $M = 0$. 
Note that with $M \neq 0$,
$K^{(M)}_{j}$ does not have the right coproduct action (\ref{23}). 
This is also quite satisfactory
from the point of view that the noncommutativity of spacetime
is an intrinsic property and should have no bearing on the mass of
the system inhabiting  it. We also point out another dissimilarity
between the relativistic and NR case. In the  relativistic
case, the generators $M_{\mu\nu}$ (eq.(\ref{4})) can be regarded as
the vector field whose integral curve generates the Rindler
trajectories, i.e. the spacetime trajectories of uniformly
accelerated particle. On the other hand, the vector field associated
with the parabolic trajectories of uniformly accelerated particle
in the NR case is given by $K^{NR}_{i}$ (eq.(\ref{12})),
which however cannot be identified with the Galileo boost
generator $K^{(M)}_{j}$  (eq.(\ref{12c})) (see Appendix), unlike
$M_{\mu\nu}$ in the relativistic case.

\noindent
At the noncommutative level, the 
action of the Galilean generator
should be applied through the twisted coproduct
\begin{eqnarray}
K^{\theta (0)}_{j} m_{\theta}\left( x^{\mu} \otimes x^{\nu} \right) =
 m_{\theta} \left(\Delta_{\theta}
\left(K^{(0)}_{j} \right)\left(x^{\mu} \otimes x^{\nu} \right)\right)\,.
\label{24}
\end{eqnarray}
Using this and noting $K^{(0)}_{j}=tP_{j}$, we have 
\begin{eqnarray}
\Delta_{\theta} \left(K^{(0)}_{j} \right)
&=&\Delta_{0} \left(K^{(0)}_{j} \right)
\label{25DELa}
\end{eqnarray}
which eventually leads to
\begin{eqnarray}
K^{\theta (0)}_{j} m_{\theta}\left( x^{\mu} \otimes x^{\nu} \right) &=& 
it\left(x^{\mu} \delta^{\nu}_{j} + \delta^{\mu}_{j} 
x^{{\nu}}\right) \nonumber \\
\Rightarrow K^{\theta (0)}_{j} m_{\theta}\left( x^{\mu} \otimes x^{\nu} 
- x^{\nu} \otimes x^{\mu} \right)  &=&0 \nonumber \\
\Rightarrow K^{\theta (0)}_{j} \left(\theta^{{\mu}{\nu}}\right) &=& 0
\label{25}
\end{eqnarray}
i.e. the antisymmetric tensor $\theta^{{\mu}{\nu}}$ is invariant
under twisted Galilean boost. The complete twisted
Galilean invariance of $\theta^{\mu \nu}$ is therefore established 
since the rest of the Galileo
generators have the same form as that of the Poincar\'{e} generators,
discussed in the previous sub-section. To put it more simply, 
eq.(\ref{25DELa}) clearly shows that the boost generator is taken
care of rather easily and the only non-triviality arises in the restoration
of rotational symmetry. 
\section{Non-Relativistic reduction in commutative space}
In this section, we discuss the NR 
reduction ($c \to \infty$) of the Klein-Gordon field
to the Schr\"{o}dinger field in 2+1 dimension\footnote{The
procedure of NR reduction holds for any
spacetime dimension.}, as this will be used
in the subsequent sections to derive the deformed algebra
of the Schr\"{o}dinger field both in the momentum and in 
the configuration space. The deformed algebra in the momentum
space for the Klein-Gordon field has already been derived in \cite{bal}.
Therefore it is advantageous to consider the NR limit
of such a deformed algebra.

\noindent
We reintroduce the speed of light `$c$' in appropriate places
from dimensional consideration to take the $c \to \infty$ limit 
at the end of the calculation, 
but we still work in the unit
$\hbar = 1$.
Let us consider the complex Klein-Gordon field 
satisfying the Klein-Gordon equation 
\begin{eqnarray}
\label{eq:45} 
\left(\frac{1}{c^2}\partial_{t}^2 - \nabla^2 
+ m^2c^2\right)\phi(x)=0
\end{eqnarray}
which follows from the extremum condition of the Klein-Gordon
action
\begin{eqnarray}
S = \int dt d^2{\bf{x}}\left[\frac{1}{c^2}\dot{\phi}^{\star}
\dot{\phi} - \phi^{\prime \star}\phi^{\prime} - 
c^2 m^2\phi^{\star}\phi\right]\,.
\label{26}
\end{eqnarray}

\noindent The Schr\"{o}dinger field is identified from the  Klein-Gordon
field by isolating the exponential factor involving rest mass energy
and eventually taking the limit $c \to \infty$.  

\noindent Hence, we set
\begin{equation}
\label{eq:46} 
\phi (\vec{x},t) = \frac{e^{-imc^2t}}{\sqrt{2m}}\psi
(\vec{x},t)
\end{equation}
which yields from eq.(\ref{eq:45}) the equation
\begin{equation}
\label{eq:47} 
-\frac{1}{2m}\nabla^2\psi =i \frac{\partial
\psi}{\partial t}-\frac{1}{2mc^2}\frac{\partial ^2\psi}{\partial
t^2}.
\end{equation}
This reduces to the Schr\"odinger equation of a free positive
energy particle
in the limit $c \to \infty$.
In this limit the action (\ref{26}) also yields the 
corresponding NR action  as
\begin{eqnarray}
S_{NR} = \int dt d^2x \, \psi^{\star}\left(i \partial_{0} + 
\frac{1}{2m}\nabla^2\right)\psi~.
\label{27w}
\end{eqnarray}
The complex scalar field $\phi({\bf{x}})$ 
can be Fourier expanded as
\begin{eqnarray}
\phi(\vec{x} , t) = \int d\mu(k)c\left[a(k)e_{k} +  
b^{\dagger}(k)e_{-k}\right]
\label{27}
\end{eqnarray}
where, $d\mu(k) = \frac{d^2\vec{k}}{2k_0 (2\pi)^2}$ is the
Lorentz invariant measure and $e_{k} = e^{- i k.x} =
e^{- i\left( Et - \vec{k}\cdot \vec{x}\right)}$.
The commutation
relation between $a_{k}$ and $a^{\dagger}_{k}$\footnote{Note 
that $k^{\mu} = \left(\frac{E}{c} , \vec{k}\right)$.} can be found
by using the well known equal time commutation relations between
$\phi$ and $\Pi_{\phi}$:
\begin{eqnarray}
\left[a(k) , a^{\dagger}(k^{\prime})\right] = 
(2\pi)^2 \frac{2 k_{0}}{c}\ \delta^2\left(
\vec{k} - \vec{k^{\prime}}\right)
\label{281}
\end{eqnarray}
and likewise for $b(k)$. In order to get the Fourier 
expansion of the field in the NR
case, we substitute eq.(\ref{eq:46})
in eq.(\ref{27}), which in the limit
$c\rightarrow \infty$ yields
\begin{eqnarray}
\psi(\vec{x} , t) = \int \frac{d^2{\vec{k}}}{(2\pi)^2}
\frac{\tilde{c}(k)}{\sqrt{2m}} \tilde{e}_{k} 
= \int \frac{d^2{\vec{k}}}{(2\pi)^2}
c(k)\tilde{e}_{k} 
\label{29}
\end{eqnarray}
where, $\tilde{e}_{k} = e^{- i \frac{ \vert \vec{k} \vert^2 t}{2m}} 
e^{ i\vec{k}\cdot \vec{x}}$, 
$\tilde{c}(k) = \lim_{c \to \infty} a(k)$ and 
$c(k) = \frac{1}{\sqrt{2m}}\tilde{c}(k)$ 
are the Schr\"{o}dinger modes.
As in eq.(\ref{eq:47}), only the positive energy part survives
in the $c \to \infty$ limit, so that this limit 
effectively projects  the positive frequency part.
The commutation relation (\ref{281}) reduces in the NR 
limit ($c \rightarrow \infty$) to
\begin{eqnarray}
\left[\tilde{c}(k) , \tilde{c}^{\dagger}(k^{\prime})\right] &=& 
(2\pi)^2 2m\ \delta^2\left(
\vec{k} - \vec{k^{\prime}}\right) \nonumber \\
\left[c(k) , c^{\dagger}(k^{\prime})\right]&=& 
(2\pi)^2 \ \delta^2\left(
\vec{k} -  \vec{k^{\prime}}\right)\,. 
\label{28}
\end{eqnarray}

\noindent
From eq.(s) (\ref{29}) and (\ref{28}), we obtain 
\begin{eqnarray}
\left[\psi(\vec{x} , t) \ ,\ \psi^{\dagger}(\vec{y} , t)\right] &=& 
\delta^2\left(
\vec{x} -  \vec{y}\right)\,.
\label{30}
\end{eqnarray}

\section{Action of twisted Galilean transformation on Fourier coefficients}
Let us consider the Fourier expansion of the relativistic 
scalar field $\phi(\vec{x} , t)$
\begin{eqnarray}
\phi(\vec{x} , t) &=& \int d\mu(k)c \tilde{\phi}(k)e_{k}~.
\label{31}
\end{eqnarray}
Here we have deliberately suppressed the negative
frequency part as it does not survive in the NR
limit $c \to \infty$, as we have seen in the previous section.
Considering the action of the Poincar\'{e} group elements on
$\phi$, we get
\begin{eqnarray}
\rho(\Lambda_{c}) \phi &=& \int d\mu(k)c\, \tilde{\phi}(k)
e_{\Lambda_{c} k} = \int d\mu(k)c\, \tilde{\phi}(\Lambda^{-1}_{c} k)
e_{k}  \\
\rho \left( e^{i P {\cdot} a} \right)  \phi &=&  \int d\mu(k)c\, 
e^{ik\cdot a}\tilde{\phi}(k)e_{k}~.
\label{32}
\end{eqnarray}
Thus the representation $\tilde{\rho}$ of the Poincar\'{e}
group on $\tilde{\phi}(k)$ is specified by
\begin{eqnarray}
\left(\tilde{\rho}(\Lambda_{c}) \tilde{\phi} \right) (k) &=&
\tilde{\phi}(\Lambda_{c}^{-1} k)  \nonumber \\
\left(\tilde{\rho}\left( e^{i P {\cdot} a} \right) \tilde{\phi}\right) (k)
&=& e^{i k {\cdot} a} \tilde{\phi}(k)~. 
\label{33}
\end{eqnarray}
Here homogeneous Lorentz transformations have been labeled by
$\Lambda_{c}$. The corresponding Galilean transformations
will be labeled by $\Lambda_{\infty}$ in the $c \to \infty$ limit.

\noindent
If $\chi$ is another scalar 
field, with Fourier expansion given by
\begin{eqnarray}
\chi(\vec{x} , t) = \int d\mu(q) c\, \tilde{\chi}(q)e_{q}\,
\label{34}
\end{eqnarray}
the tensor product of fields $\phi$ and $\chi$ is given by
\begin{equation}
\phi \otimes \chi = \int d\mu(k)d\mu(q)c^2\, \tilde{\phi}(k)
\tilde{\chi}(q)e_k \otimes e_q~.
\label{35} 
\end{equation}
Using eq.(\ref{lorentz}), one obtains the action of the
twisted Lorentz transformation on the above tensor
product of the fields
\begin{eqnarray}
\Delta_{\theta}(\Lambda_{c})(\phi \otimes \chi) =
\int d\mu(k)d\mu(q)c^2\, \tilde{\phi}(\Lambda^{-1}_{c}k)
\tilde{\chi}(\Lambda_{c}^{-1}q) e^{\frac{i}{2}k_{\mu}
\theta^{\mu \nu}q_{\nu}}e^{-\frac{i}{2}(\Lambda^{-1}_{c}k)_{\alpha}
\theta^{\alpha \beta}(\Lambda^{-1}_{c}q)_{\beta}}\left(e_k \otimes
e_q\right)\,.
\label{34f}
\end{eqnarray}
Substituting eq.(\ref{eq:46}) in the above equation, one can
write
the corresponding action of the twisted Lorentz transformations
on the tensor product of fields $\psi$ and $\xi$ (here $\xi$ is
the counterpart of $\psi$ for the field $\chi$ 
as in eq.(\ref{eq:46})) as
\begin{eqnarray}
\Delta_{\theta}(\Lambda_{c})\left(\psi \otimes \xi\right) 
&=& \int d\mu(k)d\mu(q)2m c^2\, 
\tilde{\phi}(\Lambda^{-1}_{c}k)
\tilde{\chi}(\Lambda^{-1}_{c}q) e^{\frac{i}{2}k_{i}
\theta^{i j}q_j}e^{-\frac{i}{2}(\Lambda^{-1}_{c}k)_{l}
\theta^{l n}(\Lambda^{-1}_{c}q)_{n}}\nonumber\\
&& \hspace {4.5cm}\times e^{-2i O(\frac{1}{c^2}, ....)}
\left(\tilde{e}_k \otimes
\tilde{e}_q\right) .
\label{36} 
\end{eqnarray}
Note that we have set $\theta^{0i} = 0$ in the right hand side
of the above equation. The underlying reason is that 
the substitution (\ref{eq:46}) can be carried out only in the
absence of spacetime noncommutativity ($\theta^{0i} = 0$)
as this removes any operator ordering ambiguities in eq.(\ref{eq:46}).
This should not, however, be regarded as a serious restriction
as theories with spacetime noncommutativity do not represent
a low energy limit of string theory \cite{greenberg, gomis, gomis1}

\noindent
Hence in the  limit $c \rightarrow \infty$,
we can deduce the action of the twisted Galilean transformations
($\Lambda_{\infty}$) on tensor products of the NR fields:
\begin{equation}
\Delta_{\theta}(\Lambda_{\infty})\left( \psi \otimes \xi\right)
 =  \int \frac{d^2 \vec{k}d^2 \vec{q}}{(2\pi)^4}\tilde{\psi}
(\Lambda^{-1}_{\infty}k)\tilde{\xi}(\Lambda^{-1}_{\infty}q)
e^{\frac{i}{2}m v_1\theta (k_2 - q_2)}\left(\tilde{e}_k
\otimes \tilde{e}_{q}\right)\,.
\label{37w} 
\end{equation}
Here we have  considered a boost along the $x^1$ direction with
velocity $v_1$ and $\tilde{\psi}(k) = \lim_{c \to \infty}\tilde{\phi}(k)$,
$\tilde{\xi}(q) = \lim_{c \to \infty}\tilde{\chi}(q) $.

\noindent
From the above, one can deduce the action of the twisted 
Galilean transformations ($\Lambda_{\infty}$) on the Fourier
coefficients of the NR fields
\begin{equation}
\Delta_{\theta}(\Lambda_{\infty})\left( \tilde{\psi} 
\otimes \tilde{\xi}\right)
\left(k , q\right) =  \tilde{\psi} 
\left(\Lambda^{-1}_{\infty}k\right)
\tilde{\xi} \left(\Lambda^{-1}_{\infty}q\right)
e^{\frac{i}{2}m v_1 \theta (k_2 - q_2)}
\label{37} 
\end{equation}
One can now easily generalise the above result for the case of any arbitary
direction of boost as
\begin{equation}
\Delta_{\theta}(\Lambda_{\infty})\left( \tilde{\psi} \otimes \tilde{\xi}\right)
\left(k , q\right) =  \tilde{\psi} \left(\Lambda^{-1}_{\infty}k\right)
\tilde{\xi} \left(\Lambda^{-1}_{\infty}q\right)
e^{\frac{i}{2}m \theta \vec{v} \times  (\vec{k} - \vec{q})}~.
\label{373} 
\end{equation}

\section{Quantum Fields}
In this section, we discuss the action of twisted Galilean 
transformation on NR Schr\"{o}dinger fields. 
A free relativistic complex quantum field $\phi$ of mass $m$ can be
expanded in the noncommutative plane (suppressing the negative 
frequency part) as
\begin{eqnarray}
\phi(\vec{x} , t) = \int d\mu(k) c\, d(k)e_k~.
\label{371}
\end{eqnarray}
This is just the counterpart of eq.(\ref{27}) where $a(k)$
has been replaced by $d(k)$\footnote{Note that 
$a(k) = \lim_{\theta \to 0}d(k)$.}. \\
The deformation algebra involving $d(k)$ has already been
derived in \cite{bal}. Here, we  derive the 
deformation algebra for the NR case.
The NR limit of the complex Klein-Gordon field
has already been discussed in the earlier section and the
expansion is the following:
\begin{eqnarray}
\psi(\vec{x} , t) = \int \frac{d^2{\vec{k}}}{(2\pi)^2}
\frac{\tilde{u}(k)}{\sqrt{2m}} \tilde{e}_k 
= \int \frac{d^2{\vec{k}}}{(2\pi)^2}
u(k)\tilde{e}_k \ \ ;\ \ u(k) = \frac{1}{\sqrt{2m}}\tilde{u}(k)
\label{38}
\end{eqnarray}
where, $\tilde{u}(k) = \lim_{c \to \infty} d(k)$. \\
\noindent
Note that 
$\tilde{c}(k), c(k)$ are the limits of the operators
$\tilde{u}(k) , u(k)$ respectively in the limit
$\theta^{\mu \nu} = 0$, and they satisfy the 
relations (\ref{28}).
We now argue that such relations are incompatible for $\theta^{\mu
\nu} \neq 0$. Rather, $u(k)$ and $u^\dagger(k)$
fulfill certain deformed relations which reduce to eq.(\ref{28})
for $\theta^{\mu \nu} = 0$. \\
\noindent
Suppose  that
\begin{equation}
u(k) u(q) = \tilde{T}_\theta(k,q)
u(q) u(k)\,  
\label{40}
\end{equation}
where, $\tilde{T}_\theta$ is a $\mathcal{C}$-valued function of $k$ and $q$
yet to be determined. The transformations of $u_{k} u_{l} = 
(u \otimes u) (k,l)$ and $u_{l} u_{k}$
are determined by $\Delta_\theta$. 
Applying $\Delta_\theta$ on eq.(\ref{40}) and using eq.(\ref{37}),
we get the following\footnote{Without loss
of generality, we consider the boost to be along the $x^1$ direction
for calculational convenience. Also we set $v_1 = v$.}:
\begin{equation}
u \left(\Lambda^{-1}_{\infty}k\right)
u \left(\Lambda^{-1}_{\infty}q\right)
e^{\frac{i}{2}m v \theta (k_2 - q_2)}
= \tilde{T}_\theta(k,q) u \left(\Lambda^{-1}_{\infty}q\right)
u \left(\Lambda^{-1}_{\infty}k\right)
e^{\frac{i}{2}m v \theta (q_2 - k_2)}.
\label{41}
\end{equation}
Using eq.(\ref{40}) again in the left hand side of eq.(\ref{41}), 
we get:
\begin{equation}
\tilde{T}_\theta \left(\Lambda^{-1}_{\infty}k , 
\Lambda^{-1}_{\infty}q\right)
= \tilde{T}_\theta(k,q) e^{-i m v \theta (k_2 - q_2)}.
\label{42}
\end{equation}
Note that this equation can also be obtained from the corresponding
relativistic result \cite{bal} in the $c \to \infty$ limit provided one
takes $\theta^{0 i} = 0$ right from the beginning, otherwise 
the exponential factor become rapidly
oscillating in the  $c \to \infty$ limit, yielding no
well defined NR limit. Thus in the absence of 
spacetime noncommutativity one has an appropriate NR
limit and  the above mentioned operator ordering ambiguities
can be avoided.\\
\noindent
The solution of eq.(\ref{42}) is\footnote{Note that the NR
form of the twist element also appears in \cite{luk}.}
\begin{eqnarray}
\tilde{T}_\theta(k,q) = \eta e^{ i k_{i} \theta^{i j}  q_{j} }\ ; 
\quad (i, j = 1, 2)\,
\label{43}
\end{eqnarray}
where $\eta$ is a Galilean-invariant function and
approaches the value $\pm 1$ for bosonic and fermionic fields
respectively in the limit $\theta = 0$\footnote{The value
of $\eta$ can be actually taken to be $\pm 1$ for 
bosonic and fermionic fields for all $\theta^{\mu \nu}$ \cite{bal}. An exactly
similar NR reduction of the Dirac equation can also be done
for the fermionic case.}.
Substitution of eq.(\ref{43}) in eq.(\ref{40}) yields
\begin{equation}
u(k) u(q) = \eta e^{ i k_{i} \theta^{i j}  q_{j} }
u(q) u(k).
\label{44}
\end{equation}
The adjoint of eq.(\ref{44}) gives:
\begin{equation}
u^\dagger(k) u^\dagger(q) = \eta e^{ i k_{i} \theta^{i j}  q_{j} }
u^\dagger(q) u^\dagger(k) .
\label{45}
\end{equation}
Finally the creation operator $u^\dagger(q)$ carries momentum $-q$, hence its
deformed relation reads:
\begin{equation}
u(k) u^{\dagger}(q) =  \eta e^{- i k_{i} \theta^{i j}  q_{j} } 
u^{\dagger}(q) u(k) + (2\pi)^2    \delta^2(k-q) .
\label{46}
\end{equation}
The above structure of algebra (\ref{44}, \ref{45}, \ref{46}) 
can be understood more easily 
by using the twisted projection operator $P_{\theta}$
\footnote{$P_{\theta}=F_{\theta}^{-1}P_{0}F_{\theta}$, where
$P_{0}$ is the usual projection operator for a two particle
system which projects onto the symmetric (anti-symmetric) sub-space
describing bosonic (fermionic) statistics.} (first 
introduced in \cite{vaid}) \cite{subrata}.

\noindent Now using eq.(s) (\ref{44}) and (\ref{46}), 
one can easily obtain the deformation 
algebra involving  the NR fields
$\psi(x)$ in the configuration space:
\begin{eqnarray}
\psi(x)\psi(y) &=& \int d^2x^{\prime}d^2y^{\prime}
\Gamma_{\theta}(x, y, x^{\prime}, y^{\prime})
\psi(y^{\prime})\psi(x^{\prime}) \quad ;\ \ \theta \neq 0 \nonumber \\
\psi(x)\psi(y)&=& \eta \psi(y)\psi(x) \quad ;\ \ \theta = 0\, 
\label{46k}
\end{eqnarray}
\begin{eqnarray}
\psi(x)\psi^{\dagger}(y) &=& \int d^2x^{\prime}d^2y^{\prime}
\Gamma_{\theta}(x, y, x^{\prime}, y^{\prime})\psi^{\dagger}(y^{\prime})
\psi(x^{\prime}) + \delta^2(\vec{x} - \vec{y})\ \ 
;\ \theta \neq 0 \nonumber \\
\psi(x)\psi^{\dagger}(y) &=& \eta \psi^{\dagger}(y)\psi(x) + 
\delta^2(\vec{x} - \vec{y})
\quad ;\ \ \theta = 0\,
\label{46v}
\end{eqnarray}
where,
\begin{eqnarray}
\Gamma_{\theta}(x, y, x^{\prime}, y^{\prime}) 
= \frac{\eta}{(2\pi)^2}exp\left(\frac{i}{\theta}
\left[(x^{\prime}_1 - x_1)(y_2 - y^{\prime}_2) - 
(x^{\prime}_2 - x_2)(y_1 - y^{\prime}_1)\right]\right).
\end{eqnarray}
Note at this stage that in momentum space, the twisted fermions
still satisfy $u(k)u(k)=0$ as follows from (\ref{44}), unlike what
happens in ordinary configuration space as $\psi(x)\psi(x)\neq0$.
This indicates that two identical twisted 
fermions cannot occupy the same slot
in momentum space as happens for ordinary fermions, but can
occupy the same position in configuration space for $\theta\neq0$ and
can therefore give rise to violation of Pauli's exclusion principle.
We take up this issue in the next section.
\section{Two particle correlation function}
In this section, the computation of the two particle correlation 
function $\frac{1}{Z}\langle r_1,r_2|e^{-\beta H}|
r_1,r_2\rangle$ for a free gas in 2+1 dimensions using the 
canonical ensemble is performed, 
where $Z$ is the canonical partition function
and $H$ is the NR Hamiltonian.  
This function tells us what the probability is to find particle two 
at position $r_2$, given that particle one is at $r_1$, i.e. 
it measures two particle correlations.  The relevant two 
particle state is given by
\begin{eqnarray}
|r_{1}, r_{2}\rangle&=&
\hat\psi^{\dag}(r_{1})\hat\psi^{\dag}(r_{2})|0\rangle\nonumber\\
&=&\int \frac{dq_{1}}{(2\pi)^2}\frac{dq_{2}}{(2\pi)^2}
e^{*}_{q_{1}}(r_{1})e^{*}_{q_{2}}(r_{2})u^{\dag}(q_{1})u^{\dag}(q_{2})
|0\rangle\,.
\label{rh1}
\end{eqnarray}
The two particle correlation function can therefore be written as
\begin{eqnarray}
\langle r_{1}, r_{2}|e^{-\beta H}|r_{1}, r_{2}\rangle
&=&\int dk_{1}dk_{2}e^{-\frac{\beta}{2m}(k^{2}_{1}+k^{2}_{2})}
|\langle r_{1}, r_{2}|k_{1}, k_{2}\rangle|^{2}
\label{rh2}
\end{eqnarray}
where we have introduced a complete set of momentum 
eigenstates $|k_{1}, k_{2}\rangle$.

\noindent Using eq.(\ref{46}) and noting that
\begin{eqnarray}
|k_{1}, k_{2}\rangle=u^{\dag}(k_1)u^{\dag}(k_2)|0\rangle
\label{rh3}
\end{eqnarray}
we finally obtain
\begin{eqnarray}
C(r)\equiv \frac{1}{Z}\langle r_{1}, r_{2}|e^{-\beta H}|r_{1}, r_{2}\rangle
&=&\frac{1}{A^2}\left(1\pm\frac{1}{1+\frac{\theta^2}{\lambda^4}}
e^{- 2 \pi\, r^2/(\lambda^{2}(1+\frac{\theta^2}{\lambda^4}))}\right)
\label{rh4}
\end{eqnarray}
where, $A$ is the area of the system and $\lambda$ is the mean 
thermal wavelength given by
\begin{eqnarray}
\lambda&=&\left(\frac{2\pi\beta}{m}\right)^{1/2}\quad;\quad
\beta=\frac{1}{k_{B}T}
\label{rh5}
\end{eqnarray}
and $r=r_{1}-r_{2}$. The plus and the minus signs indicate
bosons or fermions.

\noindent Although this calculation was done in 2+1 dimensions, it is 
clear that the result generalizes to higher dimensions by 
replacing $\theta^2$ by an appropriate sum of $(\theta^{ij})^2$. 
The conclusions made below, based on the general structure 
of the correlation function, will therefore also 
hold in higher dimensions. 

\noindent Expectedly, this result reduces to the standard (untwisted)
result in the limit 
$\theta\rightarrow0$ \cite{pathria}.  
Furthermore it is immediately clear that when 
$\lambda>>\sqrt{\theta}$, i.e., in the low temperature limit, 
there is virtually no deviation from the untwisted result
 as summarized in figure \ref{corr}.  This is reassuring 
as it indicates that the implied violation of Pauli's 
principle will have no observable effect at current energies.
Indeed, keeping in mind that $\sqrt{\theta}$ is probably at
the Planck length scale any deviation will only become apparent
at very high temperatures, where the NR limit 
is invalidated.  Note, however, that in contrast to the 
untwisted case the correlation function for fermions does not 
vanish in the limit $r\rightarrow 0$. Thus, there is a 
finite probability that fermions may come very 
close to each other\footnote{It should be noted 
however that this probability is determined by $\theta$ 
and therefore is very small, probably rendering it undetectable.}.
This is most clearly seen from the exchange potential 
$V(r)=-k_BT\log C(r)$ \cite{pathria, kerson} shown 
in figure \ref{exc}. This clearly demonstrates the change 
from a hardcore potential in the untwisted case to a soft 
core potential in the twisted case.  
This may have possible 
implications in astrophysical scenarios, 
although it is dubious that these 
densities are even reachable in this case.  
In any case the assumptions we made here are 
certainly violated at these extreme
conditions and a much more careful analysis is required to 
investigate the high temperature and high density consequences
of the twisted statistics. Another interesting point to note 
from figure \ref{exc} is that the twisted statistics has, 
even at these unrealistic values of $\frac{\theta}{\lambda^2}$,
virtually no effect on the bosonic correlation function at 
short separation probably suggesting that there will be no 
observable effect in Bose-Einstein condensation experiments.  
These results may also have 
interesting consequences for condensed matter systems such 
as the quantum Hall effect where the noncommutative parameter 
is related to the inverse of the magnetic field. 

\setlength{\unitlength}{1mm}
\begin{figure}
\begin{picture}(53,53)
\put(30, 0){\epsfig{file=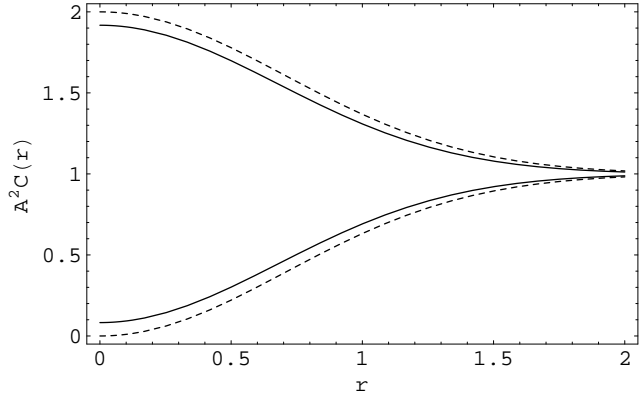, height=53mm}}
\end{picture}
\caption {Two particle correlation function $C(r)$.  
The upper two curves is the bosonic case and the 
lower curves the fermionic case.  The solid line shows the 
twisted result and the dashed line the untwisted case. This 
is shown for a schematic value of $\frac{\theta}{\lambda^2}=0.3$.  
The separation $r$ is measured in units of the thermal length $\lambda$.}
\label{corr}
\end{figure}

\setlength{\unitlength}{1mm}
\begin{figure}
\begin{picture}(53,53)
\put(30, 0){\epsfig{file=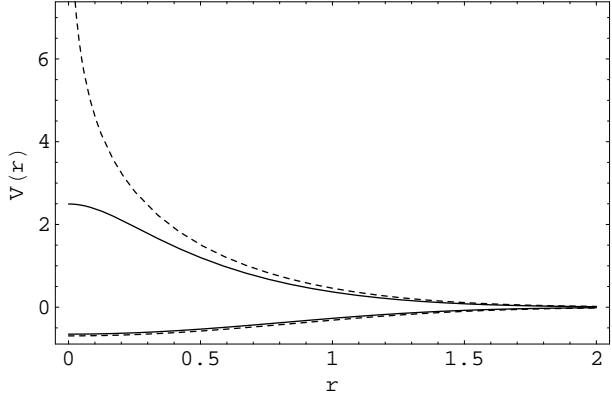, height=53mm}}
\end{picture}
\caption {Exchange potential $V(r)$ measured in units of $k_BT$. 
The irrelevant additive constant has been set zero. 
The upper two curves is 
the fermionic case and the lower curves the bosonic case.  The 
solid line shows the twisted result and the dashed line the untwisted 
case. This is shown for a schematic value of $\frac{\theta}{\lambda^2}=0.3$.  
The separation $r$ is measured in units
 of the thermal length $\lambda$.}
\label{exc}
\end{figure}

\section{Summary}
We have shown that the noncommutative parameter is twisted
Galilean invariant even in presence of spacetime noncommutativity.
This is significant in view of the fact that the usual 
Galilean symmetry is spoiled in presence of spacetime noncommutativity.

\noindent We have then derived the deformed algebra 
of the Schr\"odinger field 
in configuration and momentum space.  
This was done by studying the action of the twisted 
Galilean symmetry on the Schr\"odinger field as 
obtained from a NR reduction of the Klein-Gordon field.
The absence of any spacetime noncommutativity had to be considered
here as otherwise one cannot define a proper NR limit.

\noindent The possible consequences of 
this deformation in terms of a violation of 
the Pauli principle was studied by computing the two 
particle correlation function. From this computation,
one can infer that any possible effect of the twisted statistics
only show up at very high energies, while the effect 
at low energies should be very small, 
consistent with current experimental observations.  
Whether this effect will eventually be 
detectable through some very sensitive experiment is an open 
and enormously challenging question.  

\section*{Appendix: A brief derivation of Wigner-In$\ddot{o}$nu 
group contraction
of Poincar\'{e} group to Galilean group }
Here we summarise the well known Wigner-In$\ddot{o}$nu 
group contraction from Poincar\'{e} to Galilean algebra in order
to highlight some of the subtleties involved, as these have
direct bearings on the issues discussed in section (7.2).

\noindent
To begin with let us consider a particle undergoing uniform acceleration
`$a$', along the $x$ direction, measured in the instantaneous
 rest frame of the particle. A 
typical spacetime Rindler trajectory is given by the hyperbola
\begin{eqnarray}
x^2 - c^2t^2 = {\rho}^2 
\label{8}
\end{eqnarray}
so that the acceleration $A(t)$ w.r.t the fixed observer with
the above associated coordinates ($t , x$) measured at time
$t$ is,
\begin{eqnarray}
A(t) = \frac{dV(t)}{dt} &=& \frac{c^2}{x}\left( \frac{\rho^2}{x^2}
 \right).\nonumber 
\end{eqnarray}
Since the frame $(x , t)$ appearing in eq.(\ref{8}) coincides with that of the
fixed observer at time $t=0$, we must have
\begin{eqnarray}
\Rightarrow a = A(t = 0) &=& \frac{c^2}{\rho}  
\label{9}
\end{eqnarray}
where, $\rho$ is
the distance measured at that instant from the origin.
To take the NR limit, we have to take both 
$c \rightarrow \infty$
and $\rho \rightarrow \infty$ such that $\frac{c^2}{\rho} = a$
is held constant.
For example, the corresponding non-relativistic expression 
$\bar{x}$ for the distance travelled by the particle in time $t$
is obtained by identifying
\begin{eqnarray}
\bar{x} = \lim_{{c \to \infty}  {\rho \to \infty} } 
\left( x - \rho \right) = \frac{1}{2} a t^2
\label{10}
\end{eqnarray}
which reproduces the standard result.

\noindent Now let us consider the 
Lorentz generator along the $x$ direction
$M_{01} = i \left(x_0 \partial_{1} - x_1 \partial_{0}\right)$.
This can be rewritten in terms of $\bar{x}$ using eq.(\ref{10}),
\begin{eqnarray}
M_{01} &=& i c\left( t \frac{\partial}{\partial \bar{x}} + \frac{1}{a}
\left( 1 + \frac{\bar{x}}{\rho} \right)\frac{\partial}{\partial t}\right)
\nonumber \\
&=& cK_1 .
\label{11}
\end{eqnarray}
Note that $K_1$ by itself  does not have any $c$ dependence,
the NR limit of $K_1$ can thus be obtained by
just taking the limit $\rho \to \infty$, which yields
\begin{eqnarray}
K^{NR}_1 = \lim_{\rho \to \infty} K_1 =  
t \frac{\partial}{\partial \bar{x}} + \frac{1}{a}\frac{\partial}{\partial t}.
\label{12}
\end{eqnarray}
Although this vector field indeed generates the integral curve in
the $t$, $\bar{x}$ plane which is a parabola given by eq.(\ref{10}),
it can not be identified with the Galileo boost generator because
\begin{eqnarray}
\left[ K^{NR}_i , K^{NR}_j \right] \sim \left( P_i - P_j\right).
\label{12b}
\end{eqnarray}
The Galilean algebra on the other hand is obtained by taking
the limit $c \to \infty$ of the commutators 
involving boost in the following way:
\begin{eqnarray}
\left[ \bar{K_{1}} , \bar{K_{2}} \right] &=& 
\lim_{c \to \infty} \frac{1}{c^2}
\left[ M_{01} , M_{02} \right] = 
\lim_{c \to \infty} \frac{1}{c^2} M_{12}
= 0 \nonumber \\
\left[ P_1 , \bar{K_{1}} \right] &=& 
\lim_{c \to \infty} \frac{1}{c}
\left[ P_1 , M_{01} \right] = 
\lim_{c \to \infty} \frac{i}{c^2} P_0
= i M \nonumber \\
\left[ \bar{K_{1}} , J \right] &=& 
\lim_{c \to \infty} \frac{1}{c}
\left[ M_{01} , M_{12} \right] = i \bar{K_{2}}
\label{12a}
\end{eqnarray}
\noindent
where $M$ is identified as the mass. The rest of the commutators
have the same form as that of Poincar\'{e} algebra.
This is nothing but the famous Wigner-In$\ddot{o}$nu group 
contraction, demonstrated here in construction of the
Galilean algebra as a suitable limit of the Poincar\'{e} algebra.\\
A simple inspection, at this stage, shows the following form of the
Galileo boost generators 
\begin{eqnarray}
\bar{K_{i}} = K^{(M)}_{i} = 
it\frac{\partial}{\partial \bar{x}_{i}} + M\bar{x}_{i}
\label{12c}
\end{eqnarray}
Clearly the rest of the generators in Galilean algebra 
have the same form as Poincar\'{e} algebra. For completeness 
we enlist the full Galilean algebra in (2 +1) dimension:
\begin{eqnarray}
\left[ K^{(M)}_{i} , K^{(M)}_{j} \right] &=&
\left[ P_i , P_j \right] =  \left[ P_i , H \right] = 
\left[ J , H \right] = 0 \nonumber \\
\left[ P_i , K^{(M)}_{j} \right] &=& i \delta_{ij} M\nonumber \\
\left[ P_i , J \right] &=& i \epsilon_{ij} P_{j}\nonumber \\
\left[ K^{(M)}_{i} , J \right] &=& i \epsilon_{ij} K^{(M)}_{j}\nonumber \\
\left[ P_i , M \right] &=&  \left[ H , M \right] = 
\left[ J , M \right] = \left[ K^{(M)}_{i} , M \right] = 0.
\label{12co}
\end{eqnarray}
Finally note that, here the mass $M$ plays the role of
 central extension of the centrally extended Galilean algebra.

\chapter{Conclusions}
The main goal of this thesis is to study some aspects
of noncommutative quantum mechanics, untwisted and
twisted formulations of noncommutative quantum field
theory and applications.

\noindent There are different settings for noncommutative
field theories. The one that has been most used in all
recent applications is based on the so-called Moyal (star)
product in which for all calculational purposes 
(differentiation, integration, etc), the spacetime
coordinates are treated as ordinary (commutative)
variables and noncommutativity enters into play
in the way in which fields are multiplied.

\noindent We have first given a brief review of the star
product formalism in the thesis. Then we have moved on
to discuss a general method of obtaining both spacetime
and space-space noncommuting structures in various models
in particle mechanics exhibiting reparametrization symmetry.
A change of variables has been derived using 
gauge/reparametrization symmetry transformations
which relates the commuting algebra in the conventional
gauge to a noncommuting algebra in a non-standard gauge.

\noindent The role played by the SW map has been investigated in this work.
The map has been used to obtain an effective $U(1)$ 
gauge invariant Schr\"{o}dinger action upto order $\theta$ 
(starting from a $U(1)_{\star}$ gauge invariant 
noncommutative Schr\"{o}dinger action) 
followed by wave-function and mass renormalization. 
The effect of noncommutativity on the mass parameter appears naturally 
in our analysis. Another interesting point 
that we observe is that the external 
magnetic field has to be static and uniform in order to get 
a canonical form of Schr\"{o}dinger equation upto 
$\theta$-corrected terms, so that a natural 
probabilistic interpretation emerges. 
The Galilean symmetry of the model is 
next investigated where the translation and the rotation 
generators are seen to form a closed Euclidean 
sub-algebra of Galilean algebra. However, the boost 
is not found to be a symmetry of the system, 
even though the condition $\theta^{0i} = 0$ is Galilean invariant. 
Finally, the Hall conductivity is computed and we find that there 
is no $\theta$-correction.

\noindent Having studied this effective commutative quantum mechanical
system upto first order in $\theta$, we set out to enquire whether and how
quantum mechanics of noncommutative systems can be 
carried out for all orders in $\theta$. To that end, 
we have constructed physically equivalent 
families of noncommutative Hamiltonians. 
The implementation of this program to all orders 
in the noncommutative parameter is carried out in the case of a 
free particle and harmonic oscillator moving in a 
constant magnetic field in two dimensions. 
The role played by the SW map has also been
investigated in details.
It is found that this spectrum preserving 
map coincides with the SW map 
in the absence of interactions, but not in the presence 
of interactions. Furthermore, a new possible paradigm for 
noncommutative quantum Hall systems was demonstrated 
in a simple setting.  Here an interacting 
commutative system is traded for a weakly interacting 
noncommutative system, resulting in the same physics 
for the low energy sector. 
This provides a new rational for the introduction 
of noncommutativity in quantum Hall systems.  



\noindent We then present a very simple and elegant approach,
which is somewhat complementary to the point of view presented above,
to understand the quantum Hall system from the noncommutative
framework. The role that interactions play in 
the noncommutative structure
that arises when the relative coordinates 
of two interacting particles are
projected onto the lowest Landau level is discussed in detail. 
It is shown that the interactions
in general renormalize the noncommutative parameter away from the
non-interacting value $\frac{1}{B}$. The effective noncommutative parameter
is in general also angular momentum dependent. The filling
fractions at incompressibilty (which are in general 
renormalized by the interactions) is obtained by
an heuristic argument,
based on the noncommutative coordinates. The results are 
consistent with known results in the case of singular magnetic fields.

\noindent We have then also looked at the twisted formulation
of noncommutative quantum field theory in the context of NR framework.
This is interesting as it has been observed recently that the usual
violation of Lorentz symmetry, arising from the non-transforming
noncommutative constant matrix $\theta^{\mu\nu}$ in
$[\hat{x}^{\mu}, \hat{x}^{\nu}]=i\theta^{\mu\nu}$ can
be restored through the twisted implementation of Lorentz group
{\it{a la}} Drin'feld \cite{drinf}. So the question naturally arises
is regarding its status in NR system, where the relevant symmetry group
is the Galilean group.
Balachandran et.al
\cite{bal} have shown that for this new twisted action,
the Bose and Fermi commutation relations 
of relativistic field gets deformed as
well to render statistics as a super-selected observable. 
In this thesis, we carry out the NR version of the
above analysis. 
We have shown the twisted Galilean invariance of the noncommutative
parameter particularly under rotation, 
even in presence of spacetime noncommutativity, as we find that the
Galileo boost generators become related simply to the linear momentum
generators and thereby remain unaffected by twist.
We also obtained the deformed algebra of the Schr\"odinger field in 
configuration and momentum space by studying the action of the 
twisted Galilean group on the NR limit of the 
Klein-Gordon field, which can eventually be 
extended for a Dirac field as well in a straightforward manner. 
Using this deformed algebra we compute 
the two particle correlation function in a canonical ensemble
to show that the repulsive statistical potential between a pair
of identical (twisted) fermions can saturate to a finite value
at coincident points, thereby violating Pauli's exclusion principle.
However, it can be clearly seen that any 
possible effect is not detectable at present energies.    

\noindent Finally, we would like to mention that the issue
of braided twisted symmetry as discussed in \cite{wessfiore} has not
been investigated in this thesis.





\end{document}